\documentclass[aps,preprint,nofootinbib,eqsecnum]{revtex4}%
\usepackage{amsmath}
\usepackage{graphicx}
\usepackage{amsfonts}
\usepackage{amssymb}%
\providecommand{\U}[1]{\protect\rule{.1in}{.1in}}
\providecommand{\U}[1]{\protect\rule{.1in}{.1in}}
\providecommand{\U}[1]{\protect\rule{.1in}{.1in}}
\providecommand{\U}[1]{\protect\rule{.1in}{.1in}}

\begin{document}
\preprint{{\leftline {USC-07/HEP-B2 \hfill hep-th/0703002}}}
\title[SUSY 2T-physics]{Supersymmetric Field Theory in 2T-physics}
\author{Itzhak Bars and Yueh-Cheng Kuo}
\affiliation{Department of Physics and Astronomy, University of Southern California, Los
Angeles, CA 90089-0484, USA\vspace{1cm}}
\thanks{This work was partially supported by the US Department of Energy, grant number DE-FG03-84ER40168.}

\begin{abstract}
We construct N=1 supersymmetry in 4+2 dimensions compatible with the
theoretical framework of 2T physics field theory and its gauge symmetries. The
fields are arranged into 4+2 dimensional chiral and vector supermultiplets,
and their interactions are uniquely fixed by SUSY and 2T-physics gauge
symmetries. Many 3+1 spacetimes emerge from 4+2 by gauge fixing. Gauge degrees
of freedom are eliminated as one comes down from 4+2 to 3+1 dimensions without
any remnants of Kaluza-Klein modes. In a special gauge, the remaining physical
degrees of freedom, and their interactions, coincide with ordinary N=1
supersymmetric field theory in 3+1 dimensions. In this gauge, SUSY in 4+2 is
interpreted as superconformal symmetry SU(2,2$|$1) in 3+1 dimensions.
Furthermore, the underlying 4+2 structure imposes some interesting
restrictions on the emergent 3+1 SUSY field theory, which could be considered
as part of the predictions of 2T-physics. One of these is the absence of the
troublesome renormalizable CP violating F$\star$F terms. This is good for
curing the strong CP violation problem of QCD. An additional feature is that
the superpotential is required to have no dimensionful parameters. To induce
phase transitions, such as SUSY or electro-weak symmetry breaking, a coupling
to the dilaton is needed. This suggests a common origin of phase transitions
that is driven by the vacuum value of the dilaton, and need to be understood
in a cosmological scenario as part of a unified theory that includes the
coupling of supergravity to matter. Another interesting aspect of the proposed
theory is the possibility to utilize the inherent 2T gauge symmetry to explore
dual versions of the N=1 theory in 3+1 dimensions, such as the minimal
supersymmetric standard model (MSSM) and its duals. This is expected to reveal
non-perturbative aspects of ordinary 1T field theory.

\end{abstract}
\volumeyear{year}
\volumenumber{number}
\issuenumber{number}
\eid{identifier}
\date[]{}
\startpage{1}
\endpage{ }
\maketitle
\tableofcontents

\newpage

\section{2T-physics Field Theory}

Two Time Physics (2T-physics) \cite{2treviews}-\cite{susy2tN1} is a
unification approach for usual one time physics (1T-physics) through higher
dimensions that includes one extra timelike and one extra spacelike
dimensions. This unification is distinctly different from Kaluza-Klein theory
because in the reduction from $d+2$ to $\left(  d-1\right)  +1$ dimensions
there are no Kaluza-Klein towers of states. Instead, in the end result of the
reduction one finds a variety of $\left(  d-1\right)  +1$ emerging spacetimes
embedded in the same $d+2$ spacetime, resulting in a family of 1T-physics
systems in $\left(  d-1\right)  +1$ dimensions, with different dynamics from
each other (i.e. different Hamiltonians), obeying duality type relationships
among themselves.

Furthermore, each 1T system in the family is a holographic image of the
\textit{same parent system} in $d+2$ dimensions, and has hidden symmetries
that reflect the global symmetries of the parent theory. These hidden
symmetries, and the dualities, are reflections of the hidden extra dimensions.
Such properties of 2T-physics are summarized with some examples in Fig.1 of
ref.\cite{2tstandardM}.

The essential ingredient underlying 2T-physics is the basic gauge symmetry
Sp$(2,R)$ acting on phase space $X^{M},P_{M}$ \cite{2treviews}. The role of
Sp$\left(  2,R\right)  $ is most easily explained in the worldline description
of particles. In that context it is a generalization of the 1-parameter gauge
symmetry of worldline $\tau$ reparametrization to a 3-parameter non-Abelian
Sp$\left(  2,R\right)  $ gauge symmetry acting on phase space. This gauge
symmetry \textit{requires} the particle to live in a \textit{target} spacetime
with two timelike directions, so the 2T feature is an outcome of the gauge
symmetry rather than being an input by hand.

The extra 2 parameters in the gauge symmetry are able to remove 2 degrees of
freedom from target spacetime in many possible ways. Through such gauge fixing
one can then find many possible embeddings of \textit{phase space} in $\left(
d-1\right)  +1$ dimensions into $d+2$ dimensions. So, a given $d+2$
dimensional 2T theory descends, through Sp$\left(  2,R\right)  $ gauge fixing,
down to a family of holographic 1T images in $(d-1)+1$ dimensions. All images
are gauge equivalent (or dual) to each other, while each one is also gauge
equivalent to the same parent 2T theory in $d+2$ dimensions. However, the
various images have differing 1T-physics interpretations because of the
different definitions of \textquotedblleft time\textquotedblright\ and
\textquotedblleft Hamiltonian\textquotedblright\ inherent in the \textit{phase
space} embeddings of $\left(  d-1\right)  +1$ in $d+2$ . The rich web of
dualities among the emerging 1T-physics systems is the surprising unifying
power of the 2T-physics approach.

2T-physics includes all cases of particles moving in all possible background
fields \cite{2tbacgrounds}. It also describes particles with spin
\cite{spin2t} or with target space supersymmetry, by appropriate
generalizations of Sp$\left(  2,R\right)  $ \cite{super2t}\cite{2ttwistor}%
\cite{twistorLect}. In all such cases one finds unified families of 1T-physics
systems that emerge from a unifying parent theory directly defined in $d+2$
dimensions. So, 2T-physics appears to be sufficiently general to be able to
accommodate all 1T-physics systems as members of families of holographic
images, with each family representing some higher system with $1+1$ extra dimensions.

Recently, a field theoretic description of 2T physics has been established and
applied to the Standard Model of Particles and Forces \cite{2tstandardM}. In
the field theoretic 4+2 Standard Model (SM), the type of phenomena such as
hidden symmetries, duality, holography and emergent spacetimes are also
present owing to a newly discovered 2T gauge symmetry in field theory which is
actually a consequence of the gauge symmetry Sp$\left(  2,R\right)  $\ on the
worldline \cite{2tbrst2006}\cite{2tstandardM}. For the time being only one of
the field theoretic images, namely the \textquotedblleft massless relativistic
particle\textquotedblright\ gauge noted in Fig.1 of ref.\cite{2tstandardM}
(see footnote (\ref{embedding})), has been studied in the field theory
context. It is this $3+1$ holographic image of the $4+2$ SM that coincides
with the well known $3+1$ SM.

The underlying 4+2 structure imposes some interesting restrictions on the
emergent 3+1 Standard Model, which could be considered as part of the
predictions of 2T-physics. Some attractive features include a new solution of
the long standing strong CP problem in QCD without an axion, and novel ideas
on the origins of mass generation as briefly reiterated below.

The goal of the present paper is to formulate the general supersymmetric
version of 2T-physics field theory in $4+2$ dimensions, for fields of spins
$0,\frac{1}{2},1,$ with $N=1$ supersymmetry (SUSY). This will be a starting
point for physical applications in the form of the supersymmetric version of
the SM in $4+2$ dimensions, as well as for generalizations to higher $N=2,4,8$
supersymmetric 2T-physics field theory, which will be presented in future
papers. A summary of our results for $N=1$ SUSY has appeared as short letter
\cite{susy2tN1}.

In the following, we briefly summarize the essential features of 2T field
theory which will be the structure on which we will impose $N=1$ supersymmetry
in the coming sections.

The field theory in 4+2 dimensions with fields of spins $0,\frac{1}{2},1,$
describes a set of SO$\left(  4,2\right)  $ vectors $A_{M}^{a}\left(
X\right)  $ labeled with $M=$ SO$\left(  4,2\right)  $ vector, and $a=$ the
adjoint representation of a Yang-Mills gauge group $G$ (for example,
$G=$SU$\left(  3\right)  \times$SU$\left(  2\right)  \times$U$\left(
1\right)  $ for the Standard Model, $G=$SO$\left(  10\right)  $ for Grand
Unification); scalars $H_{i}\left(  X\right)  ,$ labeled by an internal
symmetry index $i=1,2,\cdots$ (a collection of irreducible representations of
$G$); left or right handed spinors $\psi_{L\alpha}^{I}\left(  X\right)
,\psi_{R\dot{\alpha}}^{\tilde{I}}\left(  X\right)  $ in the $4,4^{\ast}$
representations of SU$\left(  2,2\right)  =$SO$\left(  4,2\right)  ,$ labeled
with $\alpha=1,2,3,4,$ and $\dot{\alpha}=1,2,3,4,$ and internal symmetry
indices $I=1,2,\cdots$ and $\tilde{I}=1,2,\cdots$ (again, a collection of
irreducible representations of $G$).

The generic 2T-physics Lagrangian has the form of a Yang-Mills theory in $4+2$
dimensions ($G$-covariant derivatives). But it contains $4+2$ space-time
features shown explicitly in the Lagrangian below, which are needed to impose
the underlying Sp$\left(  2,R\right)  $ gauge symmetry and the related
2T-physics gauge symmetries.

There is no space here to explain the origin of the 2T-physics gauge
symmetries in field theory that are given in \cite{2tbrst2006}%
\cite{2tstandardM}. But we emphasize the basic important fact that the
equations of motion that follow from the Lagrangian below impose the
Sp$\left(  2,R\right)  $ gauge singlet conditions $X^{2}=X\cdot P=P^{2}=0$ (or
OSp$\left(  n|2\right)  $ gauge singlet conditions for a field with spin
$n/2$), but now including interactions \cite{2tstandardM}. The field theory
Lagrangian with these properties has the general form
\begin{equation}
L=\left[
\begin{array}
[c]{l}%
\delta\left(  X^{2}\right)  \left\{  -D_{M}H^{i\dagger}D^{M}H_{i}\right\}
+2~\delta^{\prime}\left(  X^{2}\right)  ~H^{i\dagger}H_{i}\\
+\delta\left(  X^{2}\right)  \left\{
\begin{array}
[c]{l}%
\frac{i}{2}\left(  \overline{\psi_{L}}^{I}X\bar{D}\psi_{IL}+\overline{\psi
_{L}}^{I}\overleftarrow{D}\bar{X}\psi_{IL}\right) \\
-\frac{i}{2}\left(  \overline{\psi_{R}}^{\tilde{I}}\bar{X}D\psi_{\tilde{I}%
R}+\overline{\psi_{R}}^{\tilde{I}}\overleftarrow{\bar{D}}X\psi_{\tilde{I}%
R}\right)
\end{array}
\right\} \\
+\delta\left(  X^{2}\right)  \left\{  y_{I}^{i\tilde{I}}\overline{\psi_{L}%
}^{I}X\psi_{\tilde{I}R}H_{i}+\left(  y_{I}^{i\tilde{I}}\right)  ^{\ast}H^{\ast
i}\overline{\psi_{R}}^{\tilde{I}}\bar{X}\psi_{IL}\right\} \\
+\delta\left(  X^{2}\right)  \left\{  -\frac{1}{4}F_{MN}^{a}F_{a}%
^{MN}-V\left(  H,H^{\ast},\Phi\right)  \right\} \\
-\frac{1}{2}\delta\left(  X^{2}\right)  ~\partial_{M}\Phi\partial^{M}%
\Phi+\delta^{\prime}\left(  X^{2}\right)  ~\Phi^{2}%
\end{array}
\right]  \label{action}%
\end{equation}
The left arrow on $\overleftarrow{D}_{M}$ means that the covariant derivative
acts on the field on its left $\overline{\psi_{L}}\overleftarrow{D}_{M}\equiv
D_{M}\overline{\psi_{L}}$. The distinctive space-time features in 4+2
dimensions include the delta function $\delta\left(  X^{2}\right)  $ and its
derivative $\delta^{\prime}\left(  X^{2}\right)  $ that impose $X^{2}%
=X^{M}X_{M}=0$ (see footnote (\ref{delta})), the kinetic terms of fermions
that include the factors $X\bar{D},\bar{X}D,$ and Yukawa couplings
proportional to $y_{I}^{i\tilde{I}},y_{I}^{i\tilde{I}}$ that include the
factors $X$ or $\bar{X},$ where $X\equiv\Gamma^{M}X_{M},$ $\bar{D}=\bar
{\Gamma}^{M}D_{M}$ etc., with $4\times4$ gamma matrices $\Gamma^{M}%
,\bar{\Gamma}^{M}$ in the $4$,$4^{\ast}$ spinor bases of SU$\left(
2,2\right)  $=SO$\left(  4,2\right)  .$ Our notation for gamma matrices for
SO$\left(  4,2\right)  =$SU$\left(  2,2\right)  $ is given in Appendix
(\ref{A}).

This Lagrangian is not invariant under translation of $X^{M},$ but is
invariant under the spacetime rotations SO$\left(  4,2\right)  .$ In fact, it
has precisely the right space-time, and gauge invariance, properties for the
$4+2$ field theory to yield the usual $3+1$ field theory$.$ The reduction from
$4+2$ dimensions $X^{M},$ to $3+1$ dimensions $x^{\mu}$ is obtained via gauge
fixing (see footnote (\ref{embedding})). The emergent $3+1$ field theory is
invariant under translations of $x^{\mu}$ and Lorentz transformations
SO$\left(  3,1\right)  $. These Poincar\'{e} symmetries are included in
SO$\left(  4,2\right)  $ that takes the non-linear form of conformal
transformations in the emergent $3+1$ dimensional space-time $x^{\mu}$. The
emergent $3+1$ theory contains just the right fields as functions of $x^{\mu}%
$: all extra degrees of freedom disappear without leaving behind any
Kaluza-Klein type modes or extra components of the vector and spinor fields in
the extra 1+1 dimensions. Furthermore, the emergent field theory has the usual
kinetic terms and Yukawa couplings in $3+1$ dimensional Minkowski space
\cite{2tstandardM}.

As in the last line of the Lagrangian, one may also include an additional
SO$\left(  4,2\right)  $ scalar, the dilaton $\Phi\left(  X\right)  ,$
classified as a singlet under the group $G.$ The dilaton is not optional if
the action is written in $d+2$ dimensions (see \cite{2tstandardM}), as it
appears in overall factors $\Phi^{\frac{2\left(  d-4\right)  }{d-2}}%
,\Phi^{-\frac{d-4}{d-2}}$ multiplying the Yang-Mills kinetic term and Yukawa
terms respectively, in order to achieve the 2T-gauge symmetry of the action.
In $4+2$ dimensions ($d=4$) these factors reduce to $1,$ but the dilaton can
still couple to the scalars $H$ in the potential $V\left(  H,H^{\ast}%
,\Phi\right)  .$

The 2T-physics field theory above is applied to construct the Standard Model
in 4+2 dimensions by choosing the gauge group $G=$SU$\left(  3\right)  \times
$SU$\left(  2\right)  \times$U$\left(  1\right)  $ and including the usual
matter representations for the Higgs, quarks and leptons (including right
handed neutrinos in singlets of $G$), but now as fields in $4+2$ dimensions.
As explained in \cite{2tstandardM} this theory descends to the usual Standard
Model in $3+1$ dimensions.

When we apply the $4+2$ approach to construct the Standard Model, almost all
of the usual terms of the $3+1$ dimensional Standard Model emerge from the
$4+2$ field theory above, except for two notable exceptions that play an
important physical role. Namely,

\begin{itemize}
\item There is no way to generate a\textit{ renormalizable} term in the
emergent $3+1$ theory that is analogous to the P and CP-violating term $\theta
F_{\mu\nu}F_{\lambda\sigma}\varepsilon^{\mu\nu\lambda\sigma}$ that is possible
in a purely 1T-physics approach in $3+1$ dimensions\footnote{Actually there
appears as if there would be a topological term of the form $\int
d^{6}X~\varepsilon^{M_{1}M_{2}M_{3}M_{4}M_{5}M_{6}}Tr\left(  F_{M_{1}M_{2}%
}F_{M_{3}M_{4}}F_{M_{5}M_{6}}\right)  $ whose density is a total divergence
for any Yang-Mills gauge group $G.$ Such a term could descend to $3+1$
dimensions $\theta F_{\mu\nu}F_{\lambda\sigma}\varepsilon^{\mu\nu\lambda
\sigma}$ with an effective $\theta\sim F^{+^{\prime}-^{\prime}}.$ However, it
can be shown that this $\theta$ is 2T gauge dependent and is gauge fixed to
zero in the process of descending from $4+2$ to $3+1$ dimensions. This and
other possible sources of the $\theta$ term are discussed and eliminated in
\cite{2tstandardM}. In this sense, the 2T gauge symmetry plays a similar role
to the Peccei-Quinn symmetry in eliminating the topological term. But one must
realize that the 2T gauge symmetry is introduced for other more fundamental
reasons and also it is not a global symmetry. Hence, unlike the Peccei-Quinn
symmetry it does not lead to an axion.}. The absence of $\theta$ in the
emergent Standard Model is due to the fact that the Levi-Civita symbol in
$4+2$ dimensions has 6 indices rather than $4,$ and also due to the
combination of 2T gauge symmetry as well as Yang-Mills gauge symmetry. The
absence of this CP-violating term in 2T-physics is of crucial importance in
the axionless resolution of the strong CP violation problem of QCD
\cite{2tstandardM}.

\item The 2T-gauge symmetry requires the potential $V\left(  H,H^{\ast}%
,\Phi\right)  $ to be purely quartic, i.e. no mass terms are permitted. Then
the emergent 3+1 theory cannot have mass terms for the scalars, and is
automatically invariant under scale transformations. This makes the mass
generation with the Higgs mechanism less straightforward since the tachyonic
mass term is not allowed. However by taking the Higgs potential of the form
$V\left(  \Phi,H\right)  =\frac{\lambda}{4}\left(  H^{\dagger}H-\alpha^{2}%
\Phi^{2}\right)  ^{2}$ we obtain the breaking of the electroweak symmetry by
the Higgs doublet $\langle H\rangle$ driven by the vacuum expectation value of
the dilaton $\langle\Phi\rangle$, thus relating the the two phase transitions
to each other. In this way the 4+2 formulation of the Standard Model provides
an appealing deeper physical basis for mass\footnote{As argued in
\cite{2tstandardM}, the dilaton driven electroweak phase transition makes a
lot more sense conceptually than the usual approach in which the electroweak
phase transition is an isolated phenomenon. This is because the Higgs vacuum
expectation value fills all space everywhere in the universe. This is a hard
concept to swallow without relating it to the evolution of the universe, which
then requires the participation of gravity. In the 2T-physics version of the
SM, the Higgs $\langle H\rangle$ has to be driven by the dilaton $\langle
\Phi\rangle$ which is a member of the gravity multiplet, so an essential part
of the physics of the Standard Model becomes intimately related to the physics
of gravity and all of its other consequences. In particular a relation is
established to other phase transitions that are expected to be dilaton driven
in the evolution of the universe, such as the vacuum selection process in
string theory, and perhaps even to inflation that is driven by a scalar field
which could be the dilaton.\label{dilatondrive}}.
\end{itemize}

Having established that 2T physics field theory introduces new
phenomenologically relevant constraints, it would be of great interest to find
out whether the SUSY version is also constrained in phenomenologically
significant ways. This is especially relevant in view of the upcoming
experimental activities at the LHC starting in 2008. It would be interesting
to formulate experimental signatures that could distinguish 2T-physics
versions of SUSY from others, due to some extra constraints rooted in the
structures of $4+2$ dimensions. The first step towards this goal is the
formulation of SUSY in 2T-physics field theory which we present in this paper.
We will establish the transformation rules for 2T-physics $N=1$ SUSY in $4+2$
dimensions, which are different than a straightforward higher dimensional
SUSY, and will build the general SUSY Lagrangian for fields with spins
$0,\frac{1}{2},1,$ with any Yang-Mills gauge group $G,$ and with any representations.

The plan of the paper is as follows. Section (\ref{summary}) gives a quick
outline of the paper for the reader who is interested in seeing the results
without the technical details. So, in section (\ref{summary}) we give a
summary of our results for the general $N=1$ supersymmetric action for a
coupled system of spin $0,\frac{1}{2},1$ fields. We give the SUSY
transformation laws that have many new features and derive the conserved SUSY
current for the fully interacting system. These fields are arranged into $N=1$
chiral and vector multiplets of SUSY in $4+2$ dimensions, consistent with the
gauge symmetries of 2T-physics, and with the gauge symmetries of a Yang-Mills
group. In section (\ref{chiral}), we deal with the chiral multiplet by itself,
discuss the SUSY symmetry in detail and derive the conserved SUSY current. In
section (\ref{vecto}), we discuss the vector multiplet by itself in detail. In
section (\ref{full}), we couple chiral multiplets with vector multiplets and
find the unique action, supersymmetry transformation, and conserved current,
justifying the outline in section (\ref{summary}). Finally in section
(\ref{conclude}) we conclude with some comments and point out future directions.

In Appendix (\ref{A}) we provide technical details on gamma matrices for
SO$\left(  4,2\right)  $. In Appendix (\ref{fierzI}) we derive some Fierz
identities that are used in the proof of SUSY including interactions. In
Appendix (\ref{B}) we discuss the closure of the SUSY algebra into the
supergroup SU$\left(  2,2|1\right)  $ when the fields are on-shell, and into a
larger algebra when the fields are off-shell.

\section{N=1 SUSY in 2T-Physics and Summary of Results \label{summary}}

In this section we will provide a summary of our results. In the following
sections, we will show how each piece in the action and the SUSY
transformations arise step by step.

To some extent the well known $3+1$ SUSY structures are a guide toward the
$4+2$ SUSY structures, since after all the $3+1$ chiral supermultiplet and
vector supermultiplet should emerge as the end result of the 2T-physics gauge
fixing. Therefore, the spin $0,\frac{1}{2},1$ fields are members of the chiral
and vector supermultiplets in the $4+2$ 2T-physics SUSY theory.

The chiral supermultiplet $\left(  \varphi,\psi_{L},F\right)  _{i}$ in $4+2$
dimensions contains a set of SO$\left(  4,2\right)  $ scalars $\varphi
_{i}\left(  X\right)  ,$ left handed spinors $\psi_{iL\alpha}\left(  X\right)
$ in the $4$ representation of SU$\left(  2,2\right)  =$SO$\left(  4,2\right)
,$ labeled with $\alpha=1,2,3,4,$ and auxiliary complex scalar fields
$F_{i}\left(  X\right)  $, all labeled by an internal symmetry index
$i=1,2,3,\cdots$ of a gauge symmetry group $G$. The internal symmetry index
$i$ is used here generically to denote any collection of \textit{several}
irreducible representations of $G$.

The vector supermultiplet $\left(  A_{M},\lambda_{L},B\right)  ^{a}$\ contains
fields that carry SO$\left(  4,2\right)  $ spacetime indices required by their
spin. Thus the spin-1 $A_{M}^{a}\left(  X\right)  $ is the Yang-Mills gauge
field, the spin-$\frac{1}{2}$ $\lambda_{\alpha L}^{a}\left(  X\right)  $ is
the gaugino and the spin-0 $B^{a}\left(  X\right)  $ is the auxiliary
field\footnote{The auxiliary field is usually called the $D$-term in $3+1$
SUSY, but we use here the letter $B$ to avoid confusion with the symbol for
covariant derivative $D.$}. They are all labeled by $a$ which belongs to the
adjoint representation of the gauge symmetry group $G$.

The SUSY transformations of the chiral and vector multiplets that leave the
action invariant will be discussed below after we make some remarks about the
significance of various terms in the action (\ref{actionsusy}). To simplify
our notation we will suppress the group $G$ indices $i,a$ in parts of the
discussion in this paper and make it explicit when it is necessary for clarity.

\subsection{Lagrangian}

In what follows, we use mostly left-handed spinors, but also find it
convenient at times to use right handed spinors as the charge conjugates of
left handed ones. The left handed spinor $\psi_{L\alpha}\left(  X\right)  ,$
in the $4$ representation of SU$\left(  2,2\right)  ,$ is labeled with
$\alpha=1,2,3,4$\ while the right handed spinor $\psi_{R\dot{\alpha}}\left(
X\right)  ,$ in the $\bar{4}$ representation of SU$\left(  2,2\right)  ,$ is
labeled with $\dot{\alpha}=1,2,3,4.$ One may also construct an 8-component
spinor of SO$\left(  4,2\right)  $ with a Majorana condition such that
$\psi_{L}$ together with $\psi_{R}$ make up the 8 components of $\psi=\left(
\genfrac{}{}{0pt}{}{\psi_{L}}{\psi_{R}}%
\right)  $ and because of the Majorana condition, $\psi_{R}$ and $\psi_{L}$
are related to each other. One could rewrite all right-handed spinors as
left-handed ones by charge conjugation which is given by
\begin{equation}
\psi_{R}\equiv C\overline{\psi_{L}}^{T}=C\eta^{T}\left(  \psi_{L}\right)
^{\ast},\;\;\text{or\ \ \ }\overline{\psi_{L}}=-\left(  \psi_{R}\right)
^{T}C. \label{cc}%
\end{equation}
Using these definitions we can also write the following relations that are
equivalent to Eq.(\ref{cc})
\begin{equation}
\psi_{L}=-C\overline{\psi_{R}}^{T}\text{, \ or\ }\overline{\psi_{R}}=\left(
\psi_{L}\right)  ^{T}C. \label{cc2}%
\end{equation}
Our SO$\left(  4,2\right)  $ gamma matrix notation in the Weyl basis, which
includes explicit forms of the antisymmetric charge conjugation matrix
$C=\tau_{1}\times\sigma_{2},$ and the symmetric SU$\left(  2,2\right)  $
metric $\eta=-i\tau_{1}\times1$ used to construct the contravariant
$\overline{\psi_{L}}^{\beta}=\left(  \left(  \psi_{L}\right)  ^{\dagger}%
\eta\right)  ^{\beta}=\left(  \psi_{L}^{\ast}\right)  _{\dot{\alpha}}%
\eta^{\dot{\alpha}\beta},$ are explained in detail in Appendix (\ref{A}).

To satisfy the gauge symmetries of 2T-physics discussed in \cite{2tstandardM},
each one of the spin $0,\frac{1}{2},1$ fields can occur only in the form of
the Lagrangian of Eq.(\ref{action}). On this structure we now impose SUSY
whose details are described in the following sections. It turns out that the
general theory of the $N=1$ chiral multiplet coupled to the $N=1$ vector
multiplet gets organized as follows%
\begin{equation}
L=L_{chiral}+L_{vector}+L_{int}+L_{dilaton} \label{actionsusy}%
\end{equation}
The vector multiplet $\left(  A_{M},\lambda_{L},B\right)  ^{a}$ with its self
interactions is described by%
\begin{equation}
L_{vector}=\delta\left(  X^{2}\right)  \left\{  -\frac{1}{4}F_{MN}^{a}%
F_{a}^{MN}+\frac{i}{2}\left[  \overline{\lambda_{L}}^{a}X\bar{D}\lambda
_{aL}+\overline{\lambda_{L}}^{a}\overleftarrow{D}\bar{X}\lambda_{aL}\right]
+\frac{1}{2}B^{a}B_{a}\right\}  \label{Lvec}%
\end{equation}
The chiral multiplet $\left(  \varphi,\psi_{L},F\right)  _{i}$, with its self
interactions are described by
\begin{align}
L_{chiral}  &  =\delta\left(  X^{2}\right)  \left\{
\begin{array}
[c]{l}%
-D_{M}\varphi^{i\dagger}D^{M}\varphi_{i}+\frac{i}{2}\left(  \overline{\psi
_{L}}^{i}X\bar{D}\psi_{iL}+\overline{\psi_{L}}^{i}\overleftarrow{D}\bar{X}%
\psi_{iL}\right)  +F^{\dagger i}F_{i}\\
+\left[  \frac{\partial W}{\partial\varphi_{i}}F_{i}-\frac{i}{2}\psi
_{iL}\left(  C\bar{X}\right)  \psi_{jL}\frac{\partial^{2}W}{\partial
\varphi_{i}\partial\varphi_{j}}\right]  +h.c.
\end{array}
\right\} \label{Lchi}\\
&  +2~\delta^{\prime}\left(  X^{2}\right)  ~\varphi^{i\dagger}\varphi
_{i}\nonumber
\end{align}
Some of the interactions of the chiral multiplet with the gauge multiplet
already appear through the gauge covariant derivatives $D^{M}\varphi_{i}$ and
$D^{M}\psi_{iL}$. Additional interactions of the vector and chiral multiplets
occur also through the auxiliary fields $B^{a}$ and the gaugino $\lambda
_{L}^{a}$ as follows%
\begin{equation}
L_{int}=\delta\left(  X^{2}\right)  \left\{  \alpha\varphi^{\dagger i}\left(
t_{a}\right)  _{i}^{~j}\varphi_{j}B^{a}+\beta\varphi^{\dagger i}\left(
t_{a}\right)  _{i}^{~j}\left(  \psi_{jL}\right)  ^{T}\left(  C\bar{X}\right)
\lambda_{L}^{a}\right\}  +~h.c. \label{Lint}%
\end{equation}
where $\alpha,\beta$ will be uniquely determined by SUSY. Finally a sketchy
description of the dilaton is given by%
\begin{equation}
L_{dilaton}=\left\{
\begin{array}
[c]{c}%
-\frac{1}{2}\delta\left(  X^{2}\right)  ~\partial_{M}\Phi\partial^{M}%
\Phi+\delta^{\prime}\left(  X^{2}\right)  \Phi^{2}+\text{superpartners of
}\Phi\\
+\delta\left(  X^{2}\right)  \left\{  \xi_{a}B^{a}\Phi^{2}+V\left(
\Phi,\varphi\right)  \right\}
\end{array}
\right\}  \label{Ldil}%
\end{equation}
We note the following points on the structure of the Lagrangian

$\left(  1\right)  $ The $W\left(  \varphi\right)  $ in $L_{chiral}$ is the
holomorphic superpotential consisting of any combination of $G$-invariant
\textit{cubic} polynomials constructed from the $\varphi_{i}$ (and excludes
the $\varphi^{i\dagger}$)%
\begin{equation}
W\left(  \varphi\right)  =y^{ijk}\varphi_{i}\varphi_{j}\varphi_{k}%
,\;y^{ijk}\text{=constants compatible with }G\text{ symmetry.}
\label{superpotential}%
\end{equation}
The purely cubic form of $W\left(  \varphi\right)  $ leads to a purely quartic
potential energy for the scalars after the auxiliary fields $F_{i}$ and
$B^{a}$ are eliminated through their equations of motion. A purely quartic
potential is required by the 2T gauge symmetry even without SUSY.

$\left(  2\right)  $ The $\bar{X}$ in the Yukawa couplings $\left(  \psi
_{iL}\right)  ^{T}\left(  C\bar{X}\right)  \psi_{jL}\frac{\partial^{2}%
W}{\partial\varphi_{i}\partial\varphi_{j}}$ or $\beta\left(  \varphi^{\dagger
}t^{a}\psi_{L}\right)  ^{T}\left(  C\bar{X}\right)  \lambda_{aL}$ is
consistent with the SU$(2,2)$=SO$\left(  4,2\right)  $ group theory property
$\left(  4\times4\right)  _{antisymmetric}=6$: namely, two left handed
fermions must be coupled to the vector $X^{M}$ to give an SO$\left(
4,2\right)  $ invariant. The $\bar{X}$ insertion is also required for the
2T-gauge invariance of the Yukawa couplings, as discussed in
\cite{2tstandardM}.

$\left(  3\right)  $ SUSY requires that the dimensionless constants
$\alpha,\beta$ are all determined in terms of the gauge coupling constants $g$
for each subgroup in $G$ as follows\footnote{There is a separate gauge
coupling $g$ for each subgroup in $G,$ so there are separate $\alpha,\beta$
proportional to the $g$ for each such subgroup.}
\begin{equation}
\alpha=g,\;\beta=\sqrt{2}g,\;
\end{equation}
The only parameters that are not fixed by the symmetries are the Yang-Mills
coupling constants $g,$ and the Yukawa couplings $y^{ijk}$ which are
restricted by invariance under $G$-symmetry, namely
\begin{equation}
\frac{\partial W}{\partial\varphi_{i}}\left(  t_{a}\varphi\right)  _{i}=0.
\label{Gsymm}%
\end{equation}

$\left(  4\right)  $ As in the non-supersymmetric case discussed in the
previous section, in the SUSY 2T-physics theory there is no way to write down
a term in $4+2$ dimensions that will reduce to the CP-violating term $\theta
F_{\mu\nu}F_{\lambda\sigma}\varepsilon^{\mu\nu\lambda\sigma}$ that is possible
in $3+1$ dimensions in the context of purely 1T-physics. The absence of this
CP-violating term is of crucial importance in the axionless resolution of the
strong CP violation problem of QCD suggested in \cite{2tstandardM}, and which
generalizes to the supersymmetric case in this paper.

$\left(  5\right)  $ Now we turn to the dilaton term $L_{dilaton}.$ As
mentioned above, the superpotential $W\left(  \varphi\right)  $ is restricted
by supersymmetry to be purely cubic in $\varphi$. So for driving the
spontaneous breakdown of the $G$ symmetry the same way as in the
non-supersymmetric case (as in footnote (\ref{dilatondrive})), as well as for
inducing soft supersymmetry breaking through the Fayet-Illiopoulos type of
term $\xi_{a}\Phi^{2}B^{a}$, it would be desirable to couple the dilaton
$\Phi$ to the chiral and vector multiplets by having interactions of the form
$V\left(  \Phi,\varphi\right)  $ and $\xi_{a}\neq0$ for U$\left(  1\right)  $
gauge subgroups. However, we have not yet included the superpartners of the
dilaton because this is still under development in the 2T-physics context, so
we are not yet in a position to discuss the SUSY constraints on the desired
couplings. So in this paper we will not be able to comment in detail on the
dilaton-driven electroweak or SUSY phase transition. However, we point out
that in agreement with footnote (\ref{dilatondrive}) this is again a
consistent message from 2T-physics, namely that the physics of the Standard
Model, in particular the electroweak phase transition that generates mass, is
not decoupled from the physics of the gravitational interactions in a complete
unified theory of all the forces. The full theory may be attained by further
pursuing these hints provided by the 2T-physics formulation of the Standard Model.

\subsection{SUSY transformations}

We now summarize the properties of the SUSY transformations for the chiral and
vector multiplets that leave invariant the action $S=\int d^{6}xL$ based on
the above Lagrangian. The supersymmetry transformation for the chiral
multiplet is (in the following $\varepsilon_{R}\equiv C\overline
{\varepsilon_{L}}^{T}$ and $\overline{\varepsilon_{R}}=\left(  \varepsilon
_{L}\right)  ^{T}C,$ and similarly for $\lambda_{R}$ or $\psi_{R},$ as in
Eqs.(\ref{cc},\ref{cc2}))%

\begin{equation}
\delta_{\varepsilon}\varphi_{i}=\left\{  \overline{\varepsilon_{R}}\bar{X}%
\psi_{iL}+X^{2}\left[
\begin{array}
[c]{c}%
-\frac{1}{2}\overline{\varepsilon_{R}}\bar{D}\psi_{iL}+\frac{1}{2}%
\frac{\partial^{2}W^{\ast}}{\partial\varphi^{\dagger i}\partial\varphi
^{\dagger j}}\overline{\psi_{L}}^{j}\varepsilon_{L}\\
-\frac{ig}{2\sqrt{2}}\left(  \overline{\varepsilon_{L}}\lambda_{L}%
^{a}+\overline{\lambda_{L}}^{a}\varepsilon_{L}\right)  \left(  t_{a}%
\varphi\right)  _{i}%
\end{array}
\right]  \right\}  \label{delphi}%
\end{equation}%
\begin{align}
\delta_{\varepsilon}\psi_{iL}  &  =i\left(  D_{M}\varphi_{i}\right)  \left(
\Gamma^{M}\varepsilon_{R}\right)  -iF_{i}\varepsilon_{L}\;\;\\
\delta_{\varepsilon}\overline{\psi_{L}}^{i}  &  =i\overline{\varepsilon_{R}%
}\bar{\Gamma}^{M}\left(  D_{M}\varphi\right)  ^{\dagger i}+i\overline
{\varepsilon_{L}}F^{\dagger i}\\
\delta_{\varepsilon}F_{i}  &  =\overline{\varepsilon_{L}}\left[  X\bar
{D}-\left(  X\cdot D+2\right)  \right]  \psi_{iL}-i\sqrt{2}g\left(
\overline{\varepsilon_{L}}X\lambda_{R}^{a}\right)  \left(  t_{a}%
\varphi\right)  _{i}. \label{delFfull}%
\end{align}
The supersymmetry transformation for the vector multiplet is%

\begin{equation}
\delta_{\varepsilon}A_{M}^{a}=\left\{  -\frac{1}{\sqrt{2}}\overline
{\varepsilon_{L}}\Gamma_{M}\bar{X}\lambda_{L}^{a}+X^{2}\left[
\begin{array}
[c]{c}%
\frac{1}{2\sqrt{2}}~\overline{\varepsilon_{L}}\Gamma_{MN}\left(  D^{N}%
\lambda_{L}^{a}\right) \\
-\frac{ig}{4}\left(  \overline{\varepsilon_{L}}\Gamma_{M}\psi_{R}^{i}\right)
\left(  t^{a}\varphi\right)  _{i}%
\end{array}
\right]  \right\}  +h.c. \label{delAa}%
\end{equation}%
\begin{align}
\delta_{\varepsilon}\lambda_{L}^{a}  &  =i\frac{1}{2\sqrt{2}}F_{MN}^{a}\left(
\Gamma^{MN}\varepsilon_{L}\right)  -\frac{1}{\sqrt{2}}B^{a}\varepsilon_{L}\\
\delta_{\varepsilon}\overline{\lambda_{L}}^{a}  &  =i\frac{1}{2\sqrt{2}%
}\left(  \overline{\varepsilon_{L}}\Gamma^{MN}\right)  F_{MN}^{a}-\frac
{1}{\sqrt{2}}\overline{\varepsilon_{L}}B\\
\delta_{\varepsilon}B^{a}  &  =\frac{i}{\sqrt{2}}\overline{\varepsilon_{L}%
}\left[  X\bar{D}-\left(  X\cdot D+2\right)  \right]  \lambda_{L}^{a}+h.c.
\label{delBfull}%
\end{align}
These SUSY transformations have some parallels to naive SUSY transformations
that one may attempt to write down as a direct generalization from $3+1$ to
$4+2$ dimensions. However, there are many features that are completely
different\footnote{Once we notice the parallels, part of the structure can be
understood from SU$\left(  2,2\right)  $ group theory. For example, consider
the gamma matrix structures $\bar{X}$, etc$.$ sandwiched between fermions,
which are absent in $3+1$ dimensions. $\overline{\varepsilon_{R}}\bar{X}%
\psi_{iL}$ is an SU$\left(  2,2\right)  $ scalar since $\overline
{\varepsilon_{R}}$ and $\psi_{iL}$ are both in the $4$ representation of
SU$\left(  2,2\right)  ,$ and the product $4\times4=6+10$ shows that when we
couple the $6$ to the SO$\left(  4,2\right)  =$SU$\left(  2,2\right)  $ vector
$X^{M}$ through the gamma matrices, we obtain a scalar.}. These include the
insertions that involve $X$ $=X^{M}\Gamma_{M}$ or $\bar{X}$ $=X^{M}\bar
{\Gamma}_{M},$ the terms proportional to $X^{2},$ and the terms proportional
to derivative terms involving $\left(  X\cdot D+2\right)  .$ These are
\textit{off-shell} SUSY transformations that include interactions and leave
invariant the off-shell action. The free field limit of our transformations
(i.e. $W=0$ and $g=0$) taken on shell (i.e. terms proportional to $X^{2}$ and
$\left(  X\cdot D+2\right)  $ set to zero) agrees with previous work which was
considered for on-shell free fields without an action principle \cite{ferrara}.

Despite all of the changes compared to naive SUSY, this SUSY symmetry provides
a representation of the supergroup SU$\left(  2,2|1\right)  $. This is
signaled by the fact that all terms are covariant under the bosonic subgroup
SU$\left(  2,2\right)  ,$ while the complex fermionic parameter $\varepsilon
_{L}$ and its conjugate $\overline{\varepsilon_{L}}$ are in the $4,4^{\ast}$
representations of SU$\left(  2,2\right)  $, as would be expected for
SU$\left(  2,2|1\right)  .$ 

The closure of these SUSY transformations is discussed in Appendix (\ref{B})
in the case of the pure chiral multiplet (i.e. gauge coupling $g=0$). The
commutator of two SUSY transformations closes to the bosonic part SU$\left(
2,2\right)  \times$U$\left(  1\right)  \subset$ SU$\left(  2,2|1\right)  $
when the fields are on-shell. More generally, when the fields are off-shell
the closure includes also a U$\left(  1\right)  $ outside of SU$\left(
2,2|1\right)  $ and a 2T-physics gauge transformation, both of which are also
gauge symmetries of the action.

When reduced to $3+1$ dimensions by choosing a gauge as prescribed in footnote
(\ref{embedding}), the SU$\left(  2,2|1\right)  $ transformations give
non-linear off-shell realization of superconformal symmetry in $3+1$ dimensions.

\subsection{Conserved supercurrent}

The Lagrangian in Eq.(\ref{actionsusy}) transforms into a total divergence
under the SUSY transformations (in the absence of the dilaton). Applying
Noether's theorem we compute the conserved SUSY current. The details are shown
step by step in sections (\ref{chiral}-\ref{full}). The result is%
\begin{equation}
J_{L}^{M}=\delta\left(  X^{2}\right)  \left\{
\begin{array}
[c]{c}%
D_{K}\left(  X_{N}\varphi_{i}\right)  \left(  \Gamma^{KN}\Gamma^{M}-\eta
^{MN}\Gamma^{K}\right)  \psi_{R}^{i}+\frac{\partial W}{\partial\varphi_{j}%
}X_{N}\Gamma^{MN}\psi_{iL}\\
+\frac{1}{2\sqrt{2}}F_{KL}^{a}X_{N}\left(  \Gamma^{KLN}\bar{\Gamma}^{M}%
-\eta^{NM}\Gamma^{KL}\right)  \lambda_{La}\\
+\frac{ig}{\sqrt{2}}\varphi_{i}\left(  t_{a}\varphi^{\dagger}\right)
^{i}X_{N}\Gamma^{MN}\lambda_{La}%
\end{array}
\right\}  .\label{JL}%
\end{equation}
where the first line comes from $L_{chiral},$ the second from $L_{vector},$
and the third from $L_{int}.$ The Hermitian conjugate of $J_{L}^{M}$ can be
written as the right-handed counterpart of the above $J_{R}^{M}=C(\overline
{J_{L}^{M}})^{T}$ (see Appendix (\ref{A}) for Hermitian and charge conjugation
properties)%
\begin{equation}
J_{R}^{M}=\delta\left(  X^{2}\right)  \left\{
\begin{array}
[c]{c}%
D_{K}\left(  X_{N}\varphi^{\dagger i}\right)  \left(  \bar{\Gamma}^{KN}%
\bar{\Gamma}^{M}-\eta^{MN}\bar{\Gamma}^{K}\right)  \psi_{iL}+\frac{\partial
W^{\ast}}{\partial\varphi^{\ast j}}X_{N}\bar{\Gamma}^{MN}\psi_{R}^{j}\\
+\frac{1}{2\sqrt{2}}F_{KL}^{a}X_{N}\left(  \bar{\Gamma}^{KLN}\Gamma^{M}%
-\eta^{NM}\bar{\Gamma}^{KL}\right)  \lambda_{Ra}\\
-\frac{ig}{\sqrt{2}}\varphi^{\dagger i}\left(  t_{a}\varphi\right)  _{i}%
X_{N}\bar{\Gamma}^{MN}\lambda_{Ra}%
\end{array}
\right\}  .\label{JR}%
\end{equation}
Using the equations of motion that follow from the action (\ref{actionsusy})
we can verify that this SUSY current is conserved
\begin{equation}
\partial_{M}J_{L}^{M}\left(  X\right)  =\partial_{M}J_{R}^{M}\left(  X\right)
=0.
\end{equation}
The conservation of the current amounts also to a proof of SUSY for the theory
of Eq.(\ref{actionsusy}) that supplies the equations of motion.

In the rest of the paper we provide the details of the theory summarized above.

\section{Chiral supermultiplet in 2T-physics \label{chiral}}

The chiral multiplet $\left(  \varphi,\psi_{L},F\right)  _{i}$ is defined in
terms of left handed spinors. As noted in Eqs.(\ref{cc},\ref{cc2}), right
handed spinors $\psi_{R}$ are treated as the charge conjugates of left handed
spinors. Hence, for each $i,$ there are only 4 independent complex fermionic
components $\left(  \psi_{L\alpha}\right)  _{i}$. If one would like to
introduce right handed independent fermions $\psi_{R}$, one may do so by
introducing more $\left(  \psi_{L}\right)  _{i}$ with different values of $i$
since these are equivalent to the $\psi_{R}$ under charge conjugation. It is
evident that the formalism in the form $\left(  \varphi,\psi_{L},F\right)
_{i},$ with a range of values for $i,$ includes all possible chiral
supermultiplets (left or right) that may be needed in various applications.

\subsection{Interacting Action for Chiral Supermultiplets}

Independent of SUSY, the free field part of the action is determined by the
field theoretic formulation of 2T-physics given in \cite{2tstandardM}%
\begin{equation}
S_{0}=\int d^{6}X~\delta\left(  X^{2}\right)  \left[
\begin{array}
[c]{l}%
\frac{1}{2}\left(  \varphi^{\dagger}\partial^{2}\varphi+\partial^{2}%
\varphi^{\dagger}\varphi\right)  +F^{\dagger}F\\
+\frac{i}{2}\left(  \overline{\psi_{L}}X\bar{\partial}\psi_{L}+\overline
{\psi_{L}}\overleftarrow{\partial}\bar{X}\psi_{L}\right)
\end{array}
\right]  \label{S0}%
\end{equation}
where $X\equiv X^{M}\Gamma_{M},$ $\bar{\partial}=\bar{\Gamma}^{M}\partial
_{M},$ etc. By using the hermiticity property
\begin{equation}
\left(  i\overline{\psi_{1L}}\Gamma^{M}\bar{\Gamma}^{N}\cdots\bar{\Gamma}%
^{K}\psi_{2L}\right)  ^{\dagger}=i\bar{\psi}_{2L}\left(  -\Gamma^{K}\right)
\cdots\left(  -\Gamma^{N}\right)  \left(  -\bar{\Gamma}^{M}\right)  \psi_{1L}%
\end{equation}
which is explained in Eqs.(\ref{herm11}-\ref{herm}), it is easily verified
that this action is Hermitian.

The delta function\footnote{Some useful properties of the delta function
include $\frac{\partial}{\partial X^{M}}\delta\left(  X^{2}\right)
=2X_{M}\delta^{\prime}\left(  X^{2}\right)  $, $X\cdot\frac{\partial}{\partial
X}\delta\left(  X^{2}\right)  =2X^{2}\delta^{\prime}\left(  X^{2}\right)
=-2\delta\left(  X^{2}\right)  $, and $\partial^{2}\delta\left(  X^{2}\right)
=2\left(  d+2\right)  \delta^{\prime}\left(  X^{2}\right)  +4X^{2}%
\delta^{\prime\prime}\left(  X^{2}\right)  =2\left(  d-2\right)
\delta^{\prime}\left(  X^{2}\right)  .$ $\;$ Here $\delta^{\prime}\left(
u\right)  ,\delta^{\prime\prime}\left(  u\right)  $ are the derivatives of the
delta function with respect to its argument $u=X^{2}.$ So we have used
$u\delta^{\prime}\left(  u\right)  =-\delta\left(  u\right)  $ and
$u\delta^{\prime\prime}\left(  u\right)  =-2\delta^{\prime}\left(  u\right)  $
as the properties of the delta function of a single variable $u$ to arrive at
the above expressions. These are to be understood in the sense of
distributions under integration with smooth functions. \label{delta}} in the
volume element $d^{6}X~\delta\left(  X^{2}\right)  $ as well as the given
structure of the kinetic terms are required by global SO$\left(  4,2\right)
=$SU$\left(  2,2\right)  $ and local 2T-physics gauge symmetries
\cite{2tstandardM}. The gauge symmetry is responsible for eliminating ghosts
and thinning out the field degrees of freedom from $4+2$ to $3+1$ dimensions
holographically \textit{without} any residual Kaluza-Klein type excitations.
It is also responsible for the unifying features of 2T-physics as a structure
above 1T-physics through various definitions of \textquotedblleft
time\textquotedblright\ in the embeddings of $3+1$ dimensions in $4+2$
dimensions (see Fig.1 in \cite{2tstandardM}).

It will be convenient to rewrite the scalar part of the free action by doing
an integration by parts so that it contains only first order derivatives. The
result is\footnote{An intermediate step in deriving Eq.(\ref{convenient}) has
the second term in the form $\int d^{6}X\delta^{\prime}\left(  X^{2}\right)
X\cdot\partial\left(  \varphi^{\dagger}\varphi\right)  .$ This differs from
the version in Eq.(\ref{convenient}) by a total derivative.}
\begin{equation}
S_{0}\left(  \varphi,F\right)  =\int d^{6}X~\left[
\begin{array}
[c]{c}%
\delta\left(  X^{2}\right)  \left(  -\partial_{M}\varphi^{\dagger}\partial
^{M}\varphi+F^{\dagger}F\right) \\
+2~\delta^{\prime}\left(  X^{2}\right)  ~\varphi^{\dagger}\varphi
\end{array}
\right]  . \label{convenient}%
\end{equation}
The term that contains $2\delta^{\prime}\left(  X^{2}\right)  $ with a
specific coefficient, is an outcome of the 2T-physics gauge symmetry.
Similarly, the fermion term is invariant under a separate 2T-physics gauge
symmetry \cite{2tstandardM}. It may also be integrated by parts. After using
$\Gamma^{M}\bar{X}=-X\bar{\Gamma}^{M}+2X^{M},$ and $\delta^{\prime}\left(
X^{2}\right)  X\bar{X}=X^{2}\delta^{\prime}\left(  X^{2}\right)
=-\delta\left(  X^{2}\right)  ,$ it takes the form%
\begin{equation}
S_{0}\left(  \psi\right)  =i\int d^{6}X~\delta\left(  X^{2}\right)
~\overline{\psi_{L}}\left[  X\bar{\partial}-\left(  X\cdot\partial+2\right)
\right]  \psi_{L}, \label{convenient2}%
\end{equation}
By using the relation $\Gamma^{MN}X_{M}\partial_{N}=X\bar{\partial}%
-X\cdot\partial$ this may be rewritten further in the spin-orbit coupling
form
\begin{equation}
S_{0}\left(  \psi\right)  =-i\int d^{6}X~\delta\left(  X^{2}\right)
~\overline{\psi_{L}}\left[  \frac{1}{2i}\Gamma^{MN}L_{MN}+2\right]  \psi_{L},
\label{convenient3}%
\end{equation}
where $L_{MN}$ is the SO$\left(  4,2\right)  $ orbital angular momentum
\begin{equation}
L_{MN}=-i\left(  X_{M}\partial_{N}-X_{N}\partial_{M}\right)  .
\end{equation}

The free field equations are derived by extremizing the action in
Eq.(\ref{S0}) or (\ref{convenient}-\ref{convenient3}) while treating carefully
the delta function as in footnote (\ref{delta}). The result is
\cite{2tstandardM}%
\begin{align}
&  \left(  \partial^{2}\varphi\right)  _{X^{2}=0}=0,\;\;\left[  \left(
X\cdot\partial+1\right)  \varphi\right]  _{X^{2}=0}=0,\;\;\left(  F\right)
_{X^{2}=0}=0,\;\\
&  \left(  X\bar{\partial}\psi_{L}\right)  _{X^{2}=0}=0,\;\;\left[  \left(
X\cdot\partial+2\right)  \psi_{L}\right]  _{X^{2}=0}=0,
\end{align}
plus their complex conjugates. Accordingly, when the fields are on-shell, they
are homogeneous with a specific degree of homogeneity, namely under rescaling
they give $\varphi\left(  tX\right)  =t^{-1}\varphi\left(  X\right)  $ and
$\psi_{L}\left(  tX\right)  =t^{-2}\psi_{L}\left(  X\right)  $. However, when
the fields are off shell they are not restricted to be homogeneous. We
emphasize that our supersymmetry transformations given below are constructed
off-shell without homogeneity restrictions on any of the fields $\left(
\varphi,\psi_{L},F\right)  _{i}$ .

With an appropriate holographic embedding of $3+1$ dimensions in $4+2$
dimensions as shown in the footnote\footnote{The \textquotedblleft
relativistic particle gauge\textquotedblright\ that provides one of the
embeddings of $3+1$ dimensions in $4+2$ dimensions is given as follows. We
choose a lightcone type basis in $4+2$ dimensions so that the flat metric
takes the form $ds^{2}=dX^{M}dX^{N}\eta_{MN}=-2dX^{+^{\prime}}dX^{-^{\prime}%
}+dX^{\mu}dX^{\nu}\eta_{\mu\nu},$ where $\eta_{\mu\nu},$ with $\mu
,\nu=0,1,2,3$ is the Minkowski metric and $X^{\pm^{\prime}}=\frac{1}{\sqrt{2}%
}\left(  X^{0^{\prime}}\pm X^{1^{\prime}}\right)  $ are the lightcone
coordinates for the extra space $X^{1^{\prime}}$and time $X^{0^{\prime}}%
$dimensions. Furthermore we choose the following parametrization
$X^{+^{\prime}}=\kappa,\;X^{-^{\prime}}=\kappa\lambda,\;X^{\mu}=\kappa x^{\mu
},$ which defines the emergent $3+1$ dimensional spacetime $x^{\mu}$ as
embedded in $4+2$. The inverse relation is $\kappa=X^{+^{\prime}}%
,\;\;\lambda=\frac{X^{-^{\prime}}}{X^{+^{\prime}}},\;x^{\mu}=\frac{X^{\mu}%
}{X^{+^{\prime}}}.$ This provides one of the many possible embedding of $3+1$
dimensions in $4+2$ dimensions. In this paper we will mainly use the $3+1$
spacetime embedding given above. This embedding, first discussed by
Dirac\cite{Dirac}, was useful to express the usual Standard Model of Particles
and Forces as a gauge fixed form of 2T-physics \cite{2tstandardM}. Therefore,
the same $3+1$ embedding will connect the supersymmetric Standard Model in
$3+1$ dimensions to the supersymmetric formulation of 2T physics in $4+2$
dimensions. In addition to the embedding described above there are many other
embeddings of $3+1$ in $4+2$. Such embeddings corresponds to Sp$\left(
2,R\right)  $ gauge choices in the underlying 2T-physics worldline theory and
lead to a variety of 1T-physics dynamical systems as summarized\ in Fig.1 of
ref. \cite{2tstandardM}. \label{embedding}}, the on-shell equations above are
equivalent to free relativistic massless fields in $3+1$ dimensions
\cite{2tstandardM}.

The general interaction among chiral supermultiplets, which we will show to be
supersymmetric directly in $4+2$ dimensions is given by
\begin{equation}
S_{int}=\int d^{6}X~\delta\left(  X^{2}\right)  ~\left[  \left(
\frac{\partial W}{\partial\varphi_{i}}F_{i}-\frac{i}{2}\left(  \overline
{\psi_{R}}\right)  _{i}X\left(  \psi_{L}\right)  _{j}\frac{\partial^{2}%
W}{\partial\varphi_{i}\partial\varphi_{j}}\right)  +h.c.\right]  \label{Sint}%
\end{equation}
where $W\left(  \varphi\right)  $ is any cubic superpotential constructed from
the scalars $\varphi_{i}$, $i=1,2,3,\cdots,$ with any desired internal
symmetry group G.

The structure of $S_{int}$ is similar to the SUSY formalism in $3+1$
dimensions, except for the fact that the Yukawa coupling in $4+2$ dimensions
involves the factor $\bar{X}$ in the expression $\left(  \overline{\psi_{R}%
}\right)  _{i}\bar{X}\left(  \psi_{L}\right)  _{j}=\left(  \psi_{L\alpha
}\right)  _{i}\left(  C\bar{X}\right)  ^{\alpha\beta}\left(  \psi_{L\beta
}\right)  _{j}$. Taking into account that $\left(  C\bar{X}\right)
^{\alpha\beta}$ is antisymmetric in $\left(  \alpha\leftrightarrow
\beta\right)  ,$ and that the fermions anticommute, this factor is symmetric
under the interchange of $i$ and $j$ and is consistent with the symmetric
$\frac{\partial^{2}W}{\partial\varphi_{i}\partial\varphi_{j}}.$

As in $3+1$ dimensions, the auxiliary fields $F_{i}$ can be solved from the
equations of motion
\begin{equation}
F^{\dagger i}=-\frac{\partial W}{\partial\varphi_{i}}=-3y^{ijk}\varphi
_{j}\varphi_{k},\;\;F_{i}=-\left(  \frac{\partial W}{\partial\varphi_{i}%
}\right)  ^{\ast}=-3y_{ijk}^{\ast}\varphi^{\dagger j}\varphi^{\dagger k},
\end{equation}
and inserted back into the action, so that $S_{total}=S_{0}+S_{int}$ can be
expressed only in terms of the dynamical fields $\left(  \varphi,\psi
_{L}\right)  _{i}$ in $4+2$ dimensions. The effective scalar potential energy
\textquotedblleft F-term\textquotedblright\ in the supersymmetric $4+2$ action
is then $V_{F}\left(  \varphi,\varphi^{\dagger}\right)  =\left\vert
\frac{\partial W}{\partial\varphi_{i}}\right\vert ^{2}$, just like $3+1$
dimensions. However, to discuss the SUSY properties of the action, it is more
convenient to keep the $F_{i}$ off-shell in the actions $S_{0},S_{int}$ as
written above. We will see that unlike $3+1$ dimensions, $S_{0},S_{int}$ are
not separately supersymmetric in $4+2$ dimensions, but after reducing the
theory to $3+1$ dimension, the mixing term will drop and the $4+2$ dimensional
supersymmetry will reduce to ordinary 3+1 \textit{superconformal} symmetry.

Renormalizability of 2T-physics field theory in $4+2$ dimensions is determined
by the renormalizability of the equivalent 1T field theory in $3+1$ dimensions
in the 2T-physics gauge described in footnote (\ref{embedding}). Hence
renormalizability restricts $W$ to be at the most cubic in $3+1$ as well as
$4+2$ dimensions. Furthermore, the \textit{2T-physics gauge symmetries}
discussed in \cite{2tstandardM} require that $S_{0}+S_{int}$ \textit{cannot
have any dimensionful couplings or mass terms} in $4+2$ dimensions. Hence
$V_{F}\left(  \varphi,\varphi^{\dagger}\right)  =\left\vert \frac{\partial
W}{\partial\varphi_{i}}\right\vert ^{2}$ must be purely quartic, or $W$ must
be \textit{purely cubic, }which is also what is required by supersymmetry.

As in the case of the non-supersymmetric Standard Model discussed in
\cite{2tstandardM}, this cubic restriction on $W$ will require that we include
a supermultiplet that includes the dilaton $\Phi$ as part of the fields in our
theory so that we can generalize to $W\left(  \varphi,\Phi\right)  $. This
would be used to drive the spontaneous breakdown of the electroweak SU$\left(
2\right)  \times$U$\left(  1\right)  $ gauge symmetry, as demanded by
phenomenology. This point is elaborated in more detail in the comments
following Eq.(\ref{actionsusy}).

\subsection{Supersymmetry Transformations \label{chiralsusy}}

SUSY transformations in 2T-physics in $4+2$ dimensions with $N$
supersymmetries were formulated for the 2T superparticle on the worldline
\cite{super2t}\cite{2ttwistor}. These form the supergroup SU$\left(
2,2|N\right)  $ as the global symmetry of the superparticle, and therefore
this is the supersymmetry in the field theory version in $4+2$ dimensions.
Here we concentrate on $N=1$ supersymmetry with the supergroup SU$\left(
2,2|1\right)  .$ The fermionic parameter $\varepsilon_{L}$ is a left-handed
spinor (just like $\psi_{L}$) in the $4$ representation of SU$\left(
2,2\right)  $. We also note the right-handed charge conjugate $\varepsilon
_{R}=C\overline{\varepsilon_{L}}^{T}$ which is not independent of
$\varepsilon_{L}$ (just like $\psi_{R}$), and classified as the $\bar{4}$ of
SU$\left(  2,2\right)  .$

We introduce the SUSY transformations of the chiral multiplet $\left(
\varphi,\psi_{L},F\right)  _{i}$ off-shell
\begin{align}
\delta_{\varepsilon}\varphi_{i}  &  =\overline{\varepsilon_{R}}\bar{X}%
\psi_{iL}-\frac{1}{2}X^{2}\overline{\varepsilon_{R}}(\bar{\partial}\psi
_{iL}+U_{ij}^{\dagger}\psi_{R}^{j})+\delta_{\varepsilon}^{1}\varphi
_{i}\label{susy1}\\
\delta_{\varepsilon}\psi_{iL}  &  =i\left(  \partial\varphi_{i}\right)
\varepsilon_{R}-iF_{i}\varepsilon_{L}\;\label{susy2}\\
\delta_{\varepsilon}F_{i}  &  =\overline{\varepsilon_{L}}\left[
X\bar{\partial}-\left(  X\cdot\partial+2\right)  \right]  \psi_{iL}%
+\delta_{\varepsilon}^{1}F_{i} \label{susy3}%
\end{align}
The additional pieces $\delta_{\varepsilon}^{1}\varphi_{i},\delta
_{\varepsilon}^{1}F_{i}$ (given in Eq.(\ref{delphi},\ref{delFfull}) and
explained in section (\ref{full})) are proportional to the gauge coupling
constants $g$ and are needed when vector supermultiplets are coupled to chiral
supermultiplets. In this section we assume the chiral supermultiplets on their
own, so we will take $\left.  \delta_{\varepsilon}^{1}\varphi_{i}\right\vert
_{g=0}=\left.  \delta_{\varepsilon}^{1}F_{i}\right\vert _{g=0}=0$. In the
absence of interactions among the chiral multiplets the coefficient
$U_{ij}^{\dagger}$ is also absent, but with interactions we will see that
$U_{ij}^{\dagger}$ must satisfy $U_{ij}^{\dagger}=\frac{\partial^{2}W^{\ast}%
}{\partial\varphi^{\dagger i}\partial\varphi^{\dagger j}}=3!y_{ijk}^{\ast
}\varphi^{\dagger k}$ where $W\left(  \varphi\right)  $ will turn out to be
the superpotential of Eq.(\ref{superpotential},\ref{Gsymm}). The last line may
be rewritten as $\delta_{\varepsilon}F_{i}=-\overline{\varepsilon_{L}}\left(
\frac{1}{2i}\Gamma^{MN}L_{MN}+2\right)  \psi_{iL}+\delta_{\varepsilon}%
^{1}F_{i}$ as in Eqs.(\ref{convenient2},\ref{convenient3}).

There are some parallels and some differences between these $4+2$ SUSY
transformations and the familiar ones in $3+1$ dimensions. In particular, the
terms proportional to $X^{2}$ and $\left(  X\cdot\partial+2\right)  $ in
$\delta_{\varepsilon}\varphi_{i}$ and $\delta_{\varepsilon}F_{i}$ respectively
have no parallels in $3+1$ dimensions, as noted following Eq.(\ref{delBfull}).

The transformation of the Hermitian conjugate fields $\left(  \varphi^{\ast
},\overline{\psi_{L}},F^{\ast}\right)  ^{i}$ is derived from above by using
the hermiticity properties of gamma matrices given in Eqs.(\ref{herm11}%
-\ref{herm})
\begin{align}
\delta_{\varepsilon}\varphi^{\dagger i} &  =\overline{\psi_{L}}^{i}%
X\varepsilon_{R}-\frac{1}{2}X^{2}\left(  \overline{\psi_{L}}^{i}%
\overleftarrow{\partial}-U^{ij}\overline{\psi_{R}}_{i}\right)  \varepsilon
_{R}+\delta_{\varepsilon}^{1}\varphi^{\dagger i}\label{susy11}\\
\delta_{\varepsilon}\overline{\psi_{L}}^{i} &  =i\overline{\varepsilon_{R}%
}\left(  \bar{\partial}\varphi^{\dagger i}\right)  +i\overline{\varepsilon
_{L}}F^{\dagger i}~\label{susy22}\\
\delta_{\varepsilon}F^{\dagger i} &  =-\overline{\psi_{L}}^{i}\left[
\overleftarrow{\partial}\bar{X}-\left(  \overleftarrow{\partial}\cdot
X+2\right)  \right]  \varepsilon_{L}+\delta_{\varepsilon}^{1}F^{\dagger
i}\label{susy33}%
\end{align}
The transformation of the charge conjugate fields $\left(  \varphi^{\ast}%
,\psi_{R},F^{\ast}\right)  ^{i},$ in terms of $\psi_{R}^{i}$ instead of
$\overline{\psi_{L}}^{i},$ are obtained by using the properties given in
Eqs.(\ref{Maj}-\ref{maj6})
\begin{align}
\delta_{\varepsilon}\varphi^{\dagger i} &  =\overline{\varepsilon_{L}}%
X\psi_{R}^{i}-\frac{1}{2}X^{2}\overline{\varepsilon_{L}}\left(  \partial
\psi_{R}^{i}+U^{ij}\psi_{iL}\right)  +\delta_{\varepsilon}^{1}\varphi^{\dagger
i}\label{susy111}\\
\delta_{\varepsilon}\psi_{R}^{i} &  =-i\left(  \bar{\partial}\varphi^{\dagger
i}\right)  \varepsilon_{L}+iF^{\dagger i}\varepsilon_{R}\label{susy222}\\
\delta_{\varepsilon}F^{\dagger i} &  =\overline{\varepsilon_{R}}\left(
\bar{X}\partial-\left(  X\cdot\partial+2\right)  \right)  \psi_{R}^{i}%
+\delta_{\varepsilon}^{1}F^{\dagger i}\label{susy333}%
\end{align}
The last line may be rewritten as $\delta_{\varepsilon}F^{\dagger
i}=-\overline{\varepsilon_{R}}\left(  \frac{1}{2i}\bar{\Gamma}^{MN}%
L_{MN}+2\right)  \psi_{R}^{i}+\delta_{\varepsilon}^{1}F^{\dagger i}.$

We will first show that the free action $S_{0}$ is invariant off-shell in the
absence of interactions provided the matrix $U,U^{\dagger}$ is dropped in the
transformation rules (\ref{susy1}-\ref{susy333}). When $S_{int}$ is included
we will show that, unlike SUSY in $3+1$ dimensions, $S_{0},S_{int}$ cannot be
made separately invariant. However, by including the $U,U^{\dagger}$ terms in
the transformation rules (\ref{susy1}-\ref{susy333}), the total action
$S_{tot}^{chiral}=S_{0}+S_{int}$ will be invariant off-shell in $4+2$ dimensions.

We begin with the free action in the form of Eqs.(\ref{convenient}%
-\ref{convenient3}). Its variation is%
\begin{equation}
\delta_{\varepsilon}S_{0}=\int d^{6}X~\left[
\begin{array}
[c]{c}%
-\delta\left(  X^{2}\right)  \partial_{M}\varphi^{\dagger}\partial^{M}\left(
\delta_{\varepsilon}\varphi\right)  +2\delta^{\prime}\left(  X^{2}\right)
~\varphi^{\dagger}\delta_{\varepsilon}\varphi\\
+\delta\left(  X^{2}\right)  \left\{  i\left(  \delta_{\varepsilon}%
\overline{\psi_{L}}\right)  \left[  X\bar{\partial}-\left(  X\cdot
\partial+2\right)  \right]  \psi_{L}+F^{\dagger}\left(  \delta_{\varepsilon
}F\right)  \right\}
\end{array}
\right]  +h.c.
\end{equation}
Inserting the SUSY transformation given above we get%
\begin{equation}
\delta_{\varepsilon}S_{0}=\int d^{6}X~\left[
\begin{array}
[c]{c}%
-\delta\left(  X^{2}\right)  \partial_{M}\varphi^{\dagger}\partial^{M}\left[
\overline{\varepsilon_{R}}\bar{X}\psi_{L}-\frac{1}{2}X^{2}\overline
{\varepsilon_{R}}(\bar{\partial}\psi_{iL}+U^{\ast}\psi_{R})\right] \\
+2\delta^{\prime}\left(  X^{2}\right)  ~\varphi^{\dagger}\left[
\overline{\varepsilon_{R}}\bar{X}\psi_{L}-\frac{1}{2}X^{2}\overline
{\varepsilon_{R}}(\bar{\partial}\psi_{L}+U^{\ast}\psi_{R})\right] \\
+i\delta\left(  X^{2}\right)  \left(  i\overline{\varepsilon_{R}}\left(
\bar{\partial}\varphi^{\dagger}\right)  +i\overline{\varepsilon_{L}}%
F^{\dagger}\right)  \left[  X\bar{\partial}-\left(  X\cdot\partial+2\right)
\right]  \psi_{L}\\
+\delta\left(  X^{2}\right)  F^{\dagger}\overline{\varepsilon_{L}}\left[
X\bar{\partial}-\left(  X\cdot\partial+2\right)  \right]  \psi_{L}%
\end{array}
\right]  +h.c. \label{delS0}%
\end{equation}
In the first two lines, after using the properties $X^{2}\delta\left(
X^{2}\right)  =0$ and $X^{2}\delta^{\prime}\left(  X^{2}\right)
=-\delta\left(  X^{2}\right)  ,$ the terms containing $X^{2}$ simplify to
$\delta\left(  X^{2}\right)  \left(  X\cdot\partial+1\right)  \varphi
^{\dagger}\overline{\varepsilon_{R}}(\bar{\partial}\psi_{iL}+U_{ij}^{\ast}%
\psi_{R}^{j}).$ In the last two lines, the terms proportional to $F^{\dagger
}\overline{\varepsilon_{L}}$ cancel each other. The surviving terms from all
lines take the form of a total divergence plus a term proportional to
$U,U^{\dagger}$ as follows
\begin{equation}
\delta_{\varepsilon}S_{0}=\int d^{6}X~\left\{
\begin{array}
[c]{c}%
\partial_{M}\left[  \delta\left(  X^{2}\right)  V_{0}^{M}\right] \\
+\delta\left(  X^{2}\right)  \left[  \left(  X\cdot\partial+1\right)
\varphi^{\dagger i}\right]  U_{ij}^{\ast}\overline{\varepsilon_{R}}\psi
_{R}^{j})
\end{array}
\right\}  +h.c=0 \label{delS0V}%
\end{equation}
Hence, in the free theory, by dropping the $U,U^{\dagger}$ terms in the
transformations laws, and dropping the total divergence with proper boundary
conditions, we have demonstrated that we have a supersymmetric action
$\delta_{\varepsilon}S_{0}=0.$ Here $V_{0}^{M}$ is given by (see footnote
\footnote{\label{Vfree}The various terms in Eq.(\ref{Vm2}) contribute to the
various terms in Eq.(\ref{delS0}) as follows. The subscripts in $\left\{
\cdots\right\}  _{n}$ denote terms that should be combined together for the
same $n$%
\begin{align*}
\partial_{M}\left[  -\delta\left(  X^{2}\right)  \left(  \partial^{M}%
\varphi^{\dagger}\right)  \overline{\varepsilon_{R}}\overline{\not X  }%
\psi_{L}\right]   &  =\left\{  -\delta\left(  X^{2}\right)  \partial
^{M}\varphi^{\dagger}\partial_{M}\left(  \overline{\varepsilon_{R}}%
\overline{\not X  }\psi_{L}\right)  \right\}  _{1}+\left\{  -2\delta^{\prime
}\left(  X^{2}\right)  \left(  X\cdot\partial\varphi^{\dagger}\right)
\overline{\varepsilon_{R}}\overline{\not X  }\psi_{L}\right\}  _{2}\\
&  +\left\{  -\delta\left(  X^{2}\right)  \left(  \partial^{2}\varphi
^{\dagger}\right)  \overline{\varepsilon_{R}}\overline{\not X  }\psi
_{L}\right\}  _{6}%
\end{align*}%
\begin{align*}
\partial_{M}\left[  \delta\left(  X^{2}\right)  \overline{\varepsilon_{R}}%
\bar{\Gamma}^{M}\left(  \left(  X\cdot\partial+1\right)  \varphi^{\dagger
}\right)  \psi_{L}\right]   &  =\left\{  \delta^{\prime}\left(  X^{2}\right)
2\left(  X\cdot\partial+1\right)  \varphi^{\dagger}\overline{\varepsilon_{R}%
}\overline{\not X  }\psi_{L}\right\}  _{2}+\left\{  \delta\left(
X^{2}\right)  \left[  \left(  X\cdot\partial+1\right)  \varphi^{\dagger
}\right]  \overline{\varepsilon_{R}}\overline{\not \partial }\psi_{L}\right\}
_{4}\\
&  +\left\{  \delta\left(  X^{2}\right)  \overline{\varepsilon_{R}}\left(
\left(  X\cdot\partial+2\right)  \left(  \overline{\not \partial }%
\varphi^{\dagger}\right)  \right)  \psi_{L}\right\}  _{7}%
\end{align*}%
\begin{align*}
\partial_{M}\left[  -\delta\left(  X^{2}\right)  \overline{\varepsilon_{R}%
}\left(  \overline{\not \partial }\varphi^{\dagger}\not X  \bar{\Gamma}%
^{M}\right)  \psi_{L}\right]   &  =\left\{  -\delta\left(  X^{2}\right)
\left[  \left(  \overline{\varepsilon_{R}}\left(  \overline{\not \partial
}\varphi^{\dagger}\right)  \not X  \overline{\not \partial }\right)  \psi
_{L}\right]  \right\}  _{3}+\left\{  \delta\left(  X^{2}\right)  \left(
\overline{\varepsilon_{R}}\left(  \partial^{2}\varphi^{\dagger}\right)
\overline{\not X  }\right)  \psi_{L}\right\}  _{6}\\
&  +\left\{  -2\delta\left(  X^{2}\right)  \left(  \overline{\varepsilon_{R}%
}\left(  \left(  X\cdot\partial+2\right)  \overline{\not \partial }%
\varphi^{\dagger}\right)  \right)  \psi_{L}\right\}  _{7}%
\end{align*}%
\[
\partial_{M}\left[  \delta\left(  X^{2}\right)  X^{M}\overline{\varepsilon
_{R}}\left(  \overline{\not \partial }\varphi^{\dagger}\right)  \psi
_{L}\right]  =\left\{  \delta\left(  X^{2}\right)  \left[  \overline
{\varepsilon_{R}}\left(  \overline{\not \partial }\varphi^{\dagger}\right)
\left(  \left(  X\cdot\partial+2\right)  \psi_{L}\right)  \right]  \right\}
_{5}+\left\{  \delta\left(  X^{2}\right)  \overline{\varepsilon_{R}}\left[
\left(  X\cdot\partial+2\right)  \overline{\not \partial }\varphi^{\dagger
}\right]  \psi_{L}\right\}  _{7}%
\]
The sum of these terms give the $\delta_{\varepsilon}S_{0}$ in Eq.(\ref{delS0}%
) after cancelling the $F^{\dagger}\overline{\varepsilon_{L}}$ terms as
follows
\[
\delta_{\varepsilon}S_{0}=\int d^{6}X~\left[
\begin{array}
[c]{c}%
\left\{  -\delta\left(  X^{2}\right)  \partial_{M}\varphi^{\dagger}%
\partial^{M}\left(  \overline{\varepsilon_{R}}\overline{\not X  }\psi
_{L}\right)  \right\}  _{1}+\left\{  ~2\delta^{\prime}\left(  X^{2}\right)
\varphi^{\dagger}\overline{\varepsilon_{R}}\overline{\not X  }\psi
_{L}\right\}  _{2}\\
+\left\{  \delta\left(  X^{2}\right)  \left(  X\cdot\partial+1\right)
\varphi^{\dagger}\overline{\varepsilon_{R}}\overline{\not \partial }\psi
_{L}\right\}  _{4}-\delta\left(  X^{2}\right)  \overline{\varepsilon_{R}%
}\overline{\not \partial }\varphi^{\dagger}\left[  \left\{  \not X
\overline{\not \partial }\right\}  _{3}-\left\{  \left(  X\cdot\partial
+2\right)  \right\}  _{5}\right]  \psi_{L}%
\end{array}
\right]  +h.c.
\]
})%

\begin{align}
V_{0}^{M}  &  =\overline{\varepsilon_{R}}\left\{
\begin{array}
[c]{c}%
-\left(  \partial^{M}\varphi^{\dagger}\right)  \bar{X}+\bar{\Gamma}^{M}\left(
X\cdot\partial+1\right)  \varphi^{\dagger}\\
-\bar{\partial}\varphi^{\dagger}X\bar{\Gamma}^{M}+X^{M}\left(  \bar{\partial
}\varphi^{\dagger}\right)
\end{array}
\right\}  \psi_{L}\\
&  =\overline{\varepsilon_{R}}\left\{  \bar{\Gamma}^{M}\varphi^{\dagger
i}+\bar{\Gamma}^{MNK}X_{N}\partial_{K}\varphi^{\dagger i}\right\}  \psi_{iL}
\label{Vm2}%
\end{align}
Now using the generalized Noether's theorem we obtain the part of the
conserved current $\left(  J_{R}^{M}\right)  _{0}$ coming from the free action
$S_{0}$
\begin{equation}
\overline{\varepsilon_{R}}\left(  J_{R}^{M}\right)  _{0}+h.c.=\left(
\begin{array}
[c]{c}%
\frac{\partial L}{\partial\left(  \partial_{M}\varphi\right)  }\delta
_{\varepsilon}\varphi+\delta\overline{\psi_{L}}\frac{\partial L}%
{\partial\left(  \partial_{M}\overline{\psi_{L}}\right)  }\\
+\frac{\partial L}{\partial\left(  \partial_{M}F\right)  }\delta_{\varepsilon
}F-\delta\left(  X^{2}\right)  V_{0}^{M}%
\end{array}
\right)  +h.c.
\end{equation}
where the last term is obtained from the total divergence in Eq.(\ref{delS0V}%
). For consistency of this computation we must use again the form of the
action in Eqs.(\ref{convenient}-\ref{convenient3}). Noting that $\frac
{\partial L}{\partial\left(  \partial_{M}\overline{\psi_{L}}\right)  }%
=\frac{\partial L}{\partial\left(  \partial_{M}F\right)  }=0,$ only the first
and last terms contribute. This gives the current which we write in several
equivalent forms
\begin{align}
\overline{\varepsilon_{R}}\left(  J_{R}^{M}\right)  _{0}  &  =\delta\left(
X^{2}\right)  ~\overline{\varepsilon_{R}}\left[  \bar{\partial}\varphi
^{\dagger}X\bar{\Gamma}^{M}-X^{M}\left(  \bar{\partial}\varphi^{\dagger
}\right)  -\bar{\Gamma}^{M}\left(  X\cdot\partial+1\right)  \varphi^{\dagger
}\right]  \psi_{L}\label{current}\\
&  =\delta\left(  X^{2}\right)  ~\overline{\varepsilon_{R}}\left[
-\bar{\Gamma}^{MPQ}X_{P}\partial_{Q}\varphi^{\dagger}-\partial^{M}\left(
\bar{X}\varphi^{\dagger}\right)  \right]  \psi_{L}\\
&  =\delta\left(  X^{2}\right)  ~\overline{\varepsilon_{R}}\left(  \bar
{\Gamma}^{QP}\bar{\Gamma}^{M}-\eta^{MP}\bar{\Gamma}^{Q}\right)  \psi
_{L}~\partial_{Q}\left(  X_{P}\varphi^{\dagger}\right) \\
&  =\delta\left(  X^{2}\right)  ~\overline{\varepsilon_{R}}\left[  \frac
{1}{2i}\left(  \bar{\Gamma}^{MPQ}L_{PQ}\varphi^{\dagger}\right)  \bar{\Gamma
}^{M}-\partial^{M}\left(  \bar{X}\varphi^{\dagger}\right)  \right]  \psi_{L}%
\end{align}
One can check explicitly that this current is conserved when we use the
equations of motion for the free action.

Now we turn to the interaction term $S_{int}$ in Eq.(\ref{Sint}) and
investigate its transformation properties under SUSY for any $W.$ Inserting
the transformation rules above we get after some simplifications
\begin{equation}
\delta_{\varepsilon}S_{int}=\int d^{6}X~\delta\left(  X^{2}\right)  \left[
\left(
\begin{array}
[c]{c}%
\frac{\partial W}{\partial\varphi_{i}}\overline{\varepsilon_{L}}\left[
X\bar{\partial}-\left(  X\cdot\partial+2\right)  \right]  \psi_{iL}.\\
-\frac{\partial}{\partial\varphi_{i}}(\frac{\partial W}{\partial\varphi_{j}%
})\partial_{M}\varphi_{i}\overline{\varepsilon_{L}}\Gamma^{M}\bar{X}\left(
\psi_{L}\right)  _{j}\\
-\left(  \frac{i}{2}\left(  \psi_{Li}\right)  ^{T}C\bar{X}\psi_{Lj}\right)
\left(  \left(  \varepsilon_{L}\right)  ^{T}C\bar{X}\psi_{Lk}\right)
\frac{\partial^{3}W}{\partial\varphi_{i}\partial\varphi_{j}\partial\varphi
_{k}}%
\end{array}
\right)  +h.c.\right]
\end{equation}
One of the crucial observations here is the gamma matrix identities for
SU$\left(  2,2\right)  =$SO$\left(  4,2\right)  $
\begin{equation}
\frac{1}{8}\left(  \Gamma^{MN}\right)  _{\alpha}^{\text{\ \ }\beta}\left(
\Gamma_{MN}\right)  _{\gamma}^{\text{ \ }\delta}=\frac{1}{4}\delta_{\alpha
}^{\text{\ \ }\beta}\delta_{\gamma}^{\text{ \ }\delta}-\delta_{\alpha
}^{\text{\ \ }\delta}\delta_{\gamma}^{\text{ \ }\beta}. \label{Fierz}%
\end{equation}
Using this identity, and the fact that $C\bar{X}$ is an antisymmetric matrix,
we see that the $\frac{\partial^{3}W}{\partial\varphi_{i}\partial\varphi
_{j}\partial\varphi_{k}}$ term drops out due to a Fierz identity proven in
Appendix (\ref{fierzI}). Then $\delta_{\varepsilon}S_{int}$ takes the form of
a total divergence plus a term proportional to $\frac{\partial W}%
{\partial\varphi_{j}}$,$\frac{\partial W^{\ast}}{\partial\varphi^{\ast j}},$
as follows%
\begin{equation}
\delta_{\varepsilon}S_{int}=\int d^{6}X~\left\{
\begin{array}
[c]{c}%
\partial_{M}\left[  \delta\left(  X^{2}\right)  V_{1}^{M}\right] \\
+\delta\left(  X^{2}\right)  \frac{\partial W^{\ast}}{\partial\varphi^{\ast
j}}\overline{\varepsilon_{R}}\left(  X\cdot\partial+2\right)  \psi_{R}^{j})
\end{array}
\right\}  +h.c. \label{delSintV}%
\end{equation}
where%
\begin{equation}
V_{1}^{M}=-\frac{\partial W^{\ast}}{\partial\varphi^{\ast j}}\overline
{\varepsilon_{R}}\left(  \bar{\Gamma}^{M}X\right)  \psi_{R}^{j}. \label{Vm3}%
\end{equation}
So, in the presence of the $U,U^{\dagger}$ terms in the transformation laws,
neither $S_{0},$ nor $S_{int}$ on their own are invariant. However, the terms
proportional to $U,U^{\dagger}$ in Eq.(\ref{delS0V}) combine to a total
divergence with the terms proportional to $\frac{\partial W}{\partial
\varphi_{j}}$,$\frac{\partial W^{\ast}}{\partial\varphi^{\ast j}}$ in
Eq.(\ref{delSintV}), in the form $\partial_{M}\left[  \delta\left(
X^{2}\right)  V_{2}^{M}+h.c.\right]  ,$\ with$\;$
\begin{equation}
V_{2}^{M}=X^{M}\frac{\partial W^{\ast}}{\partial\varphi^{\ast j}}%
\overline{\varepsilon_{R}}\psi_{R}^{j}, \label{Vm4}%
\end{equation}
provided
\begin{equation}
U^{ij}=\frac{\partial^{2}W}{\partial\varphi_{i}\partial\varphi_{j}}%
,\;U_{ij}^{\dagger}=\frac{\partial^{2}W^{\ast}}{\partial\varphi^{\ast
i}\partial\varphi^{\ast j}}.
\end{equation}
and%
\begin{equation}
U^{ij}\varphi_{j}=2\frac{\partial W}{\partial\varphi_{i}},\;U_{ij}^{\dagger
}\varphi^{\ast j}=2\frac{\partial W^{\ast}}{\partial\varphi^{\ast i}}.
\end{equation}
These conditions require $W\left(  \varphi\right)  $ to be purely cubic, but
otherwise arbitrary, function in the scalar fields $\varphi_{i}.$ Thus only in
the case of a cubic superpotential%
\begin{equation}
W\left(  \varphi\right)  =y^{ijk}\varphi_{i}\varphi_{j}\varphi_{k},
\label{cubic}%
\end{equation}
with arbitrary dimensionless constants $y^{ijk}$ which should be made
compatible with other desired symmetries, we get a total divergence for the
SUSY variation of the total action by putting together Eqs.(\ref{Vm2}%
,\ref{Vm3},\ref{Vm4})%
\begin{equation}
\delta_{\varepsilon}S_{tot}^{chiral}=\int d^{6}X~\partial_{M}\left[
\delta\left(  X^{2}\right)  \left(  V_{0}^{M}+V_{1}^{M}+V_{2}^{M}\right)
\right]  +h.c. \label{Vtot}%
\end{equation}
which implies that the total action is supersymmetric $\delta_{\varepsilon
}\left(  S_{0}+S_{int}\right)  =0$ off-shell.

The fact that the superpotential $W\left(  \varphi\right)  $ is purely cubic,
and therefore the $F$-term of the potential $V_{F}=\left\vert \partial
W/\partial\varphi_{i}\right\vert ^{2}$ is purely quartic, is in agreement with
what we should have expected on the basis of the 2T-gauge symmetry, even
without supersymmetry, as discussed in \cite{2tbrst2006}\cite{2tstandardM}.
However, it is interesting that by demanding supersymmetry we also arrive
independently at the same conclusion that only purely quartic interactions are
admitted in the field theoretic formulation of 2T-physic in $4+2$ dimensions.
This automatically implies a renormalizable field theory as easily seen from
the perspective of $3+1$ dimensions.

Since the total divergence is not trivial, the conserved current gets
contributions from both $S_{0}$ and $S_{int},$ and is given by%
\begin{equation}
\overline{\varepsilon_{R}}\left(  J_{R}^{M}\right)  ^{chiral}=\overline
{\varepsilon_{R}}\left[  \left(  J_{R}^{M}\right)  _{0}-\delta\left(
X^{2}\right)  \left(  V_{1}^{M}+V_{2}^{M}\right)  \right]
\end{equation}
where $\overline{\varepsilon_{R}}\left(  J_{R}^{M}\right)  _{0}$ is given in
Eq.(\ref{current}). Hence the supercurrent is
\begin{equation}
\left(  J_{R}^{M}\right)  ^{chiral}=\delta\left(  X^{2}\right)  \left\{
\left(  \bar{\Gamma}^{QP}\bar{\Gamma}^{M}-\eta^{MP}\bar{\Gamma}^{Q}\right)
\psi_{iL}~\partial_{Q}\left(  X_{P}\varphi^{\dagger i}\right)  +\frac{\partial
W^{\ast}}{\partial\varphi^{\ast j}}X_{N}\bar{\Gamma}^{MN}\psi_{R}^{j}\right\}
\end{equation}
By using the equations of motion for the self interacting chiral multiplets
that follow from $\left(  S_{0}+S_{int}\right)  ,$ one can verify that the
full SUSY current constructed above is conserved
\begin{equation}
\partial_{M}\left(  J_{R}^{M}\right)  _{total}=0.
\end{equation}

\section{Vector Supermultiplet in 2T-physics \label{vecto}}

We now turn to the vector supermultiplet $\left(  A_{M},\lambda_{L},B\right)
^{a}$ in the adjoint representation of the Yang-Mills gauge group $G,$ and at
first examine it by itself without coupling it to the chiral supermultiplet.

We begin with an action of the following form suggested by 2T-physics field
theory for any Yang-Mills type gauge theory in $4+2$ dimensions
\cite{2tstandardM}%

\begin{equation}
L_{vector}=\delta\left(  X^{2}\right)  \left\{
\begin{array}
[c]{c}%
-\frac{1}{4}F_{MN}^{a}F_{a}^{MN}+\frac{1}{2}B^{a}B_{a}\\
+\frac{i}{2}\left[  \overline{\lambda_{L}}^{a}X\bar{D}\lambda_{aL}%
+\overline{\lambda_{L}}^{a}\overleftarrow{D}\bar{X}\lambda_{aL}\right]
\end{array}
\right\}  . \label{Lvector}%
\end{equation}
This action is invariant under 2T-physics gauge symmetries, and has just the
right structure to get reduced to a gauge theory in $3+1$ dimensions when
gauge fixed as described in footnote (\ref{embedding}), without any
Kaluza-Klein leftovers. In this form $B^{a}$ could be integrated out and set
equal to zero through its equations of motion. But after coupling to the
chiral multiplet, integrating out the auxiliary field $B^{a}$ in the
interacting theory will give rise to the so called $D$-term which is a
$\varphi^{4}$ interaction for the scalar fields in the chiral multiplet.

We now propose the following supersymmetry transformations in $4+2$ dimensions%

\begin{align}
\delta_{\varepsilon}A_{M}^{a}  &  =\left\{  -2b\overline{\varepsilon_{L}%
}\Gamma_{M}\bar{X}\lambda_{L}^{a}+bX^{2}\overline{\varepsilon_{L}}\Gamma
_{MN}D^{N}\lambda_{L}^{a}+\delta^{1}A_{M}^{a}\right\}  +h.c.\label{delA}\\
\delta_{\varepsilon}\lambda_{L}^{a}  &  =ib^{\ast}F_{MN}^{a}\left(
\Gamma^{MN}\varepsilon_{L}\right)  -ia^{\ast}B^{a}\varepsilon_{L}%
\label{dellam}\\
\delta_{\varepsilon}\overline{\lambda_{L}}^{a}  &  =ib\left(  \overline
{\varepsilon_{L}}\Gamma^{MN}\right)  F_{MN}^{a}+ia\overline{\varepsilon_{L}%
}B\label{dellambar}\\
\delta_{\varepsilon}B^{a}  &  =a\overline{\varepsilon_{L}}\left[  X\bar
{D}-\left(  X\cdot D+2\right)  \right]  \lambda_{L}^{a}+h.c. \label{delB}%
\end{align}
Here $a$ and $b$ are complex numbers whose values remain arbitrary until they
are fixed later when we include interaction with chiral multiplets (see
(\ref{values})). Also, $\delta^{1}A_{M}^{a}$ as given in Eq.(\ref{delAa}) is
an additional piece that arises only when chiral multiplets are coupled to
vector multiplets and is determined later in Eq.(\ref{delA*}) for the coupled
theory. When vector supermultiplets are considered in isolation, as in this
section, the extra term vanishes $\left(  \delta^{1}A_{M}^{a}\right)
\rightarrow0$, however we will include it in part of the computation for later use.

The transformation of the gaugino $\lambda_{L}^{a}$ is similar to ordinary
supersymmetry transformation in $3+1$ dimension while the transformation of
the auxiliary field $B^{a}$ is similar to the transformation law of $F$ in the
chiral multiplet except that here $B^{a}$ is Hermitian. The first term in the
transformation of $A_{M}^{a}$ also resembles the one in $3+1$ dimension except
that there is a $\bar{X}$ inserted between $\Gamma_{M}$ and $\lambda_{L}^{a}%
$\ which breaks translation symmetry in $4+2$ dimension. This insertion is
required by the 2T physics structures, and has just the correct form such that
the SUSY transformations in $4+2$ dimensions reduce to ordinary superconformal
transformations in $3+1$ dimension when one fixes the 2T gauge symmetry as
described in footnote (\ref{embedding}).

Transformation of the action is%
\begin{align}
\delta_{\varepsilon}L  &  =\delta(X^{2})B^{a}\left(  \delta_{\varepsilon}%
B_{a}\right)  -\delta(X^{2})F_{a}^{MN}D_{M}\left(  \delta_{\varepsilon}%
A_{N}^{a}\right) \label{delLV1}\\
&  +\frac{i}{2}\delta(X^{2})\left[  \left(  \delta_{\varepsilon}%
\overline{\lambda_{L}}^{a}\right)  \left(  X\bar{D}+\overleftarrow{D}\bar
{X}\right)  \lambda_{aL}\right]  +h.c.\label{delLV2}\\
&  +i\delta(X^{2})f_{abc}\left(  \delta_{\varepsilon}A_{M}^{a}\right)
\overline{\lambda_{L}}^{b}\Gamma^{MN}\lambda_{L}^{c}X_{N} \label{delLV3}%
\end{align}
The last term $i\delta(X^{2})f_{abc}\left(  \delta_{\varepsilon}A_{M}%
^{a}\right)  \overline{\lambda_{L}}^{b}\Gamma^{MN}\lambda_{L}^{c}X_{N}%
$\ vanishes by itself, partly due to $X^{2}\delta\left(  X^{2}\right)  =0$ and
partly because of the non-trivial Fierz rearrangement identity proven in
Appendix (\ref{fierzI}) that follows from of Eq.(\ref{Fierz}). The first term
in line (\ref{delLV1}) proportional to $\left(  \delta_{\varepsilon}%
B_{a}\right)  $ which we define as $\left(  \delta_{\varepsilon}L\right)
_{1}$ gives%

\begin{equation}
\left(  \delta_{\varepsilon}L\right)  _{1}=a\delta(X^{2})B\overline
{\varepsilon_{L}}\left[  X\bar{D}-\left(  X\cdot D+2\right)  \right]
\lambda_{L}+h.c.
\end{equation}
In the second term in line (\ref{delLV1}) proportional to $\delta
_{\varepsilon}A_{N}^{a},$ which we define as $\left(  \delta_{\varepsilon
}L\right)  _{2},$ we first collect a total derivative (suppressing the adjoint
index $a$ for less clutter)
\begin{equation}
\left(  \delta_{\varepsilon}L\right)  _{2}=\left\{
\begin{array}
[c]{c}%
-\partial_{M}\left\{  \delta(X^{2})F^{MN}\delta_{\varepsilon}A_{N}\right\} \\
+\delta(X^{2})\left(  D_{M}F^{MN}\right)  \left(  \delta_{\varepsilon}%
A_{N}\right)  +\delta^{\prime}(X^{2})2X_{M}F^{MN}\left(  \delta_{\varepsilon
}A_{N}\right)
\end{array}
\right\}
\end{equation}
and then insert $\delta_{\varepsilon}A_{N}^{a}$ in the remainder. After using
$X^{2}\delta(X^{2})=0$ and $X^{2}\delta^{\prime}(X^{2})=-\delta(X^{2}),$ we obtain%

\begin{equation}
\left(  \delta_{\varepsilon}L\right)  _{2}=\left\{
\begin{array}
[c]{c}%
\partial_{M}\left\{  -\delta(X^{2})F^{MN}\delta_{\varepsilon}A_{N}\right\} \\
-2b\delta(X^{2})\left(  D_{M}F^{MN}\right)  \left[  \overline{\varepsilon_{L}%
}\Gamma_{N}\bar{X}\lambda_{L}^{a}+h.c\right] \\
-2b\left(  X_{M}F^{MN}\right)  \left[
\begin{array}
[c]{c}%
\delta(X^{2})\left(  \overline{\varepsilon_{L}}\Gamma_{NP}D^{P}\lambda
_{L}\right) \\
+2\delta^{\prime}(X^{2})\overline{\varepsilon_{L}}\Gamma_{NP}\lambda_{L}%
^{a}X^{P}%
\end{array}
\right]  +h.c\\
+\delta^{\prime}(X^{2})2X_{N}F^{NM}\left(  \delta^{1}A_{M}^{a}\right)
\end{array}
\right\}  .
\end{equation}
Note that since $\left(  \delta^{1}A_{M}^{a}\right)  $ is proportional to
$X^{2}$ it survives only when multiplied by $\delta^{\prime}(X^{2}).$ This
term will be dropped in this section since it is present only when there is
coupling to chiral multiplets; it will be taken into account later in
Eq.(\ref{del1a}).

The term in line (\ref{delLV2}) proportional to $\delta_{\varepsilon}%
\overline{\lambda_{L}}^{a}$ which we define as $\left(  \delta_{\varepsilon
}L\right)  _{3}$ takes the form%
\begin{equation}
\left(  \delta_{\varepsilon}L\right)  _{3}=-\frac{1}{2}\delta(X^{2}%
)[b\overline{\varepsilon_{L}}\left(  \Gamma_{MN}F^{MN}\right)  +a\overline
{\varepsilon_{L}}B](X\bar{D}\lambda_{L}+\overleftarrow{D}\bar{X}\lambda
_{L})+h.c.
\end{equation}
We do an integration by parts to change the covariant derivative hitting on
$\lambda_{L}$\ to covariant derivative hitting on $F^{MN}$\ and change the
covariant derivative hitting on $B$ to covariant derivative hitting on
$\lambda_{L},$ and in this process collect a total divergence%

\begin{equation}
\left(  \delta_{\varepsilon}L\right)  _{3}=\left\{
\begin{array}
[c]{c}%
\partial_{M}\left(
\begin{array}
[c]{c}%
-\frac{1}{2}a\delta(X^{2})B\overline{\varepsilon_{L}}\Gamma^{M}\bar{X}%
\lambda_{L}\\
-\frac{1}{2}b\delta(X^{2})\overline{\varepsilon_{L}}\left(  \Gamma_{NP}%
F^{NP}\right)  X\bar{\Gamma}^{M}\lambda_{L}%
\end{array}
\right) \\
-\delta(X^{2})aB\overline{\varepsilon_{L}}\left[  X\bar{D}-\left(  X\cdot
D+2\right)  \right]  \lambda_{L}\\
-\delta(X^{2})b\overline{\varepsilon_{L}}\left(  \Gamma_{MN}F^{MN}\right)
\Gamma_{PQ}\overleftarrow{D}^{P}X^{Q}\lambda_{L}\\
+2b\delta(X^{2})\overline{\varepsilon_{L}}\left(  \Gamma_{MN}F^{MN}\right)
\lambda_{L}%
\end{array}
\right\}  +h.c
\end{equation}
This is further developed with some gamma matrix algebra
\begin{equation}%
\begin{array}
[c]{c}%
\Gamma_{MN}\Gamma_{PQ}=\Gamma_{MNPQ}+\{\eta_{NP}\Gamma_{MQ}-\eta_{MP}%
\Gamma_{NQ}+\eta_{MQ}\Gamma_{NP}-\eta_{NQ}\Gamma_{MP}\}+\left\{  \eta_{NP}%
\eta_{MQ}-\eta_{MP}\eta_{NQ}\right\}  ,
\end{array}
\label{gidentities}%
\end{equation}
and simplifications due to the Bianchi identity $\Gamma_{MNPQ}\left(
D^{P}F^{MN}\right)  =0.$

Combining all the terms we find a total divergence after cancellations%
\begin{equation}
\left(  \delta_{\varepsilon}L\right)  _{1}+\left(  \delta_{\varepsilon
}L\right)  _{2}+\left(  \delta_{\varepsilon}L\right)  _{3}=\partial_{M}\left[
\delta(X^{2})\left(  \overline{\varepsilon_{L}}V_{L}^{M}+\overline
{\varepsilon_{R}}V_{R}^{M}\right)  ^{vector}\right]  \label{varyLvector}%
\end{equation}
where $\overline{\varepsilon_{L}}V_{L}^{M}$ is%
\begin{equation}
\left(  \overline{\varepsilon_{L}}V^{M}\right)  ^{vector}=\left\{
\begin{array}
[c]{c}%
-F^{MN}\left(  \delta_{\overline{\varepsilon_{L}}}A_{N}\right)  -\frac{1}%
{2}aB\overline{\varepsilon_{L}}\Gamma^{M}\bar{X}\lambda_{L}\\
-\frac{1}{2}bF^{PQ}\overline{\varepsilon_{L}}\Gamma_{PQ}X\bar{\Gamma}%
^{M}\lambda_{L}+2b\left(  X^{Q}F_{QP}\right)  \overline{\varepsilon_{L}}%
\Gamma^{MP}\lambda_{L}%
\end{array}
\right\}  \label{Vvector}%
\end{equation}
and similarly for $\overline{\varepsilon_{R}}V_{R}^{M}$ (obtained by replacing
left$\leftrightarrow$right) which is the Hermitian conjugate of $\overline
{\varepsilon_{L}}V_{L}^{M}$ (verified via the formulas in Appendix (\ref{A}))

Hence we have shown that the Lagrangian (\ref{Lvector}) is symmetric under the
given SUSY transformations for any complex numbers $a$ and $b$. This means
that in the transformation rules (\ref{delA}-\ref{delB}) we can replace
$a\varepsilon_{L}$ by an an independent SUSY parameter than the $b\varepsilon
_{L},$ so that the SUSY symmetry of the Lagrangian (\ref{Lvector}) is actually
twice as large. However, we will see that we will have to fix $a$ and $b$
relative to each other when there is interaction with chiral multiplets (see
Eq.(\ref{values})).

Noether's theorem for the theory with only vector supermultiplets in
Eq.(\ref{Lvector}) gives the following supercurrent
\begin{equation}
\left(  \overline{\varepsilon_{L}}J_{L}^{M}\right)  ^{vector}=\frac{\partial
L}{\partial\left(  \partial_{M}A_{N}\right)  }\left(  \delta_{\overline
{\varepsilon_{L}}}A_{N}\right)  +\left(  \delta_{\varepsilon}\overline
{\lambda_{L}}\right)  \frac{\partial L}{\partial\left(  \partial_{M}%
\overline{\lambda_{L}}\right)  }-\delta(X^{2})\left(  \overline{\varepsilon
_{L}}V_{L}^{M}\right)  ^{vector}%
\end{equation}
and similarly for $\left(  \overline{\varepsilon_{R}}J_{R}^{M}\right)
^{vector}$. The part $\frac{\partial L}{\partial\left(  \partial_{M}%
A_{N}\right)  }\left(  \delta_{\overline{\varepsilon_{L}}}A_{N}\right)
=-F^{MN}\left(  \delta_{\overline{\varepsilon_{L}}}A_{N}\right)  $ cancels
against an equal term in $\overline{\varepsilon_{L}}V_{vector}^{M}$ of
Eq.(\ref{Vvector}). The part $\left(  \delta_{\varepsilon}\overline
{\lambda_{L}}\right)  \frac{\partial L}{\partial\left(  \partial_{M}%
\overline{\lambda_{L}}\right)  }=\delta(X^{2})\left(  \delta_{\varepsilon
}\overline{\lambda_{L}}\right)  \left(  \frac{i}{2}\Gamma^{M}X\lambda
_{L}\right)  $ gives%
\begin{equation}
\left(  \delta_{\varepsilon}\overline{\lambda_{L}}\right)  \frac{\partial
L}{\partial\left(  \partial_{M}\overline{\lambda_{L}}\right)  }=\delta
(X^{2})\left\{  -\frac{1}{2}bF_{PQ}\overline{\varepsilon_{L}}\Gamma^{PQ}%
\Gamma^{M}\bar{X}\lambda_{L}-\frac{a}{2}B\overline{\varepsilon_{L}}\Gamma
^{M}\bar{X}\lambda_{L}\right\}  .
\end{equation}
The piece proportional to $a$ cancels against an equal term in $\overline
{\varepsilon_{L}}V_{vector}^{M}.$ After these simplifications we are left with
the following terms proportional only to $b\overline{\varepsilon_{L}}$%
\begin{equation}
\left(  \overline{\varepsilon_{L}}J_{L}^{M}\right)  ^{vector}=\delta
(X^{2})\left\{  -\frac{1}{2}bF_{PQ}~\overline{\varepsilon_{L}}\Gamma
^{PQ}\left(  \Gamma^{M}\bar{X}-X\bar{\Gamma}^{M}\right)  \lambda_{L}-2b\left(
X^{Q}F_{QP}\right)  ~\overline{\varepsilon_{L}}\Gamma^{MP}\lambda_{L}\right\}
\end{equation}
By using gamma matrix identities (\ref{gidentities}) and (\ref{gid1}%
-\ref{gid5}) it is convenient to bring this to the following alternative forms%
\begin{align}
\left(  \overline{\varepsilon_{L}}J_{L}^{M}\right)  ^{vector}  &
=b\delta(X^{2})F_{PQ}X_{N}~\overline{\varepsilon_{L}}\left(  \Gamma
^{PQNM}+2\eta^{MP}\Gamma^{Q}\bar{\Gamma}^{N}\right)  \lambda_{L}\\
&  =b\delta(X^{2})F_{PQ}X_{N}~\overline{\varepsilon_{L}}\left(  \Gamma
^{PQN}\bar{\Gamma}^{M}-\eta^{NM}\Gamma^{PQ}+2\eta^{MP}\eta^{NQ}\right)
\lambda_{L}. \label{JvectorL}%
\end{align}
The form in the last line makes it easy to check that this current is
conserved $\partial_{M}\left(  \overline{\varepsilon_{L}}J_{L}^{M}\right)
^{vector}=0$ as follows. After using some of the equations of motion, in
particular $X^{N}F_{MN}=0$ and\footnote{From $X^{N}F_{MN}^{a}=0$ it follows
that $\left(  X\cdot D+2\right)  F_{MN}^{a}=0$. This is needed, along with the
other equations to prove Eq.(\ref{propeom}).} $\left(  X\cdot D+2\right)
\lambda_{L}=0$ plus the Bianchi identity $D_{[M}F_{PQ]}=0$, the divergence of
the current becomes proportional to the remaining equations of motion
$X\bar{D}\lambda_{L}^{a}$ and $D^{M}F_{MQ}^{a}$ as follows
\begin{equation}
\partial_{M}\left(  \overline{\varepsilon_{L}}J_{L}^{M}\right)  ^{vector}%
=\delta(X^{2})\left\{  F_{PQ}\overline{\varepsilon_{L}}\Gamma^{PQ}X\bar
{D}\lambda_{L}+2\overline{\varepsilon_{L}}\Gamma^{QN}\lambda_{L}X_{N}\left(
D^{M}F_{MQ}\right)  \right\}  =sources~. \label{propeom}%
\end{equation}
In the absence of coupling between the vector and chiral multiplets the
sources in the equations of motion are
\begin{equation}
X\bar{D}\lambda_{L}^{a}=0,\;\;D^{M}F_{MQ}^{a}=gf^{abc}\left(  \overline
{\lambda_{L}^{b}}\Gamma_{QP}\lambda_{L}^{c}\right)  X^{P}.
\end{equation}
Therefore we obtain
\begin{equation}
\partial_{M}\left(  \overline{\varepsilon_{L}}J_{L}^{M}\right)  ^{vector}%
=\delta(X^{2})2f^{abc}X_{N}X^{P}\overline{\varepsilon_{L}}\Gamma^{QN}%
\lambda_{L}^{a}~\overline{\lambda_{L}^{b}}\Gamma_{QP}\lambda_{L}^{c}=0
\end{equation}
where in the last step we have used the Fierz identity in footnote
(\ref{fierz2}) which is valid only in special dimensions (in particular valid
for SO$\left(  4,2\right)  $). Hence, the pure vector-multiplet current is
conserved by itself, $\partial_{M}\left(  J_{L}^{M}\right)  ^{vector}=0.$ The
conservation of the current amounts also to a proof of SUSY for the theory of
Eq.(\ref{Lvector}) that supplies the equations of motion.

The Hermitian conjugate of this conserved current can be written as the right
handed current (see Appendix (\ref{A}) for Hermitian and charge conjugation
properties of gamma matrices).
\begin{equation}
\left(  \overline{\varepsilon_{R}}J_{R}^{M}\right)  ^{vector}=b^{\ast}%
\delta(X^{2})X_{N}F_{PQ}^{a}~\overline{\varepsilon_{R}}\left(  \bar{\Gamma
}^{PQN}\Gamma^{M}-\eta^{NM}\bar{\Gamma}^{PQ}\right)  \lambda_{R}%
^{a},\label{JvectorR}%
\end{equation}
Here we have dropped the term proportional to $X_{N}F^{MN}$ since the current
can be modified by terms proportional to the equations of motion, and one of
them happens to be $X_{N}F^{MN}=0$. The corresponding term should then be
dropped also from $\overline{\varepsilon_{L}}J_{L}^{M}$ in Eq.(\ref{JvectorL}).

Although we do not discuss it in detail, it is worth mentioning that the
currents $\left(  J_{L}^{M},J_{R}^{M}\right)  ^{vector}$ are invariant under
the 2T-gauge transformations \cite{2tstandardM} for the fields $\left(
A_{M}^{a},\lambda_{R}^{a},B^{a}\right)  $ and therefore they are gauge
invariant physical observables from the point of view of all the gauge symmetries.

\section{Supersymmetric 2T-physics with Fields of Spins 0,$\frac{1}{2}$,1
\label{full}}

The next step is to couple chiral supermultiplets $\left(  \varphi,\psi
_{L},F\right)  _{i}$ minimally to vector supermultiplets $\left(  \lambda
_{L},A_{M},B\right)  ^{a}$ to describe gauge interactions. This requires more
than promoting the ordinary derivatives to covariant derivatives, namely more
interaction terms also need to be added (the $\alpha,\beta$ terms) as in the
full action in Eq.(\ref{actionsusy},\ref{Lint}). Once SUSY is achieved we will
calculate the full conserved supercurrent for the coupled theory.

The full SUSY transformation rules for the chiral multiplet are the
gauge-covariantized versions of those given in Eqs.(\ref{susy1}-\ref{susy333}%
), but including also the non-zero extra terms $\left(  \delta_{\varepsilon
}^{1}F,\delta_{\varepsilon}^{1}\varphi\right)  $ which will be determined
below in Eqs.(\ref{delF*},\ref{delphi*}). Similarly, the SUSY transformation
rules for the vector multiplet are those given in Eqs.(\ref{delAa}-\ref{delB})
but with the extra non-zero term $\delta_{\varepsilon}^{1}A$ which is
determined below in Eq.(\ref{delA*}).

We will see that the parameters $\alpha,\beta,a,b$ will be fully fixed (see
Eq(\ref{values})). Eventually when we construct supergravity in the 2T
formalism, the dilaton and its partners will also contribute to the
transformation rules and restrict the possible parameters such as $\xi$ and
$V\left(  \Phi,\varphi\right)  $ which appear in $L_{dilaton}$. In this
section the dilaton and its partners will be neglected, so we will assume the
case with $L_{dilaton}$ set to zero. With all these points taken into account,
the full SUSY transformation rules will be shown to be those given in
Eqs.(\ref{delphi}-\ref{delBfull}).

Thus we consider the full Lagrangian of Eq.(\ref{actionsusy}), without a
dilaton $\Phi$ written in three pieces $L=L_{vector}+L_{chiral}+L_{int}.$ Here
$L_{vector}$ is identical to Eq.(\ref{Lvector}), $L_{chiral}$ is the gauge
covariantized version of $S_{0}+S_{int}$ of Eqs.(\ref{convenient}%
,\ref{convenient2},\ref{Sint}), and $L_{int}$ is the expression given in
Eq.(\ref{Lint}). The full SUSY variation of the parts $L_{vector}+L_{chiral}$
is almost identical to the variations discussed in the previous sections for
the uncoupled multiplets, except for replacing all derivatives by covariant
derivatives in $L_{chiral}$, and taking into account extra terms that appear
as follows

\begin{description}
\item[1.] Varying $A_{M}$ that occurs in the covariant derivatives$,\;\delta
_{\varepsilon}L_{chiral}\rightarrow\frac{\partial L_{chiral}}{\partial
A_{M}^{a}}\left(  \delta_{\varepsilon}A_{M}^{a}\right)  $.

\item[2.] The effect of the extra $\delta_{\varepsilon}^{1}F$ term
$\delta_{\varepsilon}L_{chiral}\rightarrow(\frac{\partial L_{chiral}}{\partial
F_{i}}\delta_{\varepsilon}^{1}F_{i}+h.c.).$

\item[3.] New terms that arise in $\delta_{\varepsilon}L_{chiral}$ due to
changing of orders of non-commuting covariant derivatives\footnote{The terms
proportional to $\left[  D_{M},D_{N}\right]  \sim F_{MN}$ can be obtained by
going over the computations in footnote (\ref{Vfree}) and replacing covariant
derivatives in all appropriate places. The terms that arise from commuting
covariant derivatives has the form $\delta\left(  X^{2}\right)  \left(
\left[  D_{M},D_{N}\right]  \varphi^{\dagger}\right)  ^{i}~\overline
{\varepsilon_{R}}\left[  \frac{1}{2}\bar{\Gamma}^{MN}\bar{X}-\bar{\Gamma}%
^{M}X^{N}\right]  \psi_{Li}+h.c.$, where we replace $\left[  D_{M}%
,D_{N}\right]  \varphi^{\dagger}=ig\left(  \varphi^{\dagger}F_{MN}\right)
^{i}.$ To obtain the expression in Eq.(\ref{delFA}) we prefer to use the
hermitian conjugate version of this expression $-ig\delta\left(  X^{2}\right)
\left(  F_{MN}\varphi\right)  _{i}$ $\overline{\psi_{L}}^{i}\left[  -\frac
{1}{2}X\Gamma^{MN}-\Gamma^{M}X^{N}\right]  \varepsilon_{R}+h.c.$ where we have
used Eqs.(\ref{herm}), and then use the Majorana properties of Eqs.(\ref{maj1}%
,\ref{maj5}) to rewrite it in the form $-ig\delta\left(  X^{2}\right)  \left(
F_{MN}\varphi\right)  _{i}$ $\overline{\varepsilon_{L}}\left[  \frac{1}%
{2}\Gamma^{MN}X-\Gamma^{M}X^{N}\right]  \psi_{R}^{i}+h.c.$}.

\item[4.] The effect of the extra $\delta_{\varepsilon}^{1}\varphi
,\delta_{\varepsilon}^{1}A$ terms in the variation of $\delta_{\varepsilon
}\left(  L_{vector}+L_{chiral}\right)  $. We leave these for last because they
are both proportional to $X^{2}$ so they drop out in most terms due to the
overall $\delta\left(  X^{2}\right)  $ in the Lagrangian. They can contribute
only through the variation of the kinetic terms and through the terms in the
action proportional to $\delta^{\prime}\left(  X^{2}\right)  .$
\end{description}

Hence, for $\delta_{\varepsilon}\left(  L_{vector}+L_{chiral}\right)  $ we can
use the results of the previous sections plus the extra modifications listed
above, and then add the full variation of the coupling term $\delta
_{\varepsilon}L_{int}$. So, the computation is organized as follows%
\begin{align}
\delta_{\varepsilon}L  &  =\partial_{M}\left(  \left(  \overline
{\varepsilon_{L}}V_{L}^{M}+\overline{\varepsilon_{R}}V_{R}^{M}\right)
^{chiral}+\left(  \overline{\varepsilon_{L}}V_{L}^{M}+\overline{\varepsilon
_{R}}V_{R}^{M}\right)  ^{vector}\right) \label{divs}\\
&  +\delta_{\varepsilon}L_{1+2+3}^{extra}+\delta_{\varepsilon}L_{int}%
+\delta_{\varepsilon}L_{4}^{extra}%
\end{align}
In the second line the subscripts indicate the variations that correspond to
the items listed above. The first line is the total divergence results of
Eqs.(\ref{Vtot},\ref{varyLvector}), where $\left(  V_{L,R}^{M}\right)
^{vector}$ are identical to Eq.(\ref{Vvector}), while $\left(  V_{L,R}%
^{M}\right)  ^{chiral}$ is given in Eq.(\ref{Vtot}) except for replacing all
derivatives by covariant derivatives.

The three items in $\delta_{\varepsilon}L_{1+2+3}^{extra}$ give the following
contributions
\begin{align}
\delta_{\varepsilon}L_{1+2+3}^{extra}  &  =\delta\left(  X^{2}\right)
g\left(  \delta_{\varepsilon}A_{M}^{a}\right)  \left[  -i\varphi^{\dagger
}t_{a}\overleftrightarrow{D}_{M}\varphi+X^{N}\overline{\psi_{L}}\Gamma
_{NM}t_{a}\psi_{L}\right] \label{delAextra}\\
&  +\delta\left(  X^{2}\right)  \left(  \delta_{\varepsilon}^{1}F^{\dagger
i}\right)  \left(  F_{i}+\frac{\partial W^{\ast}}{\partial\varphi^{\ast i}%
}\right)  +h.c.\label{delFextra}\\
&  +i\frac{g}{2}\delta\left(  X^{2}\right)  \left(  F_{MN}\varphi\right)
_{i}~\bar{\varepsilon}_{L}\left[  -\Gamma^{MN}X+2\Gamma^{M}X^{N}\right]
\psi_{R}^{i}+h.c. \label{delFA}%
\end{align}
It is evident that $\left(  \delta_{\varepsilon}^{1}F^{\dagger i}\right)  $
must be chosen to cancel terms proportional to $F_{i}$ coming from varying the
coupling term $\delta_{\varepsilon}L_{int}$. The only new contribution
proportional to $F_{i}$ is the variation of $\psi$ in the coupling term
$\delta_{\varepsilon}L_{int}$ given explicitly below. As will be verified
below, this fixes uniquely the extra piece in the SUSY transformation of
$F_{i},F^{\dagger i}$ as
\begin{equation}
\left(  \delta_{\varepsilon}^{1}F^{\dagger i}\right)  =i\beta\left(
\varepsilon_{L}^{T}\left(  C\bar{X}\right)  \lambda_{aL}\right)  \left(
\varphi^{\dagger}t^{a}\right)  ^{i} \label{delF*}%
\end{equation}
Inserting this $\left(  \delta_{\varepsilon}^{1}F^{\dagger i}\right)  $ and
$\left(  \delta_{\varepsilon}A_{M}^{a}\right)  $ from Eq.(\ref{delA}) into
$\delta_{\varepsilon}L_{1+2+3}^{extra}$ we obtain
\begin{align}
\delta_{\varepsilon}L_{1+2+3}^{extra}  &  =-2bg\delta\left(  X^{2}\right)
\left\{
\begin{array}
[c]{c}%
-i\left[  \overline{\varepsilon_{L}}\Gamma_{M}\bar{X}\lambda_{L}^{a}\right]
\left[  \varphi^{\dagger}t_{a}\overleftrightarrow{D}_{M}\varphi\right] \\
+\left(  \overline{\varepsilon_{L}}\Gamma^{MN}\lambda_{L}^{a}\right)  \left(
\overline{\psi_{L}}\Gamma_{PM}t_{a}\psi_{L}\right)  X_{N}X^{P}%
\end{array}
\right\}  +h.c.\label{Achiral}\\
&  +\delta\left(  X^{2}\right)  i\beta\left(  \varepsilon_{L}^{T}\left(
C\bar{X}\right)  \lambda_{aL}\right)  \left(  \varphi^{\dagger}t^{a}\right)
^{i}\left(  F_{i}+\frac{\partial W^{\ast}}{\partial\varphi^{\ast i}}\right)
+h.c.\label{Achiral3}\\
&  +i\frac{g}{2}\delta\left(  X^{2}\right)  ~\bar{\varepsilon}_{L}\left(
-\Gamma^{MN}X+2\Gamma^{M}X^{N}\right)  \psi_{R}^{i}~\left(  F_{MN}%
\varphi\right)  _{i}+h.c. \label{Achiral2}%
\end{align}
Next we compute $\delta_{\varepsilon}L_{int}$ by varying Eq.(\ref{Lint})
\begin{align}
\delta_{\varepsilon}L_{int}  &  =\alpha\delta\left(  X^{2}\right)  \left\{
\begin{array}
[c]{c}%
\left[  \left(  \varphi^{\dagger}t_{a}\varphi\right)  +\xi_{a}\Phi^{2}\right]
\left(  \delta_{\varepsilon}B^{a}\right) \\
+\left(  \delta_{\varepsilon}\varphi^{\dagger}t_{a}\varphi\right)  B^{a}+h.c.
\end{array}
\right\} \\
&  +\beta\delta\left(  X^{2}\right)  \left\{
\begin{array}
[c]{c}%
\left(  \delta_{\varepsilon}\varphi^{\dagger}\right)  t^{a}\left(  \left(
\psi_{L}\right)  ^{T}\left(  C\bar{X}\right)  \lambda_{aL}\right) \\
+\varphi^{\dagger}t^{a}\left(  \left(  \delta_{\varepsilon}\psi_{L}\right)
^{T}\left(  C\bar{X}\right)  \lambda_{aL}\right) \\
+\left(  \delta_{\varepsilon}\overline{\lambda_{aL}}\right)  X\psi_{R}%
t_{a}\varphi
\end{array}
\right\}  +h.c.
\end{align}
We insert the $\left(  \delta_{\varepsilon}\overline{\lambda_{aL}}\right)  $
in Eq.(\ref{dellambar}) and $\delta_{\varepsilon}\varphi^{\dagger i},$
$\delta_{\varepsilon}\psi_{iL}$ in Eqs.(\ref{susy1}-\ref{susy333}) with gauge
covariant derivatives replacing ordinary derivatives. After dropping the terms
$X^{2}\delta\left(  X^{2}\right)  =0$ we obtain
\begin{align}
\delta_{\varepsilon}L_{int}  &  =\alpha\delta\left(  X^{2}\right)  \left\{
\begin{array}
[c]{c}%
a\left(  \varphi^{\dagger}t_{a}\varphi\right)  \left[  \overline
{\varepsilon_{L}}\left(  \Gamma^{MN}X_{M}D_{N}-2\right)  \lambda_{L}%
^{a}\right] \\
+B^{a}\left(  \overline{\varepsilon_{L}}X\psi_{R}^{i}\right)  \left(
t_{a}\varphi\right)  _{i}%
\end{array}
\right\}  +h.c.\label{ffel0}\\
&  +\beta\delta\left(  X^{2}\right)  \left\{
\begin{array}
[c]{c}%
\left(  \overline{\varepsilon_{L}}X\psi_{R}t^{a}\right)  ^{i}\left(  \psi
_{iL}^{T}\left(  C\bar{X}\right)  \lambda_{aL}\right) \\
-i\left(  \varphi^{\dagger}t^{a}D_{M}\varphi\right)  \overline{\varepsilon
_{L}}\Gamma^{M}\bar{X}\lambda_{aL}\\
-i\left(  \varphi^{\dagger}t^{a}\right)  ^{i}F_{i}\left(  \varepsilon_{L}%
^{T}\left(  C\bar{X}\right)  \lambda_{aL}\right) \\
ibF_{MN}^{a}\left(  \overline{\varepsilon_{L}}\Gamma^{MN}X\psi_{R}^{i}\right)
\left(  t_{a}\varphi\right)  _{i}\\
+iaB^{a}\left(  \overline{\varepsilon_{L}}X\psi_{R}^{i}\right)  \left(
t_{a}\varphi\right)  _{i}%
\end{array}
\right\}  +h.c. \label{ppel}%
\end{align}
Note that the terms proportional to $F_{i},$ that appear in
Eqs.(\ref{Achiral3}) and the third line of Eq.(\ref{ppel}), cancel by the
choice of $\left(  \delta_{\varepsilon}^{1}F^{\dagger i}\right)  $ of
Eq.(\ref{delF*}) as anticipated. To cancel some of the other terms in the sum
$\delta_{\varepsilon}L_{chiral}^{extra}+\delta_{\varepsilon}L_{int}$ we fix
the unknown coefficients $\alpha,\beta,a,b$ as follows

\begin{itemize}
\item $\alpha=-i\beta a,$ cancels terms proportional to $B^{a}$ that appear in
the last lines of Eqs.(\ref{ffel0},\ref{ppel}).

\item $\beta b=\frac{g}{2},~$cancels partially terms proportional to
$F_{MN}^{a}$ that appear in Eqs.(\ref{Achiral2},\ref{ppel}). The leftover is
\begin{equation}
ig\delta\left(  X^{2}\right)  \left(  \bar{\varepsilon}_{L}\Gamma^{M}\psi
_{R}^{i}\right)  \left(  F_{MN}\varphi\right)  _{i}X^{N}+h.c. \label{remainF}%
\end{equation}

\item $\beta=4bg~$cancels partially terms proportional to $\left(
\varphi^{\dagger}t^{a}D_{M}\varphi\right)  $ that appear in Eqs.(\ref{Achiral}%
,\ref{ppel}). The leftover is $-2ibg\delta\left(  X^{2}\right)  \left(
\overline{\varepsilon_{L}}\Gamma_{M}\bar{X}\lambda_{L}^{a}\right)
D_{M}\left(  \varphi^{\dagger}t_{a}\varphi\right)  $ which can be rewritten in
the following form by using the gamma matrix identities $\Gamma_{M}\bar
{X}=\Gamma_{MN}X^{N}+X_{M}$
\begin{equation}
-2ibg\delta\left(  X^{2}\right)  \left(
\begin{array}
[c]{c}%
D_{M}\left(  \varphi^{\dagger}t_{a}\varphi\right)  \left[  \overline
{\varepsilon_{L}}\Gamma_{MN}\lambda_{L}^{a}\right]  X^{N}\\
+\left(  \overline{\varepsilon_{L}}\lambda_{L}^{a}\right)  \left[  \left(
X\cdot D+2\right)  \left(  \varphi^{\dagger}t_{a}\varphi\right)  \right] \\
-2\left(  \overline{\varepsilon_{L}}\lambda_{L}^{a}\right)  \left(
\varphi^{\dagger}t_{a}\varphi\right)
\end{array}
\right)  \label{ffel}%
\end{equation}
Furthermore, using the same $4bg=\beta,$ the charge conjugation property
$\psi_{R}=C\overline{\psi_{L}}^{T},$ and a Fierz identity, we cancel the terms
of the form $\bar{\varepsilon}\lambda\bar{\psi}\psi XX$ that appear in
(\ref{Achiral},\ref{ppel}).

\item $2\alpha a=4ibg$ cancels the terms of the form $\left(  \varphi
^{\dagger}t_{a}\varphi\right)  \left(  \overline{\varepsilon_{L}}\lambda
_{L}^{a}\right)  $ that appear in the first line of Eq.(\ref{ffel0}) and the
last line of Eq.(\ref{ffel})
\end{itemize}

From these conditions the unknown coefficients are completely fixed as $a=\pm
i\sqrt{1/2}$,\ $b=\pm^{\prime}\sqrt{1/8}$,\ $\alpha=\pm\pm^{\prime}g$%
,\ $\beta=\pm^{\prime}\sqrt{2}g.$ The $\pm,\pm^{\prime}$ signs can be absorbed
by a redefinition of the signs of $\lambda,B$ wherever they appear in the
Lagrangian and transformation rules. Therefore it is sufficiently general to
choose one set of signs for these coefficients, thus we settle with the upper
signs
\begin{equation}
a=i\frac{1}{\sqrt{2}},\;b=\frac{1}{2\sqrt{2}},\;\alpha=g,\;\beta=\sqrt{2}g
\label{values}%
\end{equation}
to agree with conventions in the case of $3+1$ dimension.

The remaining terms that have not canceled so far come from the first and
second lines of Eq.((\ref{ffel0}), the first and second lines of
Eq.((\ref{ffel}), the remainder in Eq.(\ref{remainF}) and the term
proportional to $\frac{\partial W^{\ast}}{\partial\varphi^{\ast i}}$ in
Eq.(\ref{Achiral3}). These are collected below after inserting the constants
above%
\begin{align}
&  \delta_{\varepsilon}L_{1+2+3}^{extra}+\delta_{\varepsilon}L_{int}\\
&  =\frac{ig}{\sqrt{2}}\delta\left(  X^{2}\right)  \left(  \left(
\varphi^{\dagger}t_{a}\varphi\right)  \left[  \overline{\varepsilon_{L}}%
\Gamma^{MN}X_{M}D_{N}\lambda_{L}^{a}\right]  \right)  +h.c\\
&  -\frac{ig}{\sqrt{2}}\delta\left(  X^{2}\right)  \left(
\begin{array}
[c]{c}%
D_{M}\left(  \varphi^{\dagger}t_{a}\varphi\right)  \left[  \overline
{\varepsilon_{L}}\Gamma_{MN}\lambda_{L}^{a}\right]  X^{N}\\
+\left(  \overline{\varepsilon_{L}}\lambda_{L}^{a}\right)  \left[  \left(
X\cdot D+2\right)  \left(  \varphi^{\dagger}t_{a}\varphi\right)  \right]
\end{array}
\right)  +h.c.\\
&  +ig\delta\left(  X^{2}\right)  \left(  \bar{\varepsilon}_{L}\Gamma^{M}%
\psi_{R}^{i}\right)  \left(  F_{MN}\varphi\right)  _{i}X^{N}+h.c.\\
&  -i\sqrt{2}g\delta\left(  X^{2}\right)  \left(  \varphi^{\dagger}%
t^{a}\right)  ^{i}\frac{\partial W^{\ast}}{\partial\varphi^{\ast i}}\left(
\varepsilon_{L}^{T}\left(  C\bar{X}\right)  \lambda_{aL}\right)  +h.c.
\end{align}
The last term vanishes because $W^{\ast}$ is gauge invariant, which requires
$\left(  \varphi^{\dagger}t^{a}\right)  ^{i}\frac{\partial W^{\ast}}%
{\partial\varphi^{\ast i}}=0$ as in Eq.(\ref{Gsymm}).

The remaining terms assemble into a total divergence plus terms that are
proportional to the subset of equations of motion that imply homogeneity
conditions on the fields $\left[  \left(  X\cdot D+2\right)  \left(
\varphi^{\dagger}t_{a}\varphi\right)  \right]  $ and $F_{MN}X^{N}.$ These
would vanish by the (homogeneity) subset of equations of motion on mass shell.
However, off-mass shell they can be canceled by additional pieces
$\delta_{\varepsilon}^{1}\varphi_{i}$ , $\delta_{\varepsilon}^{1}A_{M}^{a}$ in
the SUSY transformation of $\varphi,A_{M}^{a},$ by taking $\delta
_{\varepsilon}^{1}\varphi_{i}$ , $\delta_{\varepsilon}^{1}A_{M}^{a}$ to be
proportional proportional to $X^{2}$. These extra pieces generally drop out in
most terms in the SUSY variation of the Lagrangian due to $X^{2}\delta\left(
X^{2}\right)  =0,$ but survive in some of the kinetic terms and the terms that
contain $\delta^{\prime}\left(  X^{2}\right)  $. The $\delta_{\varepsilon}%
^{1}\varphi_{i}$ , $\delta_{\varepsilon}^{1}A_{M}^{a}$ variation of the
Lagrangian $\delta_{\varepsilon}\left(  L_{vector}+L_{chiral}\right)  $ are
also proportional to the subset of equations of motion $\left[  \left(  X\cdot
D+2\right)  \left(  \varphi^{\dagger}t_{a}\varphi\right)  \right]  $ or
$F_{MN}X^{N},$ as follows
\begin{equation}
\delta_{\varepsilon}L_{4}^{extra}=2\delta^{\prime}\left(  X^{2}\right)
\left(  X_{N}F_{a}^{NM}\right)  \delta_{\varepsilon}^{1}A_{M}^{a}%
+2\delta^{\prime}\left(  X^{2}\right)  \left[  \left(  X\cdot D+1\right)
\varphi^{\dagger i}\right]  \delta_{\varepsilon}^{1}\varphi_{i}%
+h.c.\label{del1a}%
\end{equation}
Therefore we can add the extra pieces $\delta_{\varepsilon}^{1}\varphi_{i}$ ,
$\delta_{\varepsilon}^{1}A_{M}^{a}$ to the variation of $\varphi_{i}$ ,
$A_{M}^{a}$ to cancel the terms noted above. So we choose
\begin{align}
\delta_{\varepsilon}^{1}A_{M}^{a} &  =-i\frac{g}{4}X^{2}\left(  \bar
{\varepsilon}_{L}\Gamma_{M}\psi_{R}^{i}\right)  \left(  t_{a}\varphi\right)
_{i}+h.c.,\label{delA*}\\
\delta_{\varepsilon}^{1}\varphi_{i} &  =-i\frac{g}{2\sqrt{2}}X^{2}\left(
\overline{\varepsilon_{L}}\lambda_{L}^{a}+\overline{\lambda_{L}}%
\varepsilon_{L}\right)  \left(  t_{a}\varphi\right)  _{i}~.\label{delphi*}%
\end{align}
Then we obtain the following expression
\begin{align}
&  \delta_{\varepsilon}L_{1+2+3}^{extra}+\delta_{\varepsilon}L_{int}%
+\delta_{\varepsilon}L_{4}^{extra}\\
&  =\frac{ig}{\sqrt{2}}\delta\left(  X^{2}\right)  \left(
\begin{array}
[c]{c}%
\left(  \varphi^{\dagger}t_{a}\varphi\right)  \left[  \overline{\varepsilon
_{L}}\Gamma^{MN}X_{M}D_{N}\lambda_{L}^{a}\right]  \\
-D_{M}\left(  \varphi^{\dagger}t_{a}\varphi\right)  \left[  \overline
{\varepsilon_{L}}\Gamma_{MN}\lambda_{L}^{a}\right]  X^{N}%
\end{array}
\right)  +h.c.\\
&  =\partial_{M}\left(  \delta\left(  X^{2}\right)  \left(  \overline
{\varepsilon_{L}}V_{L}^{M}+\overline{\varepsilon_{R}}V_{R}^{M}\right)
^{int}\right)  .\label{Vm5}%
\end{align}
In the final form we see that we have obtained a total divergence$,$ with
$\overline{\varepsilon_{L}}\left(  V_{L}^{M}\right)  ^{int}$ and its Hermitian
conjugate $\overline{\varepsilon_{R}}\left(  V_{R}^{M}\right)  ^{int}$ given
by
\begin{align}
\overline{\varepsilon_{L}}\left(  V_{L}^{M}\right)  ^{int} &  =-\frac
{ig}{\sqrt{2}}\delta\left(  X^{2}\right)  \left(  \varphi^{\dagger}%
t_{a}\varphi\right)  \left[  \overline{\varepsilon_{L}}\Gamma^{MN}X_{N}%
\lambda_{L}^{a}\right]  \label{Vm6}\\
\overline{\varepsilon_{R}}\left(  V_{R}^{M}\right)  ^{int} &  =\frac{ig}%
{\sqrt{2}}\delta\left(  X^{2}\right)  \left(  \varphi^{\dagger}t_{a}%
\varphi\right)  \left[  \overline{\varepsilon_{R}}\bar{\Gamma}^{MN}%
X_{N}\lambda_{R}^{a}\right]
\end{align}

We have shown that under the SUSY transformations the total Lagrangian (in the
absence of the dilaton) transforms into a total divergence. Using the form of
the divergence given in (\ref{Vm5}), and the previous pieces in the total
divergence noted in Eq.(\ref{divs}), we compute the conserved SUSY current by
applying Neother's theorem. The result is given by Eqs.(\ref{JL},\ref{JR}).

\subsection{Conservation of the Supercurrent}

In this section, we prove the conservation of the supercurrent obtained above.
For clarity, we separate the supercurrent into pieces and calculate one by one%
\begin{align}
\left(  J_{R}^{M}\right)  ^{total} &  =\left\{  \left(  J_{R}^{M}\right)
^{chiral}+\left(  J_{R}^{M}\right)  ^{vector}+\left(  J_{R}^{M}\right)
^{int}\right\}  \\
\left(  J_{R}^{M}\right)  ^{chiral} &  =\delta\left(  X^{2}\right)  \left\{
D_{K}\left(  X_{N}\varphi^{\dagger i}\right)  \left(  \bar{\Gamma}^{KN}%
\bar{\Gamma}^{M}-\eta^{MN}\bar{\Gamma}^{K}\right)  \psi_{iL}+\frac{\partial
W^{\ast}}{\partial\varphi^{\ast j}}X_{N}\bar{\Gamma}^{MN}\psi_{R}^{j}\right\}
\\
\left(  J_{R}^{M}\right)  ^{vector} &  =\delta\left(  X^{2}\right)  ~\frac
{1}{2\sqrt{2}}\left\{  F_{KL}^{a}X_{N}\left(  \bar{\Gamma}^{KLN}\Gamma
^{M}-\eta^{NM}\bar{\Gamma}^{KL}\right)  \lambda_{Ra}\right\}  \\
\left(  J_{R}^{M}\right)  ^{int} &  =\delta\left(  X^{2}\right)  \left\{
\begin{array}
[c]{c}%
-\frac{ig}{\sqrt{2}}\varphi^{\dagger i}\left(  t_{a}\varphi\right)  _{i}%
\bar{\Gamma}^{MN}\lambda_{Ra}X_{N}%
\end{array}
\right\}
\end{align}
By using the equations of motion that follow from (\ref{actionsusy}) one can
check explicitly that this current is conserved as follows. We first drop
terms that vanish because of the homogeneity conditions for on-shell fields.
Then we get%
\begin{align}
\partial_{M}\left(  J_{R}^{M}\right)  ^{chiral} &  =\left\{
\begin{array}
[c]{c}%
\overline{D}\varphi^{\dagger i}X\overline{D}\psi_{iL}-\left(  D^{2}%
\varphi^{\dagger i}\right)  \overline{X}\psi_{iL}\\
+\frac{-g}{2}\left(  F_{MN}\varphi^{\dagger}\right)  ^{i}\Gamma^{MN}\psi
_{iL}\\
-\frac{\partial W^{\dagger}}{\partial\varphi^{\dagger i}}\overline{X}D\psi
_{R}^{i}+\frac{\partial^{2}W^{\dagger}}{\partial\varphi^{\dagger i}%
\partial\varphi^{\dagger j}}\overline{D}\varphi^{\dagger j}X\psi_{R}^{i}%
\end{array}
\right\}  \label{dJchi}\\
\partial_{M}\left(  J_{R}^{M}\right)  ^{vector} &  =\left\{
\begin{array}
[c]{c}%
+\frac{1}{\sqrt{2}}D^{M}\left(  F_{MP}^{a}\right)  X_{N}\bar{\Gamma}%
^{PN}\lambda_{Ra}\\
+\frac{1}{2\sqrt{2}}F^{PQa}\bar{\Gamma}_{PQ}\overline{X}D\lambda_{Ra}%
\end{array}
\right\}  \label{dJvec}\\
\partial_{M}\left(  J_{R}^{M}\right)  ^{int} &  =\left\{
\begin{array}
[c]{c}%
+\frac{ig}{\sqrt{2}}\partial_{N}\left[  \varphi^{\dagger i}\left(
t_{a}\varphi\right)  _{i}\right]  \bar{\Gamma}^{MN}\lambda_{Ra}X_{M}\\
+\frac{ig}{\sqrt{2}}\varphi^{\dagger i}\left(  t_{a}\varphi\right)
_{i}\overline{X}D\lambda_{Ra}\\
+2\frac{ig}{\sqrt{2}}\varphi^{\dagger i}\left(  t_{a}\varphi\right)
_{i}\lambda_{Ra}%
\end{array}
\right\}  \label{dJint}%
\end{align}
Next we use the equations of motion to verify the conservation of the current.
All of the following equations, and their Hermitian conjugates, should be
multiplied by $\delta\left(  X^{2}\right)  ,$ so they are required to be
satisfied only at $X^{2}=0$%
\begin{equation}
\left(  X\cdot D+1\right)  \varphi_{i}=\left(  X\cdot D+2\right)
F_{i}=\left(  X\cdot D+2\right)  B^{a}=X^{N}F_{NM}^{a}=0\label{homo1}%
\end{equation}%
\begin{equation}
\left(  X\cdot D+2\right)  \psi_{R}^{i}=\left(  X\cdot D+2\right)  \lambda
_{R}^{a}=0\label{homo2}%
\end{equation}%
\begin{equation}
D^{2}\varphi^{\dagger i}+\frac{\partial^{2}W}{\partial\varphi_{i}%
\partial\varphi_{j}}F_{j}-\frac{i}{2}\overline{\psi_{Rj}}C\overline{X}%
\psi_{Lk}\frac{\partial^{3}W}{\partial\varphi_{i}\partial\varphi_{j}%
\partial\varphi_{k}}+g\left(  \varphi^{\dagger}B\right)  ^{i}+\sqrt{2}g\left(
\overline{\psi_{L}}t^{a}\right)  ^{i}X\lambda_{R}^{a}=0,
\end{equation}%
\begin{equation}
\left(  D_{M}F^{MN}\right)  ^{a}-if^{abc}\overline{\lambda}_{Lb}\Gamma
^{MN}\lambda_{Lc}X_{M}-ig\varphi^{\dagger}t^{a}\overleftrightarrow{D}%
^{N}\varphi+gX_{M}\overline{\psi_{L}}\Gamma^{MN}t^{a}\psi_{L}=0,
\end{equation}%
\begin{equation}
i\overline{X}D\psi_{R}^{i}+i\overline{X}\psi_{Lj}\frac{\partial^{2}W}%
{\partial\varphi_{i}\partial\varphi_{j}}-\sqrt{2}g\left(  \varphi^{\dagger
}t_{a}\overline{X}\lambda_{L}^{a}\right)  ^{i}=0,
\end{equation}%
\begin{equation}
B^{a}+g\varphi^{\dagger i}\left(  t_{a}\varphi\right)  _{i}=0,\;\;F_{i}%
+\frac{\partial W^{\dagger}}{\partial\varphi^{\dagger i}}=0,
\end{equation}%
\[
i\overline{X}D\lambda_{R}^{a}+\sqrt{2}g\varphi^{\dagger i}\left(
t_{a}\overline{X}\psi_{L}\right)  _{i}=0.
\]
The first two equations impose homogeneity conditions on the fields, while the
others control the dynamics. In the absence of the interaction given by
$L_{int},$ we had already proven that $\left(  J_{R}^{M}\right)  ^{chiral}$
and $\left(  J_{R}^{M}\right)  ^{vector}$ were conserved. In the presence of
the interaction $L_{int}$ the surviving terms come from the sources in the
equations of motion provided by $L_{int}.$ Indeed we find
\begin{align}
\partial_{M}\left(  J_{R}^{M}\right)  ^{chiral} &  =\text{source (interaction
with vector multiplet from }L_{int}\text{)}\\
&  =\delta\left(  X^{2}\right)  \left\{
\begin{array}
[c]{c}%
i\sqrt{2}g\overline{D}\varphi^{\dagger i}X\left(  \lambda_{R}\varphi\right)
_{i}+g\left(  \varphi^{\dagger}B\right)  ^{i}\overline{X}\psi_{iL}\\
+\sqrt{2}g\left(  \overline{\psi_{L}}X\lambda_{R}\right)  ^{i}\overline{X}%
\psi_{iL}+\frac{-g}{2}\left(  F_{MN}\varphi^{\dagger}\right)  ^{i}\Gamma
^{MN}\overline{X}\psi_{iL}%
\end{array}
\right\}  \\
&  =\delta\left(  X^{2}\right)  \left\{
\begin{array}
[c]{c}%
-i\sqrt{2}gX_{M}D_{N}\varphi^{\dagger i}\Gamma^{MN}\left(  \lambda_{R}%
\varphi\right)  _{i}+i\sqrt{2}g\left[  X\cdot D\varphi^{\dagger i}\right]
\left(  \lambda_{R}\varphi\right)  _{i}\\
+g\left(  \varphi^{\dagger}t^{a}\right)  ^{i}\left[  -g\varphi^{\dagger
i}\left(  t_{a}\varphi\right)  _{i}\right]  \overline{X}\psi_{iL}\\
+\sqrt{2}g\left(  \overline{\psi_{L}}X\lambda_{R}\right)  ^{i}\overline{X}%
\psi_{iL}+\frac{-ig}{2}\left(  F_{MN}\varphi^{\dagger}\right)  ^{i}\Gamma
^{MN}\overline{X}\psi_{iL}%
\end{array}
\right\}
\end{align}
where we have used the Fierz identity of Eq.(\ref{fierz1}), the gauge
invariance of $W\left(  \varphi\right)  $ as given in Eq.(\ref{Gsymm}), and
the equations of motion to substitute for $B^{a}.$

For the supercurrent arising from vector multiplet a similar argument gives%
\begin{align}
\partial_{M}\left(  J_{R}^{M}\right)  ^{vector}  &  =\text{source (interaction
with chiral multiplet from }L_{int}\text{)}\\
&  =\delta\left(  X^{2}\right)  \left\{
\begin{array}
[c]{c}%
\frac{1}{\sqrt{2}}\left(  ig\varphi^{\dagger}t^{a}\overleftrightarrow{D}%
_{P}\varphi-gX_{M}\overline{\psi_{L}}\Gamma_{\text{ \ \ }P}^{M}t^{a}\psi
_{L}\right)  X_{N}\bar{\Gamma}^{PN}\lambda_{Ra}\\
+\frac{ig}{2}F^{PQa}\bar{\Gamma}_{PQ}\varphi^{\dagger i}\left(  t_{a}%
\overline{X}\psi_{L}\right)  _{i}%
\end{array}
\right\}
\end{align}
where we have used the Fierz identity of Eq.(\ref{fierz2}). Inserting these
results in Eqs.(\ref{dJchi}-\ref{dJint}) to construct $\partial_{M}\left(
J_{R}^{M}\right)  ^{total}$ as the sum of these, and cancelling terms due to
the relation
\begin{equation}
\delta\left(  X^{2}\right)  \left[  \sqrt{2}g\left(  \overline{\psi_{L}%
}X\lambda_{R}\right)  ^{i}\overline{X}\psi_{iL}-\frac{1}{\sqrt{2}}%
gX_{M}\overline{\psi_{L}}\Gamma_{\text{ \ \ }P}^{M}t^{a}\psi_{L}X_{N}%
\bar{\Gamma}^{PN}\lambda_{Ra}\right]  =0,
\end{equation}
which follows from the Fierz identity in Eq.(\ref{fierz1}), we get
\begin{equation}
\partial_{M}\left(  J_{R}^{M}\right)  ^{total}=\delta\left(  X^{2}\right)
\left\{
\begin{array}
[c]{c}%
-i\sqrt{2}g\left(  D_{N}\varphi^{\dagger i}\right)  \left(  t_{a}%
\varphi\right)  _{i}\left(  X_{M}\Gamma^{MN}\lambda_{R}^{a}\right) \\
-\frac{ig}{\sqrt{2}}\left(  \left(  \varphi^{\dagger}t_{a}\right)
^{i}\overleftrightarrow{D}_{N}\varphi_{i}\right)  \left(  X_{M}\bar{\Gamma
}^{MN}\lambda_{R}^{a}\right) \\
+\frac{ig}{\sqrt{2}}\partial_{N}\left[  \varphi^{\dagger i}\left(
t_{a}\varphi\right)  _{i}\right]  \left(  X_{M}\bar{\Gamma}^{MN}\lambda
_{R}^{a}\right)
\end{array}
\right\}  =0,
\end{equation}
which is seen to sum up to zero. This proves the conservation of the total supercurrent.

\section{Physics consequences and future directions \label{conclude}}

In this paper we have explicitly constructed $N=1$ supersymmetric field theory
with fields of spin $0,\frac{1}{2},1$ in $4+2$ dimensions, which is compatible
with the theoretical framework of 2T field theory and its gauge symmetries.

This represents another significant step in demonstrating that 2T-physics is
sufficiently general to encompass all possible physical phenomena in
1T-physics. The importance of this is in the fact that 2T-physics unifies many
1T-physics systems with different dynamics in different spacetimes (so
different meanings of \textquotedblleft time\textquotedblright\ and
\textquotedblleft Hamiltonian\textquotedblright). By further pursuing this
concept in the context of supersymmetry we expect to obtain dually related 3+1
dimensional supersymmetric field theories. This could be used both as a tool
to perform possibly non-perturbative computations in supersymmetric 3+1 field
theory, as well as a new avenue to investigate what is meant by
\textquotedblleft space-time\textquotedblright\ and \textquotedblleft
unification\textquotedblright.

The 4+2 supersymmetry transformation is given here off-shell, and is shown to
leave invariant the action with all consistent interactions included. The SUSY
transformations are different from higher dimensional $N=1$ supersymmetry
transformations one would write down in 4+2 dimensions naively. If we
specialize to the on-shell and non-interacting version of our equations, we
find agreement with previous work \cite{ferrara} which was done at the level
of equations of motion without an action and only for free fields. Despite the
differences, the SUSY algebra, combined with the SU$(2,2)=$SO$\left(
4,2\right)  $ global symmetry of any 2T field theory, close to form the Lie
superalgebra of SU$\left(  2,2|1\right)  $ for on-shell fields, including
interactions. We checked this explicitly for the chiral multiplet as shown in
Appendix (\ref{B}), but we believe it to be true for the full interacting
theory. However, for off-shell fields the closure involves a tower of
additional 2T gauge transformations.

The coupling of chiral and vector multiplets is studied and is uniquely fixed
by the supersymmetry algebra. But unlike ordinary 1T supersymmetry, the
supersymmetry for 2T field theory requires the superpotential in the theory to
be purely cubic, which is consistent with what is required by 2T gauge
symmetry. In the framework of 2T field theory dimensionful parameters are not
permitted by the 2T-gauge symmetry. Therefore to induce soft supersymmetry
breaking it is desirable to couple the dilaton whose vacuum expectation value
plays the role of the desired dimensionful parameter. To maintain SUSY, the
superpartners of the dilaton should also be included.

After fixing the 2T gauge symmetry in a particular gauge as mentioned in
footnote (\ref{embedding}), the 4+2 supersymmetry transformation SU$\left(
2,2|1\right)  $ reduces to the non-linear superconformal transformation of the
corresponding massless fields in 3+1 dimensions.

The emergent 3+1 SUSY field theory \textit{in this gauge} is in most respects
similar to standard SUSY field theory. However, there are some interesting
additional constraints from the 4+2 structure which would not be present in
the general 3+1 SUSY theory. One of these is the banishing of the troublesome
\textit{renormalizable} CP violating terms \cite{pq}\cite{wittenaxion} of the
type $\theta\varepsilon_{\mu\nu\lambda\sigma}Tr\left(  F^{\mu\nu}%
F^{\lambda\sigma}\right)  .$ This is good for solving the strong CP violation
problem in QCD without an axion. This property of the emergent 3+1 theory
already occurs in the non-supersymmetric 2T field theory as described in
\cite{2tstandardM}, and continues to be true also in the supersymmetric case.

Recalling also that the superpotential can only be purely cubic, we see that
phase transitions like supersymmetry breaking and electroweak breaking need to
be driven by the dilaton vacuum expectation value, and hence according to
2T-physics such phase transitions must be intimately related to the physics of
the supergravity multiplet. The fact that these phenomena are not allowed to
be independent of each other makes the 2T-physics approach physically more
appealing as described in footnote (\ref{dilatondrive}).

It appears that to investigate phenomenological consequences of SUSY in the
context of 2T-physics, we will need to construct the 2T formulation of
supergravity which includes the dilaton and its couplings to matter along with
the graviton. This is one of our immediate projects. We will then be in a
position to describe a 2T version of the minimal supersymmetric standard model
(MSSM), or its extensions, including the dilaton. The new restrictions imposed
on it by 2T-physics, and the corresponding phenomenological consequences,
could be of great interest for phenomenological predictions at the LHC.

Generalization to extended supersymmetry with $N=2,4,8$ is another research
direction which is straightforward and will be discussed in a following paper.
This would proceed by constructing the higher $N$ theories from $N=1$ blocks
discussed generally in this paper. In this way the 2T gauge symmetry is
maintained while the higher $N$ structure puts more severe symmetry
restrictions on the theory. The higher $N$ theories in 4+2 dimensions will
then become laboratories for investigating non-perturbative phenomena both
from the point of view of the new 2T vistas as well as from the point of view
of earlier non-perturbative studies. The latter would include studies such as
the $N=2$ Seiberg-Witten solution\cite{Seiberg-Witten} or the $N=4$ AdS-CFT
phenomena\cite{AdSCFT}, but now directly in 4+2 dimensions.

Ultimately, the main impact of the 2T point of view is likely to be along the
ideas described in the first paragraph of this section, so we emphasize this
again: In coming down to 3+1 dimensions there are a variety of spacetimes that
can be obtained through the gauge fixing of the 2T gauge symmetry, and this is
expected to generate a web of dual supersymmetric field theories one of which
is the well-known 3+1 dimensional chiral multiplets coupled to the vector
multiplets. We expect that nonperturbative information can be obtained from
such dualities. The methods for performing this research will be discussed in
a following paper\cite{ibquelin}.

\begin{acknowledgments}
We gratefully acknowledge discussions with S-H. Chen, B. Orcal, and G. Quelin.
\end{acknowledgments}

\appendix

\section{Gamma matrices for SO$\left(  d,2\right)  $ and SO$\left(
4,2\right)  =$SU$\left(  2,2\right)  $ \label{A}}

We consider at first even dimensions $d+2$ in general, for a spacetime $X^{M}$
labeled by $M,$ which forms the vector basis of SO$\left(  d,2\right)  $.
There are two Weyl spinors labeled by $\alpha,\dot{\alpha}$,
\begin{equation}
\psi_{\alpha}^{L},\psi_{\dot{\alpha}}^{R},
\end{equation}
so there are two representations of gamma matrices $\left(  \Gamma^{M}\right)
_{\alpha}^{~\dot{\beta}}$ and $\left(  \bar{\Gamma}^{M}\right)  _{\dot{\alpha
}}^{~\beta}$ in the left / right Weyl bases. The gamma matrices must satisfy
the anticommutation rules%
\begin{equation}
\Gamma^{M}\bar{\Gamma}^{N}+\Gamma^{N}\bar{\Gamma}^{M}=2\eta^{MN}%
,\;\;\bar{\Gamma}^{M}\Gamma^{N}+\bar{\Gamma}^{N}\Gamma^{M}=2\eta^{MN}
\label{gid1}%
\end{equation}
where $\eta^{MN}$ is the SO$\left(  d,2\right)  $ metric with signature
$\eta^{MN}=$diag$\left(  -,-,+,+,\cdots,+\right)  .$ Then the correctly
normalized SO$\left(  d,2\right)  $ generator $J^{MN}=L^{MN}+S^{MN}$ is
represented on the two spinors by the spin $S^{MN}=\frac{1}{2i}\Gamma^{MN}$ or
$\frac{1}{2i}\bar{\Gamma}^{MN}$ where%
\begin{equation}
\Gamma^{MN}=\frac{1}{2}\left(  \Gamma^{M}\bar{\Gamma}^{N}-\Gamma^{N}%
\bar{\Gamma}^{M}\right)  ,\;\;\bar{\Gamma}^{MN}=\frac{1}{2}\left(  \bar
{\Gamma}^{M}\Gamma^{N}-\bar{\Gamma}^{N}\Gamma^{M}\right)  .
\end{equation}
Thus, when the $M,N$ indices are different one gets $\Gamma^{12}=\Gamma
^{1}\bar{\Gamma}^{2},$ etc. Similarly, antisymmetrized products of gamma
matrices applied on the two spinors are given by%

\begin{align}
\Gamma^{MNK}  &  =\frac{1}{3}\left(  \Gamma^{MN}\bar{\Gamma}^{K}+\Gamma
^{KM}\bar{\Gamma}^{N}+\Gamma^{NK}\bar{\Gamma}^{M}\right)  ,\;\\
\bar{\Gamma}^{MNK}  &  =\frac{1}{3}\left(  \bar{\Gamma}^{MN}\Gamma^{K}%
+\bar{\Gamma}^{KM}\Gamma^{N}+\bar{\Gamma}^{NK}\Gamma^{M}\right)  ,\text{ }\\
\Gamma_{MNKL}  &  =\frac{1}{4}\left(  \Gamma_{MNK}\Gamma_{L}-\Gamma
_{NKL}\Gamma_{M}+\Gamma_{KLM}\Gamma_{N}-\Gamma_{LMN}\Gamma_{K}\right)
,\;\text{etc.} \label{gid5}%
\end{align}
Thus, when the $M,N,K$ indices are different one gets $\Gamma^{123}=\Gamma
^{1}\bar{\Gamma}^{2}\Gamma^{3}$ and $\bar{\Gamma}^{123}=\bar{\Gamma}^{1}%
\Gamma^{2}\bar{\Gamma}^{3},$etc.

An explicit form of SO$\left(  d,2\right)  $ gamma matrices $\Gamma^{M}$ in
even dimensions, labeled by $M=0^{\prime},1^{\prime},\mu$ and $\mu=0,i$, is
given by
\begin{equation}
\Gamma^{0^{\prime}}=-i\tau_{1}\times1,\;\;\Gamma^{1^{\prime}}=\tau_{2}%
\times1,\;\;\Gamma^{0}=1\times1,\ \;\;\Gamma^{i}=\tau_{3}\times\gamma^{i},
\end{equation}
where $\gamma^{i}$ are the SO$\left(  d-1\right)  $ gamma matrices. The
$\bar{\Gamma}^{M}$ are the same as the $\Gamma^{M}$ for $M=0^{\prime
},1^{\prime},i,$ but for $M=0=\mu$ we have
\begin{equation}
\bar{\Gamma}^{0}=-\Gamma^{0}=-1\times1.
\end{equation}
It is useful to define a lightcone-type basis $X^{\pm^{\prime}}=\frac{1}%
{\sqrt{2}}(X^{0^{\prime}}\pm X^{1^{\prime}})$ and the corresponding gamma
matrices
\begin{equation}
\Gamma^{\pm^{\prime}}=\frac{1}{\sqrt{2}}\left(  \Gamma^{0^{\prime}}\pm
\Gamma^{1^{\prime}}\right)  =-i\sqrt{2}\tau^{\pm}\times1.
\end{equation}
In this basis the metric takes the form $ds^{2}=dX^{M}dX^{N}\eta
_{MN}=-2dX^{+^{\prime}}dX^{-^{\prime}}+dX^{\mu}dX^{\nu}\eta_{\mu\nu}$ where
$\eta_{\mu\nu}$ is the Minkowski metric for SO$\left(  d,1\right)  .$
Explicitly we write%
\begin{align}
\Gamma^{+^{\prime}}  &  =\left(
\genfrac{}{}{0pt}{}{0}{0}%
\genfrac{}{}{0pt}{}{-i\sqrt{2}}{0}%
\right)  ,\;\;\Gamma^{-^{\prime}}=\left(
\genfrac{}{}{0pt}{}{0}{-i\sqrt{2}}%
\genfrac{}{}{0pt}{}{0}{0}%
\right)  ,\;\Gamma^{\mu}=\left(
\genfrac{}{}{0pt}{}{\bar{\gamma}^{\mu}}{0}%
\genfrac{}{}{0pt}{}{0}{-\gamma^{\mu}}%
\right) \\
\bar{\Gamma}^{+^{\prime}}  &  =\left(
\genfrac{}{}{0pt}{}{0}{0}%
\genfrac{}{}{0pt}{}{-i\sqrt{2}}{0}%
\right)  ,\;\;\bar{\Gamma}^{-^{\prime}}=\left(
\genfrac{}{}{0pt}{}{0}{-i\sqrt{2}}%
\genfrac{}{}{0pt}{}{0}{0}%
\right)  ,\;\bar{\Gamma}^{\mu}=\left(
\genfrac{}{}{0pt}{}{\gamma^{\mu}}{0}%
\genfrac{}{}{0pt}{}{0}{-\bar{\gamma}^{\mu}}%
\right)
\end{align}
where
\begin{equation}
\gamma_{\mu}=\left(  1,\gamma_{i}\right)  ,\;\;\bar{\gamma}_{\mu}=\left(
-1,\gamma_{i}\right)  ,\;\text{or\ }\gamma^{\mu}=\left(  -1,\gamma^{i}\right)
,\;\;\bar{\gamma}^{\mu}=\left(  1,\gamma^{i}\right)  ,
\end{equation}
paying attention to the lower or upper $\mu$ indices since the SO$\left(
d,1\right)  $ metric is $\eta_{\mu\nu}=$diag$\left(  -1,1,1,\cdots,1\right)
$. It should be emphasized that the $\gamma^{\mu},\bar{\gamma}^{\mu}$ in
$\bar{\Gamma}^{\mu}$ are switched relative to $\Gamma^{\mu}.$ We can further
write
\begin{equation}
\gamma^{1}=\sigma^{1}\times1,\;\;\gamma^{2}=\sigma^{2}\times1,\;\;\gamma
^{r}=\sigma^{3}\times\rho^{r},
\end{equation}
where the $\rho^{r}$ are the gamma matrices for SO$\left(  d-3\right)  $. With
the explicit form of the gamma matrices above we have
\begin{gather}
\Gamma^{+^{\prime}-^{\prime}}=\left(
\genfrac{}{}{0pt}{}{-1}{0}%
\genfrac{}{}{0pt}{}{0}{1}%
\right)  ,\ \;\Gamma^{+^{\prime}\mu}=i\sqrt{2}\left(
\genfrac{}{}{0pt}{}{0}{0}%
\genfrac{}{}{0pt}{}{\bar{\gamma}^{\mu}}{0}%
\right)  ,\\
\Gamma^{\mu\nu}=\left(
\genfrac{}{}{0pt}{}{\bar{\gamma}^{\mu\nu}}{0}%
\genfrac{}{}{0pt}{}{0}{\gamma^{\mu\nu}}%
\right)  ,\;\Gamma^{-^{\prime}\mu}=-i\sqrt{2}\left(
\genfrac{}{}{0pt}{}{0}{\gamma^{\mu}}%
\genfrac{}{}{0pt}{}{0}{0}%
\right)  ,\
\end{gather}
similarly
\begin{gather}
\bar{\Gamma}^{+^{\prime}-^{\prime}}=\left(
\genfrac{}{}{0pt}{}{-1}{0}%
\genfrac{}{}{0pt}{}{0}{1}%
\right)  ,\ \;\bar{\Gamma}^{+^{\prime}\mu}=i\sqrt{2}\left(
\genfrac{}{}{0pt}{}{0}{0}%
\genfrac{}{}{0pt}{}{\gamma^{\mu}}{0}%
\right)  ,\ \;\\
\bar{\Gamma}^{\mu\nu}=\left(
\genfrac{}{}{0pt}{}{\gamma^{\mu\nu}}{0}%
\genfrac{}{}{0pt}{}{0}{\bar{\gamma}^{\mu\nu}}%
\right)  ,\;\bar{\Gamma}^{-^{\prime}\mu}=-i\sqrt{2}\left(
\genfrac{}{}{0pt}{}{0}{\bar{\gamma}^{\mu}}%
\genfrac{}{}{0pt}{}{0}{0}%
\right)  ,\ \;
\end{gather}
Then $X=X^{M}\Gamma_{M},$ $\bar{X}=X^{M}\bar{\Gamma}_{M},$ $\frac{1}{2}%
\Gamma_{MN}J^{MN}$, $\frac{1}{2}\bar{\Gamma}_{MN}J^{MN}$ etc. take explicit
matrix forms, such as%
\begin{equation}
X^{M}\Gamma_{M}=-X^{+^{\prime}}\Gamma^{-^{\prime}}-X^{-^{\prime}}%
\Gamma^{+^{\prime}}+X^{\mu}\Gamma_{\mu}=\left(
\genfrac{}{}{0pt}{}{X^{\mu}\bar{\gamma}_{\mu}}{i\sqrt{2}X^{+^{\prime}}}%
\genfrac{}{}{0pt}{}{i\sqrt{2}X^{-^{\prime}}}{-X^{\mu}\gamma_{\mu}}%
\right)
\end{equation}
and
\begin{align}
\frac{1}{2}\Gamma_{MN}J^{MN}  &  =-\Gamma^{+^{\prime}-^{\prime}}J^{+^{\prime
}-^{\prime}}+~\frac{1}{2}J_{\mu\nu}\Gamma^{\mu\nu}-\Gamma_{~\mu}^{+^{\prime}%
}J^{-^{\prime}\mu}-\Gamma_{~\mu}^{-^{\prime}}J^{+^{\prime}\mu}\\
&  =\left(
\begin{array}
[c]{cc}%
\frac{1}{2}J_{\mu\nu}\bar{\gamma}^{\mu\nu}+J^{+^{\prime}-^{\prime}} &
-i\sqrt{2}\bar{\gamma}_{\mu}J^{-^{\prime}\mu}\\
i\sqrt{2}\gamma_{\mu}J^{+^{\prime}\mu} & \frac{1}{2}J_{\mu\nu}\gamma^{\mu\nu
}-J^{+^{\prime}-^{\prime}}%
\end{array}
\right)  .
\end{align}
If we specialize to SO$\left(  4,2\right)  =$SU$\left(  2,2\right)  $ with
$d+2=6.$ Then the $\rho^{r}$ are replaced just by the number $1$ and then the
$\gamma_{\mu},\bar{\gamma}_{\mu}$ are given in terms of the $2\times2$ Pauli
matrices
\begin{equation}
\gamma_{\mu}=\left(  1,\vec{\sigma}\right)  ,\;\;\bar{\gamma}_{\mu}=\left(
-1,\vec{\sigma}\right)  ,\;\;\text{or}\ \;\;\gamma^{\mu}=\left(
-1,\vec{\sigma}\right)  ,\;\;\bar{\gamma}^{\mu}=\left(  1,\vec{\sigma}\right)
.
\end{equation}

\subsection{Metric, Hermitian conjugation}

To be specific we now specialize to SO$\left(  4,2\right)  =$SU$\left(
2,2\right)  $ with $d+2=6.$ The gamma matrices we have defined are consistent
with the metric $\eta^{\dot{\alpha}\beta}$ or $\eta^{\alpha\dot{\beta}}$ in
spinor space given as follows
\begin{align}
\eta^{\dot{\alpha}\beta} &  =-i\tau_{1}\times1=\Gamma^{0^{\prime}}=\bar
{\Gamma}^{0^{\prime}},\;\;\eta^{\alpha\dot{\beta}}=\Gamma^{0^{\prime}}%
=\bar{\Gamma}^{0^{\prime}}\\
\overline{\psi_{L}}^{\beta} &  =\left(  \psi_{L}^{\dagger}\right)
_{\dot{\alpha}}\eta^{\dot{\alpha}\beta}=\left(  \psi_{L}^{\dagger}\bar{\Gamma
}^{0^{\prime}}\right)  ^{\beta},\;\;\\
\overline{\psi_{R}}^{\dot{\beta}} &  =\left(  \psi_{R}^{\dagger}\right)
_{\alpha}\eta^{\alpha\dot{\beta}}=\left(  \psi_{R}^{\dagger}\Gamma^{0^{\prime
}}\right)  ^{\dot{\beta}}.
\end{align}
The metric $\eta$ has the following properties
\begin{equation}
\eta=\Gamma^{0^{\prime}},\;\eta^{2}=-1,\;\eta^{-1}=-\eta,\;\eta^{T}%
=\eta,\;\eta^{\dagger}=-\eta.
\end{equation}
We then note that
\begin{equation}
\Gamma^{0^{\prime}}\Gamma^{M}\Gamma^{0^{\prime}}=\left(  \overset{0^{\prime}%
}{i\tau_{1}\times1},\;\overset{1^{\prime}}{\tau_{2}\times1},\;\overset
{0}{-1\times1},\;\overset{i}{\tau_{3}\times\sigma^{i}},\;\right)  =\left(
\bar{\Gamma}^{M}\right)  ^{\dagger}\text{ .}\label{g0gg0}%
\end{equation}
From this we obtain the following properties
\begin{align}
\eta\Gamma^{M}\eta^{-1} &  =-\left(  \bar{\Gamma}^{M}\right)  ^{\dagger
}\;\text{and\ \ }\eta\bar{\Gamma}^{M}\eta^{-1}=-\left(  \bar{\Gamma}%
^{M}\right)  ^{\dagger}\label{herm11}\\
\eta\Gamma^{MN}\left(  \eta\right)  ^{-1} &  =-\left(  \Gamma^{MN}\right)
^{\dagger},\;\text{and\ \ }\eta\bar{\Gamma}^{MN}\eta^{-1}=-\left(  \bar
{\Gamma}^{MN}\right)  ^{\dagger}\label{herm22}\\
\eta\Gamma^{MN}\Gamma^{K}\eta^{-1} &  =\left(  \bar{\Gamma}^{K}\Gamma
^{MN}\right)  ^{\dagger},\;\text{and\ \ }\eta\Gamma^{K}\bar{\Gamma}^{MN}%
\eta^{-1}=\left(  \Gamma^{MN}\bar{\Gamma}^{K}\right)  ^{\dagger}%
\label{herm33}\\
\eta\Gamma^{MNK}\eta^{-1} &  =\left(  \bar{\Gamma}^{MNK}\right)  ^{\dagger
},\;\text{and\ \ }\eta\bar{\Gamma}^{MNK}\eta^{-1}=\left(  \Gamma^{MNK}\right)
^{\dagger}%
\end{align}
The second line is derived from the first line: $\eta\left(  \Gamma^{M}%
\bar{\Gamma}^{N}\right)  \eta^{-1}=\left(  \bar{\Gamma}^{M}\right)  ^{\dagger
}\left(  \Gamma^{N}\right)  ^{\dagger}=\left(  \Gamma^{N}\bar{\Gamma}%
^{M}\right)  ^{\dagger}$ which leads to $\left(  \eta\Gamma^{M}\bar{\Gamma
}^{N}\right)  =-\left(  \eta\Gamma^{N}\bar{\Gamma}^{M}\right)  ^{\dagger}$.
Similarly the third line is derived from the first and second lines
$\eta\Gamma^{MN}\Gamma^{K}\eta^{-1}=\left[  -\left(  \Gamma^{MN}\right)
^{\dagger}\right]  \left[  -\left(  \bar{\Gamma}^{K}\right)  ^{\dagger
}\right]  =\left(  \bar{\Gamma}^{K}\Gamma^{MN}\right)  ^{\dagger}$ etc., while
the fourth line follows from the third. Note that the patterns of $\Gamma
,\bar{\Gamma}$ on the left or right are not the same in each line. From these
we obtain the following properties of the matrices $\eta$, $\eta\Gamma^{M}$,
$\eta\Gamma^{MN}$, $\eta\Gamma^{MNK},$ etc. under Hermitian conjugation
\begin{equation}
\eta=-\eta^{\dagger},\;\eta\Gamma^{M}=\left(  \eta\bar{\Gamma}^{M}\right)
^{\dagger},\;\eta\Gamma^{MN}=\left(  \eta\Gamma^{MN}\right)  ^{\dagger}%
,\;\eta\Gamma^{MNK}=-\left(  \eta\bar{\Gamma}^{MNK}\right)  ^{\dagger
}\label{herm11a}%
\end{equation}
and similarly for $\eta\bar{\Gamma}^{M}$, $\eta\bar{\Gamma}^{MN}$, $\eta
\bar{\Gamma}^{MNK},$ etc.

Using these properties of the metric we obtain the following hermiticity
properties for pairs of fermions%
\begin{equation}%
\begin{array}
[c]{ll}%
\left(  \overline{\psi_{1L}}\psi_{2L}\right)  ^{\dagger}=-\overline{\psi_{2L}%
}\psi_{1L}, & \left(  \overline{\psi_{1L}}\Gamma^{M}\psi_{2R}\right)
^{\dagger}=\overline{\psi_{2R}}\bar{\Gamma}^{M}\psi_{1L},\\
\left(  \overline{\psi_{1L}}\Gamma^{M}\bar{\Gamma}^{N}\psi_{2L}\right)
^{\dagger}=-\overline{\psi_{2L}}\Gamma^{N}\bar{\Gamma}^{M}\psi_{1L},\; &
\left(  \overline{\psi_{1L}}\Gamma^{MN}\psi_{2L}\right)  ^{\dagger}%
=\overline{\psi_{2L}}\Gamma^{MN}\psi_{1L}\\
\left(  \overline{\psi_{1L}}\Gamma^{MN}\Gamma^{K}\psi_{2R}\right)  ^{\dagger
}=-\overline{\psi_{2R}}\bar{\Gamma}^{K}\Gamma^{MN}\psi_{1L}, & \left(
\overline{\psi_{1L}}\Gamma^{MNK}\psi_{2R}\right)  ^{\dagger}=-\overline
{\psi_{2R}}\Gamma^{MNK}\psi_{1L}%
\end{array}
\label{herm}%
\end{equation}
These are used to verify the hermiticity of the action, transformation
properties, and consistency of SU$\left(  2,2|1\right)  $ group theoretical
structures that appear in the text.

\subsection{Charge conjugation, transposition, Majorana spinors}

Next we define the charge conjugation matrix $C$ for SU$\left(  2,2\right)  $
by
\begin{equation}
C=\tau_{1}\times\sigma_{2}=-\tilde{C}\eta,\;\text{where }\ \tilde{C}\equiv
C\Gamma^{0^{\prime}}=-1\times i\sigma_{2}.
\end{equation}
It has the following properties
\begin{equation}
C=\tau_{1}\times\sigma_{2},\;\;C^{2}=1,\;C^{-1}=C,\;C^{T}=-C,\;C^{\dagger}=C.
\end{equation}
Then we see explicitly that%
\[
C\Gamma^{M}=\left(  \overset{0^{\prime}}{-1\times i\sigma_{2}},\;\overset
{1^{\prime}}{\tau_{3}\times i\sigma_{2}},\;\overset{0}{\tau_{1}\times
\sigma_{2}},\;\overset{i}{-\tau_{2}\times i\sigma_{2}\sigma^{i}}\right)
\]
which shows that $C\Gamma^{M}$ are all antisymmetric matrices. Similarly,
$C\bar{\Gamma}^{M}$ are also antisymmetric. Therefore we derive the following
properties%
\begin{equation}%
\begin{array}
[c]{ll}%
C\Gamma^{M}C^{-1}=\left(  \Gamma^{M}\right)  ^{T},\;\; & C\bar{\Gamma}%
^{M}C^{-1}=\left(  \bar{\Gamma}^{M}\right)  ^{T}\;\;\\
C\Gamma^{MN}C^{-1}=-\left(  \bar{\Gamma}^{MN}\right)  ^{T},\; & C\bar{\Gamma
}^{MN}C^{-1}=-\left(  \Gamma^{MN}\right)  ^{T}\;\\
C\Gamma^{MNK}C^{-1}=\left(  \Gamma^{MNK}\right)  ^{T},\;\; & C\bar{\Gamma
}^{MNK}C^{-1}=\left(  \bar{\Gamma}^{MNK}\right)  ^{T}\;
\end{array}
\label{Cgammas}%
\end{equation}
The second and third lines are derived from the first line: $C\left(
\Gamma^{M}\bar{\Gamma}^{N}\right)  C^{-1}=\left(  \Gamma^{M}\right)
^{T}\left(  \bar{\Gamma}^{N}\right)  ^{T}=\left(  \bar{\Gamma}^{N}\Gamma
^{M}\right)  ^{T}$ which leads to $\left(  C\Gamma^{M}\bar{\Gamma}^{N}\right)
=-\left(  C\bar{\Gamma}^{N}\Gamma^{M}\right)  ^{T},$ etc. Note that the
patterns of $\Gamma,\bar{\Gamma}$ on the left or right of each equation are
not the same in each line. From these we obtain the following properties of
$C\Gamma^{M},$ $C\Gamma^{MN},$ $C\Gamma^{MNK}$ etc. under transposition%
\begin{equation}
\left(  C\Gamma^{M}\right)  ^{T}=-\left(  C\Gamma^{M}\right)  ,\;\left(
C\Gamma^{MN}\right)  ^{T}=\left(  C\bar{\Gamma}^{MN}\right)  ,\;\left(
C\Gamma^{MNK}\right)  ^{T}=\left(  C\Gamma^{MNK}\right)  \label{Cgamma}%
\end{equation}
and similarly for $C\bar{\Gamma}^{M},$ $C\bar{\Gamma}^{MN},$ $C\bar{\Gamma
}^{MNK}$ .

The charge conjugate spinor of a left-handed spinor $\left(  \psi_{L}\right)
_{\alpha}$ (a $4$ of SU$\left(  2,2\right)  $) is a right handed spinor which
we denote as $\left(  \psi_{R}^{c}\right)  _{\dot{\alpha}}$ (a $\bar{4}$ of
SU$\left(  2,2\right)  $) and define it by%
\begin{equation}
\left(  \psi_{R}^{c}\right)  _{\dot{\alpha}}=\left(  C\overline{\psi_{L}}%
^{T}\right)  _{\dot{\alpha}}=C_{\dot{\alpha}\beta}\left(  \overline{\psi_{L}%
}^{T}\right)  ^{\beta}=\left(  C\left(  \Gamma^{0^{\prime}}\right)  ^{T}%
\psi_{L}^{\ast}\right)  _{\dot{\alpha}}=\left(  \tilde{C}\psi_{L}^{\ast
}\right)  _{\dot{\alpha}},\;
\end{equation}
with $\tilde{C}\equiv-1\times i\sigma_{2}.$ From this we extract $\left(
\psi_{R}^{c}\right)  ^{\ast}=\left(  \tilde{C}^{\ast}\psi_{L}\right)  =\left(
\tilde{C}\psi_{L}\right)  $ which gives the following form after multiplying
both sides with $\tilde{C}$
\begin{equation}
\psi_{L}=-\tilde{C}\left(  \psi_{R}^{c}\right)  ^{\ast}=-C\left(
\Gamma^{0^{\prime}}\right)  ^{T}\left(  \psi_{R}^{c}\right)  ^{\ast
}=-C\overline{\left(  \psi_{R}^{c}\right)  }^{T}.
\end{equation}
So, for consistency with these equations, the charge conjugate spinors need to
be defined with the following patterns of chiralities and signs%
\begin{align}
\left(  \psi_{R}^{c}\right)   &  =\left(  C\overline{\psi_{L}}^{T}\right)
,\;\;\;\psi_{L}=-C\overline{\left(  \psi_{R}^{c}\right)  }^{T},\\
\left(  \psi_{L}^{c}\right)   &  =-\left(  C\overline{\psi_{R}}^{T}\right)
,\;\;\;\psi_{R}=C\overline{\left(  \psi_{L}^{c}\right)  }^{T}.
\end{align}

We now define a Majorana fermion for SO$(4,2)$ as one that satisfies the
following condition
\begin{equation}
\psi_{L,R}^{c}\overset{Majorana}{=}\psi_{L,R}.
\end{equation}
Then from the above definition of $\psi_{L,R}^{c}$ we derive consistently that
a Majorana fermion has the following properties
\begin{gather}
\psi_{R}\overset{Majorana}{=}C\overline{\psi_{L}}^{T},\;\;\;\;\;\psi
_{L}\overset{Majorana}{=}-C\overline{\psi_{R}}^{T}\label{Maj}\\
\overline{\psi_{R}}\overset{Majorana}{=}\left(  \psi_{L}\right)
^{T}C,\;\;\;\overline{\psi_{L}}\overset{Majorana}{=}-\left(  \psi_{R}\right)
^{T}C \label{Majbar}%
\end{gather}
Using this and Eq.(\ref{Cgamma}) we now see the following permutation
properties of Majorana spinors when $\psi_{1},\psi_{2}$ are interchanged
(treated as anticommuting Grassmann numbers). Thus we find $\overline
{\psi_{1L}}\Gamma^{M}\psi_{2R}$ and $\overline{\psi_{1R}}\bar{\Gamma}^{M}%
\psi_{2L}$ are symmetric under the interchange of $1\leftrightarrow2$
\begin{align}
&  \overline{\psi_{1L}}\Gamma^{M}\psi_{2R}\overset{Majorana}{=}-\left(
\psi_{1R}\right)  ^{T}C\Gamma^{M}\psi_{2R}\label{maj1}\\
&  =-\left(  \psi_{2R}\right)  ^{T}C\Gamma^{M}\psi_{1R}=\overline{\psi_{2L}%
}\Gamma^{M}\psi_{1R},
\end{align}
With similar manipulations we establish the following properties under the
interchange of $1\leftrightarrow2$ for Majorana spinors. These follow from
Eqs.(\ref{Cgammas},\ref{Cgamma})
\begin{equation}%
\begin{array}
[c]{ll}%
\text{For Majorana~fermions only} & \overline{\psi_{1L}}\psi_{2L}%
=-\overline{\psi_{2R}}\psi_{1R}\\
\overline{\psi_{1L}}\Gamma^{M}\psi_{2R}=\overline{\psi_{2L}}\Gamma^{M}%
\psi_{1R}\;,\; & \overline{\psi_{1R}}\bar{\Gamma}^{M}\psi_{2L}=\overline
{\psi_{2R}}\bar{\Gamma}^{M}\psi_{1L},\\
\overline{\psi_{1L}}\Gamma^{M}\bar{\Gamma}^{N}\psi_{2L}=-\overline{\psi_{2R}%
}\bar{\Gamma}^{N}\Gamma^{M}\psi_{1R},\; & \overline{\psi_{1L}}\Gamma^{MN}%
\psi_{2L}=\overline{\psi_{2R}}\bar{\Gamma}^{MN}\psi_{1R}\\
\overline{\psi_{1L}}\Gamma^{MN}\Gamma^{K}\psi_{2R}=-\overline{\psi_{2L}}%
\Gamma^{K}\bar{\Gamma}^{MN}\psi_{1R},\; & \overline{\psi_{1R}}\bar{\Gamma
}^{MN}\bar{\Gamma}^{K}\psi_{2L}=-\overline{\psi_{2R}}\bar{\Gamma}^{K}%
\Gamma^{MN}\psi_{1L},\\
\overline{\psi_{1L}}\Gamma^{MNK}\psi_{2R}=-\overline{\psi_{2L}}\Gamma
^{MNK}\psi_{1R},\;\; & \overline{\psi_{1R}}\bar{\Gamma}^{MNK}\psi
_{2L}=-\overline{\psi_{2R}}\bar{\Gamma}^{MNK}\psi_{1L},
\end{array}
\label{maj5}%
\end{equation}
Note that for the gamma matrices $\Gamma^{M},$ $\Gamma^{MNK}$ the interchange
$1\leftrightarrow2$ is symmetric or antisymmetric, but for the gamma matrices
$1,\Gamma^{MN}$ the interchange $1\leftrightarrow2$ is neither symmetric nor
antisymmetric since left handed fermions are replaced by right handed ones,
and vice versa. These properties are used to manipulate various terms in
proving the SUSY properties of the action and to check the consistency of
SU$\left(  2,2|1\right)  $ group theoretical structures that appear in the
text. In particular, from the third line above we deduce the following
properties of the fermion kinetic term%
\begin{equation}
i\left(  \overline{\psi_{L}}X\bar{D}\psi_{L}+\overline{\psi_{L}}%
\overleftarrow{D}\bar{X}\psi_{L}\right)  \overset{Majorana}{=}-i\left(
\overline{\psi_{R}}\bar{X}D\psi_{R}+\overline{\psi_{R}}\overleftarrow{\bar{D}%
}X\psi_{R}\right)  .~ \label{maj6}%
\end{equation}
This agrees with the correct overall signs of the kinetic terms for fermions
of left/right chiralities.

\subsection{8-component SO(4,2) Majorana spinor}

Although we prefer to use the 4-component left or right SU$(2,2)$=SO$(4,2)$
spinor notation in this paper, for completeness and for future reference we
also discuss the 8-component spinor in this Appendix. The left or right
SU$(2,2)$=SO$(4,2)$ spinor can be rewritten as the 8-component Majorana spinor
of SO$(4,2)$. To see this we now introduce an $8$ dimensional Majorana spinor
$\psi$ and its conjugate $\bar{\psi}$ through the following definitions in
which $\psi_{L},\psi_{R}$ are related to each other as shown in Eqs.(\ref{Maj}%
,\ref{Majbar})%
\begin{align}
\psi &  =\left(
\begin{array}
[c]{c}%
\psi_{L}\\
\psi_{R}%
\end{array}
\right)  ,\;\;\bar{\psi}=\left(  \psi_{L}^{\dagger},\;\psi_{R}^{\dagger
}\right)  \left(
\begin{array}
[c]{cc}%
\eta & 0\\
0 & \bar{\eta}%
\end{array}
\right)  =\left(  \overline{\psi_{L}},\;\overline{\psi_{R}}\right)  ,\;\;\\
\psi^{c}  &  =\left(
\begin{array}
[c]{cc}%
0 & -C\\
C & 0
\end{array}
\right)  \left(
\begin{array}
[c]{c}%
\overline{\psi_{L}}^{T}\\
\overline{\psi_{R}}^{T}%
\end{array}
\right)  \overset{Majorana}{=}\left(
\begin{array}
[c]{c}%
\psi_{L}\\
\psi_{R}%
\end{array}
\right)  =\psi
\end{align}
Note that $\psi$ satisfies the Majorana condition $\psi^{c}=\psi.$ We can also
write (see Eq.(\ref{Majbar}))%
\begin{equation}
\bar{\psi}=\psi^{T}c,\text{ with }c=-\left(
\begin{array}
[c]{cc}%
0 & -C\\
C & 0
\end{array}
\right)  ,\;c^{T}=c \label{c}%
\end{equation}
Then $\bar{\psi}_{1}\gamma^{M_{1}\cdots M_{n}}\psi_{2}$ take the form%
\begin{align}
\bar{\psi}_{1}\gamma^{M}\psi_{2}  &  =\left(  \overline{\psi_{1L}}%
,\;\overline{\psi_{1R}}\right)  \left(
\begin{array}
[c]{cc}%
0 & \Gamma^{M}\\
\bar{\Gamma}^{M} & 0
\end{array}
\right)  \left(
\begin{array}
[c]{c}%
\psi_{2L}\\
\psi_{2R}%
\end{array}
\right) \\
\bar{\psi}_{1}\gamma^{MN}\psi_{2}  &  =\left(  \overline{\psi_{1L}%
},\;\overline{\psi_{1R}}\right)  \left(
\begin{array}
[c]{cc}%
\Gamma^{MN} & 0\\
0 & \bar{\Gamma}^{MN}%
\end{array}
\right)  \left(
\begin{array}
[c]{c}%
\psi_{2L}\\
\psi_{2R}%
\end{array}
\right)
\end{align}
etc. For 8-component Majorana spinors we obtain the following permutation
properties when $\psi_{1},\psi_{2}$ are interchanged%
\begin{align}
\bar{\psi}_{1}\psi_{2}  &  =\overline{\psi_{1L}}\psi_{2L}+\overline{\psi_{1R}%
}\psi_{2R}\\
&  :\overset{Majorana}{=}-\overline{\psi_{2R}}\psi_{1R}-\overline{\psi_{2L}%
}\psi_{1L}=-\bar{\psi}_{2}\psi_{1}\nonumber
\end{align}%
\begin{align}
\bar{\psi}_{1}\gamma^{M}\psi_{2}  &  =\overline{\psi_{1L}}\Gamma^{M}\psi
_{2R}+\overline{\psi_{1R}}\bar{\Gamma}^{M}\psi_{2L}\\
&  :\overset{Majorana}{=}\overline{\psi_{2L}}\Gamma^{M}\psi_{1R}%
+\overline{\psi_{2R}}\bar{\Gamma}^{M}\psi_{1L}=\bar{\psi}_{2}\gamma^{M}%
\psi_{1}\nonumber
\end{align}%
\begin{align}
\bar{\psi}_{1}\gamma^{MN}\psi_{2}  &  =\overline{\psi_{1L}}\Gamma^{MN}%
\psi_{2L}+\overline{\psi_{1R}}\bar{\Gamma}^{MN}\psi_{2R}\\
&  :\overset{Majorana}{=}\overline{\psi_{2R}}\bar{\Gamma}^{MN}\psi
_{1R}+\overline{\psi_{2L}}\Gamma^{MN}\psi_{1L}=\bar{\psi}_{2}\gamma^{MN}%
\psi_{1}\nonumber
\end{align}%
\begin{align}
\bar{\psi}_{1}\gamma^{MNK}\psi_{2}  &  =\overline{\psi_{1L}}\Gamma^{MNK}%
\psi_{2R}+\overline{\psi_{1R}}\bar{\Gamma}^{MNK}\psi_{2L}\\
&  :\overset{Majorana}{=}-\overline{\psi_{2L}}\Gamma^{MNK}\psi_{1R}%
-\overline{\psi_{2R}}\bar{\Gamma}^{MNK}\psi_{1L}=-\bar{\psi}_{2}\gamma
^{MNK}\psi_{1}\nonumber
\end{align}
In summary, $\bar{\psi}_{i}\gamma^{M_{1}\cdots M_{n}}\psi_{j}$ have the
following symmetry or antisymmetry properties under the interchange of $i,j$%
\begin{equation}%
\begin{array}
[c]{ll}%
\text{symmetric:} & \bar{\psi}_{i}\left(  \gamma^{M}\right)  \psi_{j}%
,\;\bar{\psi}_{i}\left(  \gamma^{MN}\right)  \psi_{j}\\
\text{antisymmetric:~} & \bar{\psi}_{i}\left(  1\right)  \psi_{j},\;\bar{\psi
}_{i}\left(  \gamma^{MNK}\right)  \psi_{j}%
\end{array}
\label{symasyA}%
\end{equation}

We can also introduce an additional $8\times8$ gamma matrix $\gamma
^{7}=\left(
\genfrac{}{}{0pt}{}{1}{0}%
\genfrac{}{}{0pt}{}{0}{-1}%
\right)  $ which anticommutes with the other six gamma matrices $\left\{
\gamma_{7},\gamma_{M}\right\}  =0,$ and construct $\gamma^{7}\gamma^{M}%
,\gamma^{7}\gamma^{MN}$ and $\gamma^{7}$ as the additional traceless
$8\times8$ gamma matrices that complete the set of all $8\times8$ matrices.
For these we have the following permutation properties%
\begin{align}
\bar{\psi}_{1}\gamma^{7}\psi_{2}  &  =\overline{\psi_{1L}}\psi_{2L}%
-\overline{\psi_{1R}}\psi_{2R}\\
&  :\overset{Majorana}{=}-\overline{\psi_{2R}}\psi_{1R}+\overline{\psi_{2L}%
}\psi_{1L}=\bar{\psi}_{2}\gamma^{7}\psi_{1}\nonumber
\end{align}%
\begin{align}
\bar{\psi}_{1}\gamma^{7}\gamma^{M}\psi_{2}  &  =\overline{\psi_{1L}}\Gamma
^{M}\psi_{2R}-\overline{\psi_{1R}}\bar{\Gamma}^{M}\psi_{2L}\nonumber\\
&  :\overset{Majorana}{=}\overline{\psi_{2L}}\Gamma^{M}\psi_{1R}%
-\overline{\psi_{2R}}\bar{\Gamma}^{M}\psi_{1L}=\bar{\psi}_{2}\gamma^{7}%
\gamma^{M}\psi_{1}%
\end{align}%
\begin{align}
\bar{\psi}_{1}\gamma^{7}\gamma^{MN}\psi_{2}  &  =\overline{\psi_{1L}}%
\Gamma^{MN}\psi_{2L}-\overline{\psi_{1R}}\bar{\Gamma}^{MN}\psi_{2R}\nonumber\\
&  :\overset{Majorana}{=}\overline{\psi_{2R}}\bar{\Gamma}^{MN}\psi
_{1R}-\overline{\psi_{2L}}\Gamma^{MN}\psi_{1L}=-\bar{\psi}_{2}\gamma^{7}%
\gamma^{MN}\psi_{1}%
\end{align}
In summary, $\bar{\psi}_{i}\gamma^{7}\gamma^{M_{1}\cdots M_{n}}\psi_{j}$ have
the following symmetry or antisymmetry properties under the interchange of
$\psi_{i},\psi_{j}$
\begin{equation}%
\begin{array}
[c]{ll}%
\text{symmetric:} & \bar{\psi}_{i}\left(  \gamma^{7}\right)  \psi_{j}%
,\;\bar{\psi}_{i}\left(  \gamma^{7}\gamma^{M}\right)  \psi_{j},\\
\text{antisymmetric:~} & \bar{\psi}_{i}\left(  \gamma^{7}\gamma^{MN}\right)
\psi_{j}%
\end{array}
\label{symasyB}%
\end{equation}

The symmetry or antisymmetry properties given above can be related to the
properties of SO$\left(  5,2\right)  $ gamma matrices given by $\gamma
^{m}=\left(  \gamma^{M},\gamma^{7}\right)  .$ Specifically we note the
$8\times8$ SO$\left(  5,2\right)  $ gamma matrices $c$ and $c\gamma^{mnk}$ are
symmetric while $c\gamma^{m},c\gamma^{mn}$ are antisymmetric, where $c$ is
given in Eq.(\ref{c}). This is easily understood by simple counting of
dimensions in spinor and vector spaces of SO(5,2). That is, the symmetric
products in spinor space gives $\left(  8\times8\right)  _{s}=\frac{8\cdot
9}{1\cdot2}=36,$ while for gamma matrices $c\oplus c\gamma^{mnk}$ we count the
same dimension, namely $1+\frac{7\cdot6\cdot5}{1\cdot2\cdot3}=36.$ Similarly
for the antisymmetric product in spinor space we have $\left(  8\times
8\right)  _{a}=\frac{8\cdot7}{1\cdot2}=28,$ while for $c\gamma^{m}\oplus
c\gamma^{mn}$ we count the same dimension, namely $7+\frac{7\cdot6}{1\cdot
2}=28.$ From this we immediately conclude that the SO$\left(  5,2\right)  $
gamma matrices have definite symmetry properties, namely $c\oplus
c\gamma^{mnk}$ are symmetric and $c\gamma^{m}\oplus c\gamma^{mn}$ are
antisymmetric. Taking into account that $\bar{\psi}=\psi^{T}c$ (see
Eq.(\ref{c})), and an extra minus sign due to the interchange of Grassmann
numbers, we obtain the permutation properties of fermion bilinears $\bar{\psi
}_{i}\left(  \gamma^{m_{1}\cdots m_{n}}\right)  \psi_{j}=\psi_{i}^{T}\left(
c\gamma^{m_{1}\cdots m_{n}}\right)  \psi_{j}$ under the interchange of
$\psi_{i},\psi_{j}$ as follows%
\begin{equation}%
\begin{array}
[c]{ll}%
\text{symmetric:} & \bar{\psi}_{j}\left(  \gamma^{m}\right)  \psi_{i}%
,\;\bar{\psi}_{i}\left(  \gamma^{mn}\right)  \psi_{j}\\
\text{antisymmetric:~} & \bar{\psi}_{i}\left(  1\right)  \psi_{j},\;\bar{\psi
}_{i}\left(  \gamma^{mnk}\right)  \psi_{j}%
\end{array}
\end{equation}
These SO$\left(  5,2\right)  $ properties reduce to the SO$\left(  4,2\right)
$ properties for Majorana fermions as given in Eqs.(\ref{symasyA}%
,\ref{symasyB}) by specializing the indices $m=\left(  M,7\right)  .$

The charge conjugation or Majorana properties described in this Appendix are
used to verify the SU$\left(  2,2|1\right)  $ group theoretical consistency of
the fermion bilinears that appear in the transformation laws and other
structures given in the text.

\section{Fierz Identities \label{fierzI}}

In this appendix we prove the two Fierz identities%
\begin{equation}
\delta\left(  X^{2}\right)  \frac{\partial^{3}W}{\partial\varphi_{i}%
\partial\varphi_{j}\partial\varphi_{k}}\left(  \overline{\psi_{Ri}}\bar{X}%
\psi_{Lk}\right)  \left(  \overline{\varepsilon_{R}}\bar{X}\psi_{Lj}\right)
=0. \label{fierz1}%
\end{equation}
and
\begin{equation}
\delta(X^{2})~f_{abc}\left(  \overline{\varepsilon_{L}}\left[  \Gamma_{M}%
,\bar{X}\right]  \lambda_{L}^{a}\right)  ~\left(  \overline{\lambda_{L}}%
^{b}\left[  \Gamma^{M},\bar{X}\right]  \lambda_{L}^{c}\right)  =0.
\label{fierz2}%
\end{equation}

We start with the gamma matrix identity of Eq.(\ref{Fierz}), which allows us
to write
\begin{align}
&  \left[  \left(  \overline{\psi_{Ri}}\bar{X}\right)  ^{\alpha}\delta
_{\alpha}^{\text{\ \ }\delta}\left(  \psi_{Lj}\right)  _{\delta}\right]
\left[  \left(  \overline{\varepsilon_{R}}\bar{X}\right)  ^{\gamma}%
\delta_{\gamma}^{\text{ \ }\beta}\left(  \psi_{Lk}\right)  _{\beta}\right]
\nonumber\\
&  =-\frac{1}{4}\left(  \overline{\psi_{Ri}}\bar{X}\psi_{Lk}\right)  \left(
\overline{\varepsilon_{R}}\bar{X}\psi_{Lj}\right)  +\frac{1}{8}\left(
\overline{\psi_{Ri}}\bar{X}\Gamma_{MN}\psi_{Lk}\right)  \left(  \overline
{\varepsilon_{R}}\bar{X}\Gamma^{MN}\psi_{Lj}\right)  .
\end{align}
This equation is rearranged by moving the first term on the right side to the
left side. After multiplying both sides with the the totally symmetric
$\frac{\partial^{3}W}{\partial\varphi_{i}\partial\varphi_{j}\partial
\varphi_{k}}$ and summing over $i,j,k$ we derive%
\begin{equation}
\frac{5}{4}\frac{\partial^{3}W}{\partial\varphi_{i}\partial\varphi_{j}%
\partial\varphi_{k}}\left(  \overline{\psi_{Ri}}\bar{X}\psi_{Lk}\right)
\left(  \overline{\varepsilon_{R}}\bar{X}\psi_{Lj}\right)  =\frac{1}{8}%
\frac{\partial^{3}W}{\partial\varphi_{i}\partial\varphi_{j}\partial\varphi
_{k}}\left(  \overline{\psi_{Ri}}\bar{X}\Gamma_{MN}\psi_{Lk}\right)  \left(
\overline{\varepsilon_{R}}\bar{X}\Gamma^{MN}\psi_{Lj}\right)  .
\end{equation}
We focus on the term $\overline{\psi_{Ri}}\bar{X}\Gamma_{MN}\psi_{Lk}$ on the
right hand side which can be rewritten by using gamma matrix identities as
$\overline{\psi_{Ri}}\bar{X}\Gamma_{MN}\psi_{Lk}=X^{P}\overline{\psi_{Ri}%
}\left(  \Gamma_{PMN}+\eta_{PM}\Gamma_{N}-\eta_{PN}\Gamma_{M}\right)
\psi_{Lk}.$ The $\Gamma_{PMN}$ term can be rewritten as $X^{P}\left(
\psi_{Li}\right)  ^{T}\left(  C\Gamma_{PMN}\right)  \psi_{Lk}$ by using the
charge conjugation property $\overline{\psi_{Ri}}=\left(  \psi_{Li}\right)
^{T}C$. We argue that the $\Gamma_{PMN}$ term can be dropped due to the
symmetric property of $\frac{\partial^{3}W}{\partial\varphi_{i}\partial
\varphi_{j}\partial\varphi_{k}}$\ under the interchange of $i$ and $k,$ the
symmetric property of $\left(  C\Gamma_{PMN}\right)  ^{\alpha\beta}$ under the
interchange of $\alpha$ and $\beta$, and the antisymmetry under the
interchange of two fermions. The remaining terms on the right hand side take
the form $\frac{1}{8}\frac{\partial^{3}W}{\partial\varphi_{i}\partial
\varphi_{j}\partial\varphi_{k}}\left(  \overline{\psi_{Ri}}X_{[M}\Gamma
_{N]}\psi_{Lk}\right)  \left(  \overline{\varepsilon_{R}}X^{[M}\Gamma^{N]}%
\psi_{Lj}\right)  .$ We drop the term proportional to $X^{2}$ since there is
an overall $\delta\left(  X^{2}\right)  $. Then we obtain the relation%
\begin{equation}
\frac{5}{4}\delta\left(  X^{2}\right)  \frac{\partial^{3}W}{\partial
\varphi_{i}\partial\varphi_{j}\partial\varphi_{k}}\left(  \overline{\psi_{Ri}%
}\bar{X}\psi_{Lk}\right)  \left(  \overline{\varepsilon_{R}}\bar{X}\psi
_{Lj}\right)  =-\frac{1}{2}\delta\left(  X^{2}\right)  \frac{\partial^{3}%
W}{\partial\varphi_{i}\partial\varphi_{j}\partial\varphi_{k}}\left(
\overline{\psi_{Ri}}\bar{X}\psi_{Lk}\right)  \left(  \overline{\varepsilon
_{R}}\bar{X}\psi_{Lj}\right)
\end{equation}
Pulling all terms to the same side of the equation and rearranging, we obtain
the desired Fierz identity of Eq.(\ref{fierz1}).

Next we prove the Fierz identity of Eq.(\ref{fierz2}) where $\left[
\Gamma_{M},\bar{X}\right]  $ is \textit{defined} as
\begin{equation}
\left[  \Gamma_{M},\bar{X}\right]  \equiv\Gamma_{M}\bar{X}-X\bar{\Gamma}%
_{M}=2\Gamma_{MN}X^{N}.
\end{equation}
We start again with the gamma matrix identity of Eq.(\ref{Fierz}) to write%
\begin{align}
&  f_{abc}\left(  \overline{\varepsilon_{L}}\left[  \Gamma_{M},\bar{X}\right]
\lambda_{L}^{a}\right)  \;\left(  \overline{\lambda_{L}}^{b}\left[  \Gamma
^{M},\bar{X}\right]  \lambda_{L}^{c}\right) \nonumber\\
&  =f_{abc}\left[
\begin{array}
[c]{c}%
\frac{1}{8}(\overline{\lambda}_{bL}\left[  \Gamma^{M},\bar{X}\right]
\Gamma^{RQ}\lambda_{aL})(\overline{\varepsilon_{L}}\left[  \Gamma_{M},\bar
{X}\right]  \Gamma_{RQ}\lambda_{cL})\\
-\frac{1}{4}(\overline{\lambda}_{bL}\left[  \Gamma^{M},\bar{X}\right]
\lambda_{aL})(\overline{\varepsilon_{L}}\left[  \Gamma_{M},\bar{X}\right]
\lambda_{cL})
\end{array}
\right]  \label{f1}%
\end{align}
The last term on the right side has the same form as the left side, so the
equation is rearranged as%
\begin{align}
&  \frac{3}{4}f_{abc}\overline{\varepsilon_{L}}\left[  \Gamma_{M},\bar
{X}\right]  \lambda_{L}^{a}\;\overline{\lambda_{L}}^{b}\left[  \Gamma^{M}%
,\bar{X}\right]  \lambda_{L}^{c}\nonumber\\
&  =\frac{1}{8}f_{abc}(\overline{\lambda}_{bL}\left[  \Gamma^{M},\bar
{X}\right]  \Gamma^{RQ}\lambda_{aL})(\overline{\varepsilon_{L}}\left[
\Gamma_{M},\bar{X}\right]  \Gamma_{RQ}\lambda_{cL})
\end{align}
By using gamma matrix identities (\ref{gidentities}), and setting $X^{2}$
terms to zero since there is an overall $\delta\left(  X^{2}\right)  ,$ the
right hand side can be rewritten as
\begin{equation}
\frac{4}{8}f^{abc}(\overline{\lambda}_{bL}\Gamma^{MNRQ}\lambda_{aL}%
)(\overline{\varepsilon_{L}}\Gamma_{MPRQ}\lambda_{cL})X_{N}X^{P}+\frac{4}%
{8}f^{abc}(\overline{\lambda}_{bL}\left[  \Gamma^{M},\bar{X}\right]
\lambda_{aL})(\overline{\varepsilon_{L}}\left[  \Gamma_{M},\bar{X}\right]
\lambda_{cL}). \label{f2}%
\end{equation}
The last term in Eq.(\ref{f2}) is similar to the left side of Eq.(\ref{f1}),
so they combine, and we are left with%
\begin{equation}
\frac{1}{4}\delta\left(  X^{2}\right)  f_{abc}\overline{\varepsilon_{L}%
}\left[  \Gamma_{M},\bar{X}\right]  \lambda_{L}^{a}\;\overline{\lambda_{L}%
}^{b}\left[  \Gamma^{M},\bar{X}\right]  \lambda_{L}^{c}=\frac{4}{8}%
\delta\left(  X^{2}\right)  f^{abc}X_{N}X^{P}(\overline{\lambda}_{bL}%
\Gamma^{MNRQ}\lambda_{aL})(\overline{\varepsilon_{L}}\Gamma_{MPRQ}\lambda
_{cL}). \label{fierz3}%
\end{equation}
Next we use the fact that $\Gamma^{M_{1}M_{2}M_{3}M_{4}}=\frac{1}%
{2}\varepsilon^{M_{1}M_{2}M_{3}M_{4}M_{5}M_{6}}\Gamma_{M_{5}M_{6}}$ and we
perform the sum
\begin{equation}
\left(  \frac{1}{2}\right)  ^{2}X_{N}X^{P}\varepsilon^{MNRQM_{5}M_{6}%
}\varepsilon_{MPRQN_{5}N_{6}}=\frac{3!}{4}\delta_{\lbrack P}^{N}\delta_{N_{5}%
}^{M_{5}}\delta_{N_{6}]}^{M_{6}}X_{N}X^{P}.
\end{equation}
Here we drop the terms proportional to $X^{2}$ since there is an overall delta
function $\delta\left(  X^{2}\right)  .$ Inserting this in Eq.(\ref{fierz3}),
we find the right hand side of (\ref{fierz3}) looks the same as the left side
of (\ref{fierz3}), but with the numerical coefficient $-3/4$ on the right
versus $1/4$ on the left. Pulling all terms to the same side of
Eq.(\ref{fierz3}) we obtain the desired Fierz identity of Eq.(\ref{fierz2}).

\section{Off-shell Closure for Chiral Supermultiplet \label{B}}

We now consider the closure of the SUSY transformations $\left(
\delta_{\varepsilon_{1}}\delta_{\varepsilon_{2}}-\delta_{\varepsilon_{2}%
}\delta_{\varepsilon_{1}}\right)  $ applied on each field in the chiral
multiplet $\left(  \varphi,\psi_{L},F\right)  _{i},$ in the absence of
interactions with the vector multiplet (i.e. $g=0$), with each $\delta
_{\varepsilon}$ given in Eqs.(\ref{susy1}-\ref{susy3}).

\subsection{Closure for Scalars}%

\begin{align}
\delta_{\lbrack\varepsilon_{1}}\delta_{\varepsilon_{2}]}\varphi &
=\overline{\varepsilon_{R}}_{[2}\bar{X}\left(  \delta_{\varepsilon_{1]}}%
\psi_{L}\right)  -\frac{1}{2}X^{2}\left(  \overline{\varepsilon_{R}}_{[2}%
\bar{\partial}\left(  \delta_{\varepsilon_{1]}}\psi_{L}\right)  +\overline
{\varepsilon_{R}}_{[2}U^{\dagger}\left(  \delta_{\varepsilon_{1]}}\psi
_{R}\right)  \right) \\
&  =\left\{
\begin{array}
[c]{c}%
i\overline{\varepsilon_{R}}_{[2}\bar{\Gamma}^{MN}\varepsilon_{R1]}%
X_{M}\partial_{N}\varphi+i\overline{\varepsilon_{R}}_{[2}\varepsilon
_{R1]}\left[  X\cdot\partial\varphi-\frac{1}{2}X^{2}\partial^{2}\varphi\right]
\\
-i\overline{\varepsilon_{R}}_{[2}\bar{X}\varepsilon_{L1]}F+\frac{1}{2}%
X^{2}i\overline{\varepsilon_{R}}_{[2}\bar{\Gamma}^{M}\varepsilon_{L1]}%
\partial_{M}F\\
-\frac{1}{2}X^{2}i\overline{\varepsilon_{R}}_{[2}\varepsilon_{R1]}U^{\dagger
}F^{\dagger}+\frac{1}{2}X^{2}i\overline{\varepsilon_{R}}_{[2}\bar{\Gamma}%
^{M}\varepsilon_{L1]}U^{\dagger}\partial_{M}\varphi
\end{array}
\right\} \\
&  =\left\{
\begin{array}
[c]{c}%
-\frac{1}{2}\overline{\varepsilon_{R}}_{[2}\bar{\Gamma}^{MN}\varepsilon
_{R1]}L_{MN}\varphi-i\overline{\varepsilon_{R}}_{[2}\varepsilon_{R1]}\varphi\\
+i\overline{\varepsilon_{R}}_{[2}\varepsilon_{R1]}\left(  X\cdot
\partial+1\right)  \varphi-\frac{1}{2}X^{2}i\overline{\varepsilon_{R}}%
_{[2}\varepsilon_{R1]}\left(  \partial^{2}\varphi+U^{\dagger}F^{\dagger
}\right)
\end{array}
\right\}  \label{closurescalar}%
\end{align}
where we have used
\begin{equation}
\overline{\varepsilon_{R}}_{[2}\bar{\Gamma}^{M}\varepsilon_{L1]}=0
\end{equation}
from the second line to the third line.

From Appendix (\ref{A}), one can conclude that $\overline{\varepsilon_{R}%
}_{[2}\Gamma^{MN}\varepsilon_{R1]}$ and $i\overline{\varepsilon_{R}}%
_{[2}\varepsilon_{R1]}$ are imaginary numbers. These effective parameters in
the first line of (\ref{closurescalar}) are interpreted as the closure to the
global bosonic subgroup SU$(2,2)\times$U$\left(  1\right)  \subseteq
$SU$\left(  2,2|1\right)  $, where the U$\left(  1\right)  $ is the so-called
$R$-symmetry. The second line (\ref{closurescalar}) is proportional to the
2T-gauge symmetry generators connected to the phase space constraints $X\cdot
P$ and $X^{2}$. So these terms in the closure are 2T-gauge transformations of
the scalar field \cite{2tstandardM}. If the field is partially on shell by
setting $X^{2}=0$ and $\left(  X\cdot\partial+1\right)  \varphi=0,$ to satisfy
these constraints (derived as equations of motion in Eq.(\ref{homo1})), then
the closure for such fields is purely into the bosonic subgroup of SU$\left(
2,2|1\right)  .$

This makes it clear that for fields that satisfy the Sp$\left(  2,R\right)  $
gauge invariance conditions (i.e. partially on-shell), the closure is into
SU$(2,2)\times$U$\left(  1\right)  \subseteq$SU$\left(  2,2|1\right)  .$
However, for general off-shell fields the closure of two SUSY transformations
is into the global SU$(2,2)\times$U$\left(  1\right)  \subseteq$SU$\left(
2,2|1\right)  ,$ plus 2T-gauge transformations connected to the underlying
Sp$\left(  2,R\right)  $ \cite{2tstandardM}. The same pattern is observed for
the other components of the chiral multiplet as follows.

\subsection{Closure for auxiliary fields}

The closure of the auxiliary field $F$ works as usual,%
\begin{align}
\delta_{\lbrack\varepsilon_{1}}\delta_{\varepsilon_{2}]}F  &  =-\overline
{\varepsilon_{R}}_{[2}\left(  \frac{1}{2i}\Gamma^{MN}L_{MN}+2\right)
\delta_{\varepsilon_{1}]}\psi_{L}\\
&  =\left\{
\begin{array}
[c]{c}%
\frac{1}{2}\overline{\varepsilon_{R}}_{[2}\Gamma^{MN}\varepsilon_{L1]}%
L_{MN}F+2i\overline{\varepsilon_{R}}_{[2}\varepsilon_{L1]}F-2i\overline
{\varepsilon_{R}}_{[2}\bar{\Gamma}^{M}\varepsilon_{R1]}\partial_{M}\varphi\\
-\frac{1}{2}\overline{\varepsilon_{R}}_{[2}\Gamma^{MNP}\varepsilon_{L1]}%
L_{MN}\partial_{p}\varphi-\frac{1}{2}\overline{\varepsilon_{R}}_{[2}%
\bar{\Gamma}^{[M}\varepsilon_{R1]}\eta^{N]P}L_{MN}\partial_{p}\varphi
\end{array}
\right\} \\
&  =-\frac{1}{2}\overline{\varepsilon_{R}}_{[2}\bar{\Gamma}^{MN}%
\varepsilon_{R1]}L_{MN}F+2i\overline{\varepsilon_{R}}_{[2}\varepsilon_{R1]}F
\end{align}
where we have used (see Appendix (\ref{A}))%
\begin{equation}
\overline{\varepsilon_{R}}_{[2}\varepsilon_{R1]}=\overline{\varepsilon_{R}%
}_{[2}\varepsilon_{L1]} \label{exchangeLR0}%
\end{equation}%
\begin{equation}
\overline{\varepsilon_{R}}_{[2}\bar{\Gamma}^{MN}\varepsilon_{R1]}%
=-\overline{\varepsilon_{R}}_{[2}\Gamma^{MN}\varepsilon_{L1]}
\label{exchangeLR2}%
\end{equation}
The closure on $F$ consists again of the global bosonic subgroup
SU$(2,2)\times$U$\left(  1\right)  \subseteq$SU$\left(  2,2|1\right)  .$

\subsection{Closure for Spinors}

To calculate the closure on the spinor, we use the following Fierz identities
which will be derived later in this subsection.%
\begin{equation}
\overline{\varepsilon_{R}}_{[1}\bar{\Gamma}^{M}\psi_{L}\Gamma_{M}%
\varepsilon_{R2]}=-\frac{3}{2}\overline{\varepsilon_{R}}_{[1}\varepsilon
_{R2]}\psi_{L}+\frac{1}{4}\overline{\varepsilon_{R}}_{[1}\bar{\Gamma}%
^{MN}\varepsilon_{R2]}\Gamma_{MN}\psi_{L}, \label{closurespinor1}%
\end{equation}
and%

\begin{equation}
\overline{\varepsilon_{R}}_{[1}\bar{\Gamma}^{M}\left(  X_{M}\partial_{N}%
\psi_{L}\right)  \Gamma^{N}\varepsilon_{R2]}=\left[
\begin{array}
[c]{c}%
-\frac{1}{4}\overline{\varepsilon_{R}}_{[1}\bar{\Gamma}^{MN}\varepsilon
_{R2]}X_{M}\partial_{N}\psi_{L}-\frac{1}{4}\overline{\varepsilon_{R}}%
_{[1}\varepsilon_{R2]}X\cdot\partial\psi_{L}\\
+\frac{1}{8}\overline{\varepsilon_{R}}_{[1}\bar{\Gamma}^{PRMN}\varepsilon
_{R2]}\Gamma_{PR}X_{M}\partial_{N}\psi_{L}\\
+\frac{1}{4}\overline{\varepsilon_{R}}_{[1}\bar{\Gamma}^{MR}\varepsilon
_{R2]}\Gamma_{R}X_{M}\partial\psi_{L}-\frac{1}{4}\overline{\varepsilon_{R}%
}_{[1}\bar{\Gamma}^{MN}\varepsilon_{R2]}X_{M}\partial_{N}\psi_{L}\\
+\frac{1}{4}\overline{\varepsilon_{R}}_{[1}\bar{\Gamma}^{PN}\varepsilon
_{R2]}X\Gamma_{P}\partial_{N}\psi_{L}-\frac{1}{4}\overline{\varepsilon_{R}%
}_{[1}\bar{\Gamma}^{MN}\varepsilon_{R2]}X_{M}\partial_{N}\psi_{L}\\
+\frac{1}{8}\overline{\varepsilon_{R}}_{[1}\bar{\Gamma}^{MN}\varepsilon
_{R2]}\Gamma_{MN}X\cdot\partial\psi_{L}\\
+\frac{1}{4}\overline{\varepsilon_{R}}_{[1}\varepsilon_{R2]}X\partial\psi
_{L}-\frac{1}{4}\overline{\varepsilon_{R}}_{[1}\varepsilon_{R2]}X\cdot
\partial\psi_{L}%
\end{array}
\right]  , \label{closurespinor2}%
\end{equation}
and%
\begin{equation}
\overline{\varepsilon_{L}}_{[1}\psi_{L}\varepsilon_{L2]}=-\frac{1}{4}%
\overline{\varepsilon_{R}}_{[1}\varepsilon_{R2]}\psi_{L}-\frac{1}{8}%
\Gamma_{PR}\psi_{L}\overline{\varepsilon_{R}}_{[1}\bar{\Gamma}^{PR}%
\varepsilon_{R2]}, \label{closurespinors3}%
\end{equation}
and%
\begin{equation}
\overline{\varepsilon_{L}}_{[1}\Gamma^{MN}\left(  X_{M}\partial_{N}\psi
_{L}\right)  \varepsilon_{L2]}=\left\{
\begin{array}
[c]{c}%
\frac{1}{4}\overline{\varepsilon_{R}}_{[1}\bar{\Gamma}^{MN}\varepsilon
_{R2]}\left(  X_{M}\partial_{N}\psi_{L}\right)  +\frac{1}{8}\overline
{\varepsilon_{R}}_{[1}\bar{\Gamma}^{MNPR}\varepsilon_{R2]}\Gamma_{PR}\left(
X_{M}\partial_{N}\psi_{L}\right) \\
-\frac{1}{4}\overline{\varepsilon_{R}}_{[1}\bar{\Gamma}^{MR}\varepsilon
_{R2]}\Gamma_{\text{ \ }R}^{P}X_{M}\partial_{P}\psi_{L}+\frac{1}{4}%
\overline{\varepsilon_{R}}_{[1}\bar{\Gamma}^{MR}\varepsilon_{R2]}%
\Gamma_{\text{ \ }R}^{P}X_{P}\partial_{M}\psi_{L}\\
-\frac{1}{4}\overline{\varepsilon_{R}}_{[1}\varepsilon_{R2]}\Gamma^{MN}%
X_{M}\partial_{N}\psi_{L}%
\end{array}
\right\}  . \label{closurespinor4}%
\end{equation}
Then we compute%
\begin{align}
\delta_{\lbrack\varepsilon_{1}}\delta_{\varepsilon_{2}]}\psi_{L}  &
=i\partial(\delta_{\lbrack\varepsilon_{1}}\varphi)\varepsilon_{R2]}%
-i\delta_{\lbrack\varepsilon_{1}}F\varepsilon_{L2]}\\
&  =\left\{
\begin{array}
[c]{c}%
i\overline{\varepsilon_{R}}_{[1}\partial\left[  \bar{X}\psi_{L}-\frac{1}%
{2}X^{2}(\bar{\partial}\psi_{L}+U^{\dagger}\psi_{R})\right]  \varepsilon
_{R2]}\\
-i\overline{\varepsilon_{L}}_{[1}(\Gamma^{MN}X_{M}\partial_{N}-2)\psi
_{L}\varepsilon_{R2]}%
\end{array}
\right\}
\end{align}
This becomes%
\begin{equation}
=\left\{
\begin{array}
[c]{c}%
\left[  i\overline{\varepsilon_{R}}_{[1}\bar{\Gamma}^{M}\psi_{L}\Gamma
_{M}\varepsilon_{R2]}\right]  +\left[  i\overline{\varepsilon_{R}}_{[1}%
\bar{\Gamma}^{M}\left(  X_{M}\partial_{N}\psi_{L}\right)  \Gamma
^{N}\varepsilon_{R2]}\right] \\
\left[  -i\overline{\varepsilon_{L}}_{[1}\Gamma^{MN}X_{M}\partial_{N}\psi
_{L}\varepsilon_{R2]}\right]  +\left[  2i\overline{\varepsilon_{L}}_{[1}%
\psi_{L}\varepsilon_{R2]}\right]  +\left(  X\zeta+X^{2}\varrho\right)
\end{array}
\right\}
\end{equation}
Here we note that everything of the form $X\zeta+X^{2}\varrho$ in the
transformation of $\psi_{L}$ is a 2T-gauge transformation of the spinor
\cite{2tstandardM}.

This can further be put into the form%
\begin{align}
\delta_{\lbrack\varepsilon_{1}}\delta_{\varepsilon_{2}]}\psi_{L}  &  =\left\{
-\frac{1}{2}\overline{\varepsilon_{R}}_{[2}\bar{\Gamma}^{MN}\varepsilon
_{R1]}\left(  L_{MN}+\frac{1}{2i}\Gamma_{MN}\right)  \psi_{L}+\frac{i}%
{2}\overline{\varepsilon_{R}}_{[2}\varepsilon_{R1]}\psi_{L}\right\} \\
&  +\left\{
\begin{array}
[c]{c}%
-\frac{i}{8}\overline{\varepsilon_{R}}_{[2}\bar{\Gamma}^{MN}\varepsilon
_{R1]}\Gamma_{MN}\left(  X\cdot\partial+2\right)  \psi_{L}\\
\frac{3}{4}i\overline{\varepsilon_{R}}_{[2}\varepsilon_{R1]}\left(
X\cdot\partial+2\right)  \psi_{L}+\left(  X\zeta+X^{2}\varrho\right)
\end{array}
\right\}
\end{align}
In this form, we see that the first bracket represents the closure into the
bosonic subgroup SU$(2,2)\times$U$\left(  1\right)  \subseteq$SU$\left(
2,2|1\right)  ,$ with the correct SU$\left(  2,2\right)  $ generator $\left(
L_{MN}+\frac{1}{2i}\Gamma_{MN}\right)  $ for the spin $1/2$ fermion. The
second bracket is again a 2T-gauge transformation since $\left(
X\cdot\partial+2\right)  \psi_{L}$ is the action of the Sp$\left(  2,R\right)
$ generator $X\cdot P$ on the fermion \cite{2tstandardM}. For a partially
on-shell homogeneous field $\left(  X\cdot\partial+2\right)  \psi_{L}=0$ that
is Sp$\left(  2,R\right)  $ gauge invariant (which is an equation of motion at
$g=0$ as in Eq.(\ref{homo2})), the second bracket drops out. Hence for
Sp$\left(  2,R\right)  $ gauge invariant fields the closure is purely into the
bosonic subgroup of SU$\left(  2,2|1\right)  $.

If we gauge fix the 2T gauge symmetry as in footnote (\ref{embedding}), the
transformations become the familiar hidden superconformal symmetry of $N=1$
chiral multiplet.

\subsubsection{Proof of the identities C10-C13}

The first two identities are proved as follows. Using the Fierz identity in
Eq.(\ref{Fierz}), we can write%
\begin{equation}
\overline{\varepsilon_{R}}_{[1}\bar{\Gamma}^{M}\psi_{L}\Gamma^{N}%
\varepsilon_{R2]}=-\frac{1}{4}\overline{\varepsilon_{R}}_{[1}\bar{\Gamma}%
^{M}\Gamma^{N}\varepsilon_{R2]}\psi_{L}+\frac{1}{8}\overline{\varepsilon_{R}%
}_{[1}\bar{\Gamma}^{M}\Gamma^{PR}\Gamma^{N}\varepsilon_{R2]}\Gamma_{PR}%
\psi_{L}.
\end{equation}
Using the commutation relation $[\bar{\Gamma}^{M},\Gamma^{PR}]=2\eta^{MP}%
\bar{\Gamma}^{R}-2\eta^{MR}\bar{\Gamma}^{P}$ we change the order of
$\bar{\Gamma}^{M}$ and $\Gamma^{PR}$ for the second term on the right hand
side. After that, use $\bar{\Gamma}^{M}\Gamma^{N}=\bar{\Gamma}^{MN}-\eta^{MN}$
and (\ref{gidentities}), to get,%
\begin{equation}
\overline{\varepsilon_{R}}_{[1}\bar{\Gamma}^{M}\psi_{L}\Gamma^{N}%
\varepsilon_{R2]}=\left\{
\begin{array}
[c]{c}%
-\frac{1}{4}\overline{\varepsilon_{R}}_{[1}\bar{\Gamma}^{MN}\varepsilon
_{R2]}\psi_{L}-\frac{1}{4}\overline{\varepsilon_{R}}_{[1}\varepsilon_{R2]}%
\eta^{MN}\psi_{L}+\frac{1}{8}\overline{\varepsilon_{R}}_{[1}\bar{\Gamma
}^{PRMN}\varepsilon_{R2]}\Gamma_{PR}\psi_{L}\\
-\frac{1}{4}\overline{\varepsilon_{R}}_{[1}\bar{\Gamma}^{P\{M}\varepsilon
_{R2]}\Gamma_{P}^{\text{ \ \ }N\}}\psi_{L}+\frac{1}{8}\eta^{MN}\overline
{\varepsilon_{R}}_{[1}\bar{\Gamma}^{PR}\varepsilon_{R2]}\Gamma_{PR}\psi
_{L}+\frac{1}{4}\overline{\varepsilon_{R}}_{[1}\varepsilon_{R2]}\Gamma
^{MN}\psi_{L}%
\end{array}
\right\}  \label{closurespinor}%
\end{equation}
Then we can use (\ref{closurespinor}) to derive the first two identities
\[
\overline{\varepsilon_{R}}_{[1}\bar{\Gamma}^{M}\psi_{L}\Gamma_{M}%
\varepsilon_{R2]}=-\frac{3}{2}\overline{\varepsilon_{R}}_{[1}\varepsilon
_{R2]}\psi_{L}+\frac{1}{4}\overline{\varepsilon_{R}}_{[1}\bar{\Gamma}%
^{MN}\varepsilon_{R2]}\Gamma_{MN}\psi_{L},
\]
and%
\begin{align}
&  \overline{\varepsilon_{R}}_{[1}\bar{\Gamma}^{M}\left(  X_{M}\partial
_{N}\psi_{L}\right)  \Gamma^{N}\varepsilon_{R2]}\nonumber\\
&  =\left[
\begin{array}
[c]{c}%
-\frac{1}{4}\overline{\varepsilon_{R}}_{[1}\bar{\Gamma}^{MN}\varepsilon
_{R2]}X_{M}\partial_{N}\psi_{L}-\frac{1}{4}\overline{\varepsilon_{R}}%
_{[1}\varepsilon_{R2]}X\cdot\partial\psi_{L}\\
+\frac{1}{8}\overline{\varepsilon_{R}}_{[1}\bar{\Gamma}^{PRMN}\varepsilon
_{R2]}\Gamma_{PR}X_{M}\partial_{N}\psi_{L}\\
-\frac{1}{4}\overline{\varepsilon_{R}}_{[1}\bar{\Gamma}^{PM}\varepsilon
_{R2]}\Gamma_{PN}X_{M}\partial^{N}\psi_{L}\\
+\frac{1}{4}\overline{\varepsilon_{R}}_{[1}\bar{\Gamma}^{PN}\varepsilon
_{R2]}\Gamma_{MP}X^{M}\partial_{N}\psi_{L}\\
+\frac{1}{8}\overline{\varepsilon_{R}}_{[1}\bar{\Gamma}^{MN}\varepsilon
_{R2]}\Gamma_{MN}X\cdot\partial\psi_{L}\\
+\frac{1}{4}\overline{\varepsilon_{R}}_{[1}\varepsilon_{R2]}\Gamma^{MN}%
X_{M}\partial_{N}\psi_{L}%
\end{array}
\right]  ,.
\end{align}%
\begin{equation}
=\left[
\begin{array}
[c]{c}%
-\frac{1}{4}\overline{\varepsilon_{R}}_{[1}\bar{\Gamma}^{MN}\varepsilon
_{R2]}X_{M}\partial_{N}\psi_{L}-\frac{1}{4}\overline{\varepsilon_{R}}%
_{[1}\varepsilon_{R2]}X\cdot\partial\psi_{L}\\
+\frac{1}{8}\overline{\varepsilon_{R}}_{[1}\bar{\Gamma}^{PRMN}\varepsilon
_{R2]}\Gamma_{PR}X_{M}\partial_{N}\psi_{L}\\
+\frac{1}{4}\overline{\varepsilon_{R}}_{[1}\bar{\Gamma}^{MR}\varepsilon
_{R2]}\Gamma_{R}X_{M}\partial\psi_{L}-\frac{1}{4}\overline{\varepsilon_{R}%
}_{[1}\bar{\Gamma}^{MN}\varepsilon_{R2]}X_{M}\partial_{N}\psi_{L}\\
+\frac{1}{4}\overline{\varepsilon_{R}}_{[1}\bar{\Gamma}^{PN}\varepsilon
_{R2]}X\Gamma_{P}\partial_{N}\psi_{L}-\frac{1}{4}\overline{\varepsilon_{R}%
}_{[1}\bar{\Gamma}^{MN}\varepsilon_{R2]}X_{M}\partial_{N}\psi_{L}\\
+\frac{1}{8}\overline{\varepsilon_{R}}_{[1}\bar{\Gamma}^{MN}\varepsilon
_{R2]}\Gamma_{MN}X\cdot\partial\psi_{L}\\
+\frac{1}{4}\overline{\varepsilon_{R}}_{[1}\varepsilon_{R2]}X\partial\psi
_{L}-\frac{1}{4}\overline{\varepsilon_{R}}_{[1}\varepsilon_{R2]}X\cdot
\partial\psi_{L}%
\end{array}
\right]
\end{equation}
On the other hand, using the Fierz identities (\ref{exchangeLR0}) and
(\ref{exchangeLR2}) we can easily derive%
\begin{align}
\overline{\varepsilon_{L}}_{[1}\psi_{L}\varepsilon_{L2]} &  =-\frac{1}%
{4}\overline{\varepsilon_{L}}_{[1}\varepsilon_{L2]}\psi_{L}+\frac{1}{8}%
\Gamma_{PR}\psi_{L}\overline{\varepsilon_{L}}_{[1}\Gamma^{PR}\varepsilon
_{L2]},\\
&  =-\frac{1}{4}\overline{\varepsilon_{R}}_{[1}\varepsilon_{R2]}\psi_{L}%
-\frac{1}{8}\Gamma_{PR}\psi_{L}\overline{\varepsilon_{R}}_{[1}\bar{\Gamma
}^{PR}\varepsilon_{R2]}.
\end{align}
Now let's tackle the last identity. First, we use the Fierz identity,%
\begin{align}
&  \overline{\varepsilon_{L}}_{[1}\Gamma^{MN}\left(  X_{M}\partial_{N}\psi
_{L}\right)  \varepsilon_{L2]}\\
&  =-\frac{1}{4}\overline{\varepsilon_{L}}_{[1}\Gamma^{MN}\varepsilon
_{L2]}\left(  X_{M}\partial_{N}\psi_{L}\right)  +\frac{1}{8}\overline
{\varepsilon_{L}}_{[1}\Gamma^{MN}\Gamma^{PR}\varepsilon_{L2]}\Gamma
_{PR}\left(  X_{M}\partial_{N}\psi_{L}\right)
\end{align}
Then we use (\ref{gidentities}) to get%
\begin{equation}
\overline{\varepsilon_{L}}_{[1}\Gamma^{MN}\left(  X_{M}\partial_{N}\psi
_{L}\right)  \varepsilon_{L2]}=\left\{
\begin{array}
[c]{c}%
\frac{1}{4}\overline{\varepsilon_{R}}_{[1}\Gamma^{MN}\varepsilon_{R2]}\left(
X_{M}\partial_{N}\psi_{L}\right)  +\frac{1}{8}\overline{\varepsilon_{L}}%
_{[1}\Gamma^{MNPR}\varepsilon_{L2]}\Gamma_{PR}\left(  X_{M}\partial_{N}%
\psi_{L}\right)  \\
+\frac{1}{4}\overline{\varepsilon_{L}}_{[1}\Gamma^{MR}\varepsilon_{L2]}%
\Gamma_{\text{ \ }R}^{P}X_{M}\partial_{P}\psi_{L}-\frac{1}{4}\overline
{\varepsilon_{L}}_{[1}\Gamma^{MR}\varepsilon_{L2]}\Gamma_{\text{ \ }R}%
^{P}X_{P}\partial_{M}\psi_{L}\\
-\frac{1}{4}\overline{\varepsilon_{L}}_{[1}\varepsilon_{L2]}\Gamma^{MN}%
X_{M}\partial_{N}\psi_{L}%
\end{array}
\right\}  .
\end{equation}
Finally, we use (\ref{exchangeLR0}, \ref{exchangeLR2}) to change left-handed
spinors to the charge-conjugated right-handed spinors, and similarly%
\begin{align}
\overline{\varepsilon_{L}}_{[1}\Gamma^{MNPR}\varepsilon_{L2]} &
=-i\epsilon^{MNPRQS}\overline{\varepsilon_{L}}_{[1}\Gamma_{QS}\varepsilon
_{L2]},\\
&  =i\epsilon^{MNPRQS}\overline{\varepsilon_{R}}_{[1}\bar{\Gamma}%
_{QS}\varepsilon_{R2]},\\
&  =\overline{\varepsilon_{R}}_{[1}\bar{\Gamma}^{MNPR}\varepsilon_{R2]},
\end{align}
to obtain the last identity.

\end{document}